\documentclass{aa}
\usepackage{graphicx}
\usepackage{natbib}
\bibpunct{(}{)}{;}{a}{}{,}
\newcommand{\bm}{\begin{math}}
\newcommand{\enm}{\end{math}}
\newcommand{\bdm}{\begin{displaymath}}
\newcommand{\edm}{\end{displaymath}}
\newcommand{\be}{\begin{equation}}
\newcommand{\ee}{\end{equation}}
\newcommand{\bea}{\begin{eqnarray}}
\newcommand{\eea}{\end{eqnarray}}
\newcommand{\ba}{\begin{array}}
\newcommand{\ea}{\end{array}}
\newcommand{\btbb}{\begin{tabbing}}
\newcommand{\etbb}{\end{tabbing}}
\newcommand{\btab}{\begin{tabular}}
\newcommand{\etab}{\end{tabular}}
\newcommand{\bfi}{\begin{figure}}
\newcommand{\efi}{\end{figure}}

\newcommand{\bfid}{\begin{figure*}}
\newcommand{\efid}{\end{figure*}}
\newcommand{\btabl}{\begin{table}}
\newcommand{\etabl}{\end{table}}
\newcommand{\btabld}{\begin{table*}}
\newcommand{\etabld}{\end{table*}}
\newcommand{\bc}{\begin{center}}
\newcommand{\ec}{\end{center}}
\newcommand{\bab}{\begin{abstract}}
\newcommand{\eab}{\end{abstract}}
\newcommand{\bi}{\begin{itemize}}
\newcommand{\ei}{\end{itemize}}
\newcommand{\pa}{\partial}
\newcommand{\bbib}{\begin{thebibliography}}
\newcommand{\ebib}{\end{thebibliography}}
\newcommand{\ed}{\end{document}}
\def\gapprox{\;\rlap{\lower 2.5pt            % > ungefaehr =
 \hbox{$\sim$}}\raise 1.5pt\hbox{$>$}\;}       
\def\lapprox{\;\rlap{\lower 2.5pt            % < ungefaehr =
 \hbox{$\sim$}}\raise 1.5pt\hbox{$<$}\;} 
\begin{document}
%
%
%\thesaurus{02(02.19.1;02.09.1;02.20.1;02.08.1;09.11.1;08.23.3)}
%
%
\title{Supersonic turbulence in
       shock-bound interaction zones I: symmetric settings}
\author{Doris Folini\inst{1}  \and Rolf Walder\inst{2,3} }
\institute{Institut f\"ur Astronomie, ETH Z\"urich, CH-8092 Z\"urich,
           Switzerland; 
           \hspace{0.3cm} E-mail: folini@astro.phys.ethz.ch,
           \and 
           Observatoire de Strasbourg,
           67000 Strasbourg, France; \hspace{0.3cm}  E-mail: walder@astro.phys.ethz.ch
           \and 
           Max-Planck-Institut f\"{u}r Astrophysik, 85741 Garching, Germany;
           }
\offprints{D. Folini}
\date{Received ... ; accepted ...}
%
%
%
%%%%%%%%%%%%%%%%%%%%%%%%%%%%%  Abstract  %%%%%%%%%%%%%%%%%%%%%%%%%%%%%%
%
%
\abstract{
  Colliding hypersonic flows play a decisive role in many astrophysical
  objects. They contribute, for example, to the molecular cloud structure, the
  X-ray emission of O-stars, differentiation of galactic sheets, appearance of
  wind-driven structures, or, possibly, to the prompt emission of $\gamma$-ray
  bursts. Our intention is thorough investigation of the turbulent interaction
  zone of such flows, the cold dense layer (CDL).  In this paper, we focus on
  the idealized model of a 2D plane parallel isothermal slab and on symmetric
  settings, where both flows have equal parameters.  We performed a set of
  high-resolution simulations with upwind Mach-numbers, $5 < M_{\mathrm{u}} <
  90$.
  
  We find that the CDL is irregularly shaped and has a patchy and filamentary
  interior. The size of these structures increases with $\ell_{\mathrm{cdl}}$,
  the extension of the CDL. On average, but not at each moment, the solution
  is nearly self-similar and only depends on $M_{\mathrm{u}}$. We give the
  corresponding analytical expressions, with numerical constants derived from
  the simulation results. In particular, we find the root-mean-square
  Mach-number to scale as $M_{\mathrm{rms}} \approx 0.2 M_{\mathrm{u}}$.  The
  mean density, $\rho_{\mathrm{m}} \approx 30 \rho_{\mathrm{u}}$ is
  independent of $M_{\mathrm{u}}$. The fraction $f_{\mathrm{eff}}$ of the
  upwind kinetic energy that survives shock passage scales as
  $f_{\mathrm{eff}}= 1 - M_{\mathrm{rms}}^{-0.6}$. This dependence persists if
  the upwind flow parameters differ from one side to the other of the CDL,
  indicating that the turbulence within the CDL and its driving are mutually
  coupled. Another finding points in the same direction, namely that the
  auto-correlation length of the confining shocks and the characteristic
  length scale of the turbulence within the CDL are proportional.  Larger
  upstream Mach-numbers lead to a faster expanding CDL, confining interfaces
  that are less inclined with respect to the upstream flow direction, more
  efficient driving, and finer interior structure with respect to the
  extension of the CDL.
}
\authorrunning{Folini \& Walder }
\titlerunning{Compressible turbulence in shock-bounded interaction zones}
\maketitle
\section{Introduction}
\label{sec:intro}
Supersonically turbulent, shock-bound interaction zones are important
for a variety of astrophysical objects. They contribute, for example,
to structure formation in molecular clouds~\citep{hunter-et-al:86,
  ballesteros-hartmann-vazquez:99, hartmann-et-al:01, hueckstaedt:03,
  heyer-brunt:04, vazquez-semadeni:04} and to galaxy
formation~\citep{anninos-norman:96, kang-et-al:05}.  They affect the
X-ray emission of line-driven hot-star winds~\citep{owocki-et-al:88,
  feldmeier-et-al:97, feldmeier-owocki:98, oskinova-et-al:04} and
contribute substantially to the physics and emitted spectrum of
colliding wind binaries~\citep{stevens-et-al:92, nussbaumer-walder:93,
  folini-walder:00, marchenko-et-al:03, corcoran-et-al:05}. The
currently most promising model for the prompt emission of $\gamma$-ray
bursts is based on internal shocks~\citep{rees-meszaros:94,
  panaitescu-et-al:99, piran:04, fan-wei:04}. A similar mechanism
has been proposed for micro-quasars~\citep{kaiser-et-al:00}, BL Lacs
and Blazars~\citep{ghisellini-et-al:02, mimica-et-al:04}, and 
Herbig-Haro objects~\citep{matzner-mckee:99}.

So far, the shape and turbulent interior of shock-bound interaction zones have
been mostly studied separately. In this paper we focus on the system as a
whole, stressing that upwind flows, confining interfaces of the interaction
zone, and the interior structure of this zone form a tightly coupled system.
The turbulence within the interaction zone affects the shape of the confining
shocks, which in turn determines how much energy is thermalized at these
shocks and how much energy remains available for driving the turbulence.

A variety of papers have been written on the shape and stability of 2D
interaction zones, of which we mention only a few.  \citet{vishniac:94} shows
by analytical means that geometrically thin, isothermal, 2D, planar,
shock-bounded slabs are non-linearly unstable, coining the term non-linear
thin shell instability, or NTSI, for this instability.
\citet{blondin-marks:96} essentially reproduce these analytical predictions
numerically, also mentioning the occurrence of supersonic turbulence within
the slab. Performing 2D radiative and isothermal simulations of colliding
molecular clouds, \citet{klein-woods-tod:98} observe the complex shaping and
instability of the collision zone. The role of a radiative cooling layer has
been addressed by several authors.  \citet{strickland-blondin:95} numerically
investigated flows against a wall in 2D, finding that an unstable cooling
layer introduces disturbances in the interface separating the cooling layer
from the cooled matter.  Looking at colliding flows instead of a flow against
a wall, \citet{walder-folini:98} show that one unstable cooling layer is
sufficient to destabilize both confining interfaces of the cooled matter. In
addition, the cooled matter becomes supersonically turbulent. If self-gravity
is included fragmentation of the interaction zone is observed
\citep{anninos-norman:96, hunter-et-al:86}.

An overwhelming amount of literature meanwhile exists on supersonic
turbulence. At least part of this attention arises because it is thought that
supersonic turbulence can explain the structuring and support of molecular
clouds and thus that it plays a decisive role in star formation. A
comprehensive view of this issue can be found in the recent reviews
by~\citet{maclow-klessen:04}, \citet{elmegreen-scalo:04},
and~\citet{scalo-elmegreen:04}. Of particular interest for the work we present
here is the paper by \citet{maclow:99}, where Fig. 4 shows that the
wave length of the driving is apparent in the spatial scale of the turbulent
structure for monochromatically driven turbulence in a 3D periodic box.  The
possible importance of the finite size of the slab was recently pointed out by
\citet{burkert-hartmann:04}.

We are trying to make four points with this paper. First, we argue that,
within the frame of isothermal Euler equations and in infinite space, the
solution may be self-similar and dependent only on the upstream Mach-number,
at least to first approximation. Based on this assumption, we give expressions
for average quantities of the slab.  Second, we show that the numerical
solution, which is defined only on a finite computational domain and includes
(implicit) numerical dissipation, remains close to self-similar, as long as the
width of the slab is small and the root-mean-square Mach-number larger than
one.  Third, we stress the tight mutual coupling between the turbulence and
its driving. Fourth, we point out that spatial scales generally grow with 
extension $\ell_{\mathrm{cdl}}$ of the interaction zone, but decrease with
increasing upstream Mach-number $M_{\mathrm{u}}$.

Results are based on a set of simulations that differ only in their upwind
Mach-numbers. In this paper we restrict the analysis of these simulations to
the above-mentioned three objectives. We postpone a more detailed analysis of
the interior structure of the interaction zone to a subsequent paper.

In the following, we first give the details of our physical model and
numerical method in Sect.~\ref{sec:runs_and_tools} . In
Sect.~\ref{sec:anal-scaling} we derive the self-similar scaling relations. The
numerical results are present in Sect.~\ref{sec:num_results}.  Discussion
follows in Sect.~\ref{sec:discussion}, and conclusions in
Sect.~\ref{sec:conc}.
\section{Physical model and numerical method}
\label{sec:runs_and_tools}
The numerical treatment of supersonic turbulence is an issue in its
own right, so we start this section with a brief summary of some
results that are relevant to the present work.  We then specify the
physical model we consider, explain the numerical method we use and
the simulations we perform.

\subsection{Simulating supersonic turbulence}
\label{sec:simulating}
The shock-compressed layer studied in this paper is supersonically turbulent
with root-mean-square Mach-numbers between about 1 and 10.  An important
fraction of the kinetic energy is dissipated in shocks.  Euler equations are
sufficient for describing this part of the problem. A cascade transfers the
remaining energy to higher and higher wave numbers until it is finally
destroyed on the viscous dissipation scale. To also capture this part of the
problem, the compressible Navier-Stokes equations should be used;  however,
the range of spatial scales associated with the energy cascade exceeds the
capacity of any computer by far.
  
In subsonic turbulence, one way out is to use a suitable sub-grid scale model.
The model is used to compute an effective viscosity coefficient, which should
mimic the cascading between the smallest scale still resolved by the numerical
grid and the viscous dissipation scale as precisely as possible. This
coefficient is then used in the Navier-Stokes equations instead of the
physical viscosity~\citep{lesieur:99}. For the approach to work it
is essential that the effective viscosity obtained from the sub-grid scale
model exceeds the (implicit) numerical viscosity of the overall numerical
scheme. This can be achieved in subsonic turbulence by the use of
low-dissipation schemes~\citep{lele:92}.

In supersonic turbulence, explicit sub-grid scale modeling so far does not
exist in the above sense. The basic reason is that the numerical treatment of
supersonic turbulence requires schemes that can treat shocks appropriately,
such as the widely used shock capturing schemes. The (implicit) numerical
viscosity of such schemes is, however, much too large to match the above
requirement, even if the schemes are of a high order \citep{garnier-et-al:99,
  porter-et-al:92}. One strategy for this case, the so called MILES approach
(monotone integrated large-eddy simulation), was proposed
by~\citet{boris-et-al:92} and further explored
by~\citet{porter-et-al:92,porter-et-al:94}. The basic claim is that the
numerical viscosity inherent to shock capturing schemes \citep{hirsch:95,
  leveque:02} acts already as a physically correct sub-grid scale model.
Solving the Euler equations by means of a shock capturing scheme thus should
yield the correct physical answer.

The validity of the claim that implicit numerical viscosity alone leads to a
correct physical solution was investigated by~\citet{garnier-et-al:99} for a
selection of shock capturing schemes, among them a MUSCL-scheme (monotone
upwind scheme for conservation laws) similar to the one we use (see
Sect.~\ref{sec:num_meth}). For the cases considered (essentially decaying
subsonic), they find that the scheme indeed acts as a (very dissipative)
sub-grid scale model in that it preserves the flow from energy accumulation on
small spatial scales. However, they also find that structures defined on less
than 5 grid points are affected by substantial numerical damping.
\citet{porter-et-al:94} find, in addition, that the dissipation properties of
their scheme (MUSCL with PPM) are highly non-linear, and also they depend not
only on the grid spacing but also on the wave length of the flow structure.
Structures on less than 32 grid points are affected by numerical damping.

We rely on the MILES approach in this paper for the lack of a better model,
although, to our knowledge, the validity and quality of the approach has never
been tested for supersonic turbulence. The numerical solutions we obtain are
thus rather solutions of the Navier-Stokes equations.  Nevertheless, as
dissipation in shocks by far dominates numerical dissipation, we expect the
'Euler character' of the solution to prevail.
\subsection{The model problem}
\label{sec:phys_model}
\begin{figure}[tp]
\centerline{\includegraphics[width=8.5cm]{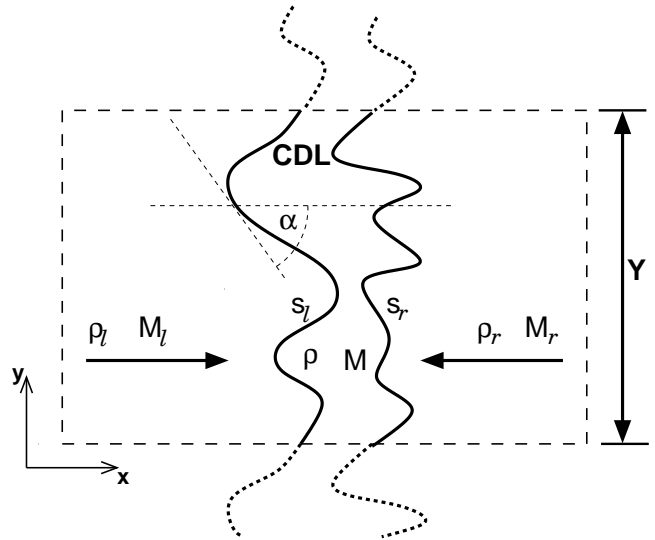}}
\caption{Sketch of physical model problem. $\rho_{\mathrm{i}}$,  
  $M_{\mathrm{i}}$, and $s_{\mathrm{i}}$ denote the density, Mach-number, and
  confining shock of the left ($i=l$) and right ($i=r$) flow. $\rho$ and $M$
  denote the density and Mach-number of the CDL.  $\alpha$ is the absolute
  value of the angle between the x-axis and the tangent to the shock. CDL is
  the shock-compressed interaction zone. The dashed rectangle indicates the
  computational domain with y-extension Y.  Periodic boundary conditions in
  y-direction imply periodic continuation of the solution (dotted continuation
  of left and right shock).}
\label{fig:sketch2d}
\end{figure}

The model problem we consider consists of a 2D, plane-parallel, infinitely
extended, isothermal, shock compressed slab. A sketch is given in
Fig.~\ref{fig:sketch2d}. Two high Mach-number flows, oriented parallel (left
flow, subscript $l$) and anti-parallel (right flow, subscript $r$) to the
x-direction, collide head on. The resulting high-density interaction zone, the
shock compressed slab, is oriented in the y-direction. We denote this
interaction zone by CDL for `cold dense layer' to remain consistent with
notation used already in~\citet{walder-folini:96,walder-folini:98}.  We
investigated this system within the frame of Euler equations (but see also
Sect.~\ref{sec:simulating}), together with a polytropic equation of state,
\begin{eqnarray}
\frac{\pa  \rho}{\pa t} + \vec{\nabla} \left( \rho \vec{v} \right) & = & 0, \\
\frac{\pa \rho \vec{v}}{\pa t} + \vec{\nabla} \left( \rho \vec{v} \otimes \vec{v} + \frac{p}{\mu} I \right) & = & 0, \\
\frac{\pa E}{\pa t} + \vec{\nabla} ( \vec{v} \left(E + p \right) ) & = & 0, \\
e & = & p/(\gamma - 1) .
\end{eqnarray}
Here, $\rho$ is the particle density, $\mu$ the average mass per particle,
$\vec{v} = (v_\mathrm{x},v_\mathrm{y})$ is the velocity vector, $p$ thermal
pressure, $I$ the identity tensor, $e$ the thermal energy density, and $E=\rho
\vec{v}^{2}/2 + e$ the total energy density. For the polytropic exponent, we
choose $\gamma = 1.000001$.  This value guarantees that jump conditions and
wave speeds of a Mach-90 shock are within 0.01 per cent of the isothermal
values.

Within the frame of this paper we consider only symmetric settings,
where the left (subscript $l$) and right (subscript $r$) colliding
flow have identical parameters (subscript $u$ for upstream):
$\rho_{\mathrm{l}} = \rho_{\mathrm{r}} \equiv \rho_{\mathrm{u}}$ and
$|v_{\mathrm{l}}| = |v_{\mathrm{r}}| \equiv v_{\mathrm{u}}$.

We look at the problem in a dimensionless form and express velocities in units
of the isothermal sound speed $a=\sqrt{T k_{\mathrm{B}}/\mu}$, with $T$ the
temperature and $k_{\mathrm{B}}$ the Boltzmann constant.  Densities we express
in terms of the upstream density $\rho_{\mathrm{u}}$. Finally, we express
lengths in units of $\mathrm{Y}_{\mathrm{0}}$, the smallest y-extent of the
computational domain we used. This artificial choice is necessary as
there is no natural time-independent length scale to the problem (see
Sect.~\ref{sec:anal-scaling}).
\subsection{Numerical method}
\label{sec:num_meth}
Our results were with the AMRCART-code\footnote{AMRCART is
  part of the A-MAZE code-package~\citep{amaze:00}, which contains 3D
  adaptive MHD and radiative transfer codes. The package, along with a
  brief description, is
  publicly available at \\
  http://www.astro.phys.ethz.ch/staff/folini/folini.html or \\
  http://www.astro.phys.ethz.ch/staff/walder/walder.html.}. We used
the multidimensional high-resolution finite-volume-integration scheme
developed by~\citet{colella:90} on the basis of a Cartesian mesh.
Tests showed that this algorithm, compared to dimensional splitting
schemes, is significantly more accurate in capturing flow features not
aligned with the axis of the mesh. In all our simulations we used a
version of the scheme that is (formally) second order accurate in
space and in time for smooth flows.

We combine this integration scheme with the adaptive mesh algorithm
by~\citet{berger:85}. While a rather coarse mesh was sufficient for the upwind
flows, the turbulent CDL was resolved on a much finer scale.

We found it useful to have our CDL moving in positive x-direction at a speed
of about Mach 20-40. If the CDL was essentially stationary with respect to the
computational grid, we observed alignment effects of strong shocks that were
nearly parallel to a cell interface (in y-direction). Through the global
motion of the CDL, which implied supersonic motion of the confining shocks
with respect to the computational grid, we got rid of this problem.  We
checked that this procedure introduced no systematic effects into the
solution. The problem of alignment effects when dealing with high Mach-number
flows, nearly stationary shocks, and high order upwind schemes is well known
and not particular to our scheme~\citep{colella-woodward:84, quirk:94,
  jasak-weller:95}. Other work arounds exist, such as smoothing of interfaces
by additional viscosity, which is often applied in PPM implementations.

\subsubsection{Numerical settings and integration time}
\label{sec:num_settings}
In the x-direction, our computational domain extended over $200\,
\mathrm{Y}_{\mathrm{0}}$.  The y-extent $\mathrm{Y}$ of our domain
varied between simulations, $\mathrm{Y}_{\mathrm{0}} \le \mathrm{Y}
\le 6\, \mathrm{Y}_{\mathrm{0}}$ (see Table~\ref{tab:list_of_runs}).
Boundary conditions at the left and right boundaries (x-direction) were
`supersonic inflow'. In the y-direction we had periodic boundary
conditions.  The cell size at the coarsest level was $0.2 \,
\mathrm{Y}_{\mathrm{0}}$. The cells at the finest level, covering the
CDL, were smaller by a factor $2^{6}$ to $2^{9}$, yielding between 320
and 2560 cells over a distance $\mathrm{Y}_{\mathrm{0}}$ (depending
on the simulation, see Table~\ref{tab:list_of_runs}).

As will be shown, the relevant time-dependent quantity for the
evolution of CDL mean quantities is the average x-extension of the
CDL, $\ell_{\mathrm{cdl}}$.  We defined it as $\ell_{\mathrm{cdl}}
\equiv V / \mathrm{Y}$, where $V$ is the 2D volume of the CDL. For
later use we also introduce the volume integrated density
$m_{\mathrm{cdl}} \equiv \int_{\mathrm{V}} \rho$, the mean density
$\rho_{\mathrm{m}} \equiv m_{\mathrm{cdl}} / V$, and the average
column density $ N \equiv m_{\mathrm{cdl}}/\mathrm{Y} =
\rho_{\mathrm{m}} \ell_{\mathrm{cdl}}$. The last quantity was made
dimensionless by division through $ N_{\mathrm{0}} \equiv
\rho_{\mathrm{u}}\mathrm{Y}_{\mathrm{0}}$. We stopped most simulations 
at $\ell_{\mathrm{cdl}} = \mathrm{Y}/2$.
\subsubsection{Initial conditions}
\label{sec:initial_conditions}
We investigated three different initial conditions, I=0,1,2.

{\bf I=0:} No CDL exists at $t=0$. The left and right flows are
initially separated by a single interface. The interface is wiggled
with a single, sinusoidal mode of wave length $0.1\, \mathrm{Y}$ and amplitude
$0.0195\, \mathrm{Y}_{\mathrm{0}}$ (about 3 to 25 grid cells, depending on the
discretization).

{\bf I=1:} A CDL is present at time $t=0$. It has a column density of
$ N  = 14 \,  N_{\mathrm{0}}$ and a thickness of
$0.03125\, \mathrm{Y}_{\mathrm{0}}$. The confining shocks are both wiggled, with
the same sinusoidal mode and amplitude as the interface in the case
I=0. The mass within the CDL is at rest and of constant density, $\rho
= \rho_{\mathrm{u}} M_{\mathrm{u}}^{2}$, the density the CDL would
have in 1D.  Note that this initialization implies some violation of
the Rankine-Hugoniot jump conditions at the interfaces.

{\bf I=2:} A CDL is present at time $t=0$, with column density $N =
56\, N_{\mathrm{0}}$ and a thickness of $0.125\,
\mathrm{Y}_{\mathrm{0}}$. The right shock is wiggled as for I=1, the
left shock is straight. The density and velocity in the CDL are set as
for I=1.

We stress that the initial wiggling of the shocks is not compelling.  The only
effect of this wiggling is to speed up the initial phase of the evolution.
Test cases using another wiggling or starting from straight shocks end up like
the simulations we are going to present in the following.

We would like to add a side note on this last point, from our observation that
the slab is also destabilized when bound by straight shocks. This has already
been reported by~\citet{blondin-marks:96}, who ascribed the destabilization to
'numerical noise'.  Meanwhile, \citet{robinet-et-al:00} have investigated what
is called the carbuncle phenomenon in some more detail. They showed that -
contrary to what has been believed so far - a single straight shock is linearly
unstable for exactly one mode associated to the upstream Mach-number of
$M_{\mathrm{crit}} = [(5+\gamma) / (3-\gamma)]^{1/2}$. For isothermal
conditions, this yields $M_{\mathrm{crit}}= \sqrt{3}$.  They also showed that
this single unstable mode is sufficient for making straight shocks aligned with
the mesh numerically unstable at all Mach-numbers if the computation is done
with a low-viscosity, high-order, shock-capturing scheme. To what degree this
instability for a straight shock of any Mach-number is really physical seems
an open question to us.
\subsection{The different runs}
The runs we performed differ in their upwind Mach-numbers, which lie in a
range $5 \lapprox M_{\mathrm{u}} \lapprox 90 $, as well as in their
initialization, numerical discretization, and the y-extent of the domain.  The
labels of the different runs are built up as M\_I.R.Y. Here, M is the upwind
Mach-number, I the initialization (0, 1, or 2), R gives the refinement of the
spatial discretization, relative to the coarsest grid simulation we performed
(1, 2, 4, or 8).  R=1 corresponds to a finest cell size of about $3 \cdot
10^{-3} \, \mathrm{Y}_{\mathrm{0}}$, R=2 indicates a twice smaller cell size.
Y is the domain size (1, 2, 4, or 6) in units of $\mathrm{Y}_{\mathrm{0}}$.
For example, R22\_0.2.4 denotes a run with $M_{\mathrm{u}}=22$, initialization
I=0, finest cell size about $1.5 \cdot 10^{-3} \mathrm{Y}_{\mathrm{0}}$, and
y-extent $4 \, \mathrm{Y}_{\mathrm{0}}$.

The runs we performed are listed in Table~\ref{tab:list_of_runs}.  Individual
columns in Table~\ref{tab:list_of_runs} contain (column number in square
brackets): label of run [1], following the scheme
label=\,M$_{\mathrm{u}}$\_\,I.R.Y, where I is the initial condition, R the
refinement factor such that cell size = $3.125 \cdot 10^{-3}
\mathrm{Y}_{\mathrm{0}} / \mathrm{R}$, and Y is the y-extension of the
computational domain in units of $\mathrm{Y}_{\mathrm{0}}$; Mach-number of
upstream flow, $M_{\mathrm{u}}$ [2]; stopping time of simulation in terms of
$\ell( N )$ [3]; y-averaged x-extension of CDL at stopping time, relative to
y-extent of computational domain, $\ell_{\mathrm{cdl}}/\mathrm{Y}$ [4];
average quantities [5-9] of: rms Mach-number, $M_{\mathrm{rms}}$ [5]; mean
density in units of upstream density, $\rho_{\mathrm{m}}/\rho_{\mathrm{u}}$
[6]; shock length in units of y-domain, $\ell_{\mathrm{sh}}/Y$ [7]; driving
efficiency, $f_{\mathrm{eff}}$ [8]; averages taken over $10 \le \ell( N ) \le
70$ for I=0 and over $60 \le \ell( N ) \le 120$ for I=1, for I=2 we give the
values at the end of the simulation in parentheses instead.
\section{Scaling properties of the model problem}
\label{sec:anal-scaling}
\begin{figure}[tbp]
\centerline{
  \includegraphics[width=4.5cm]{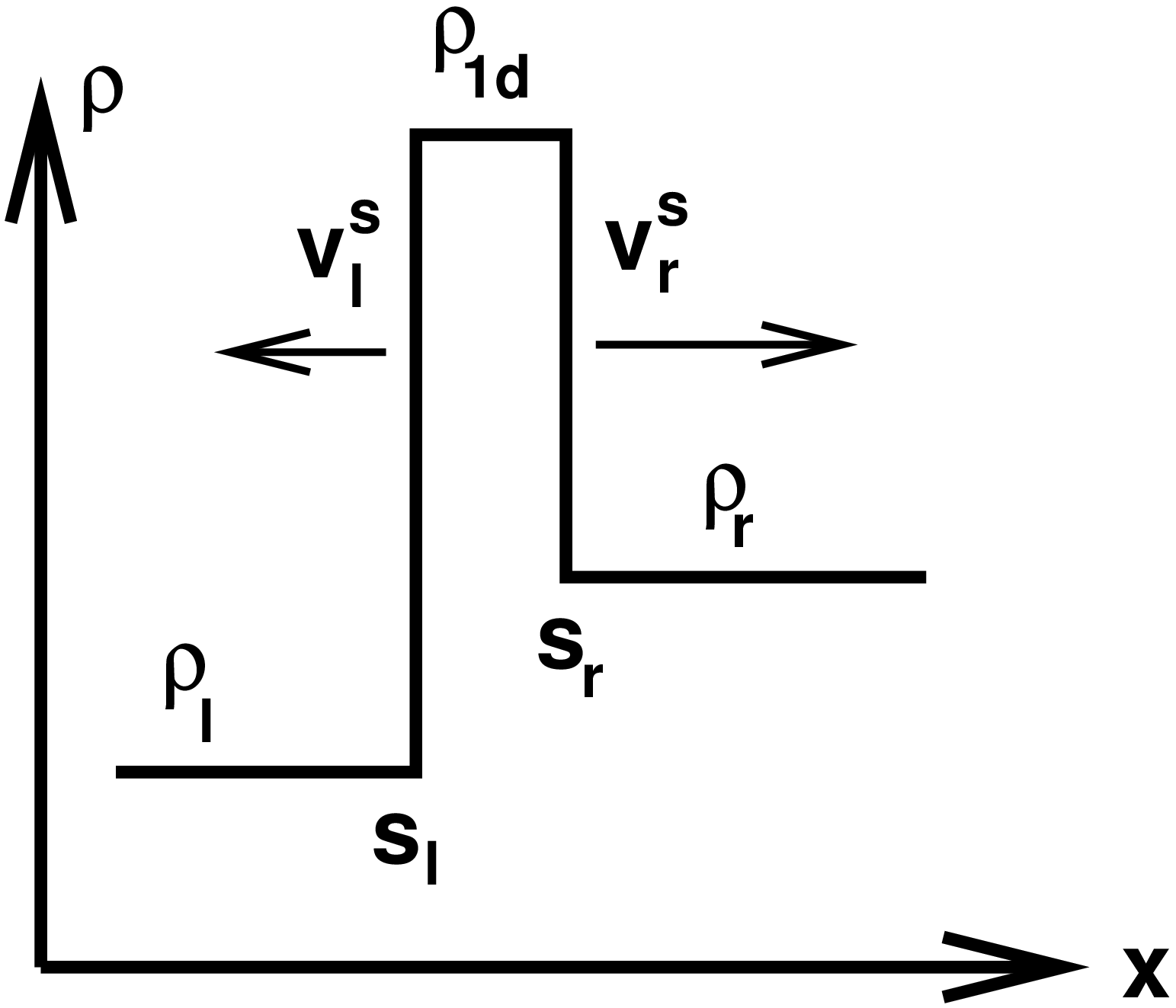}
   }
\centerline{
  \includegraphics[width=4.5cm]{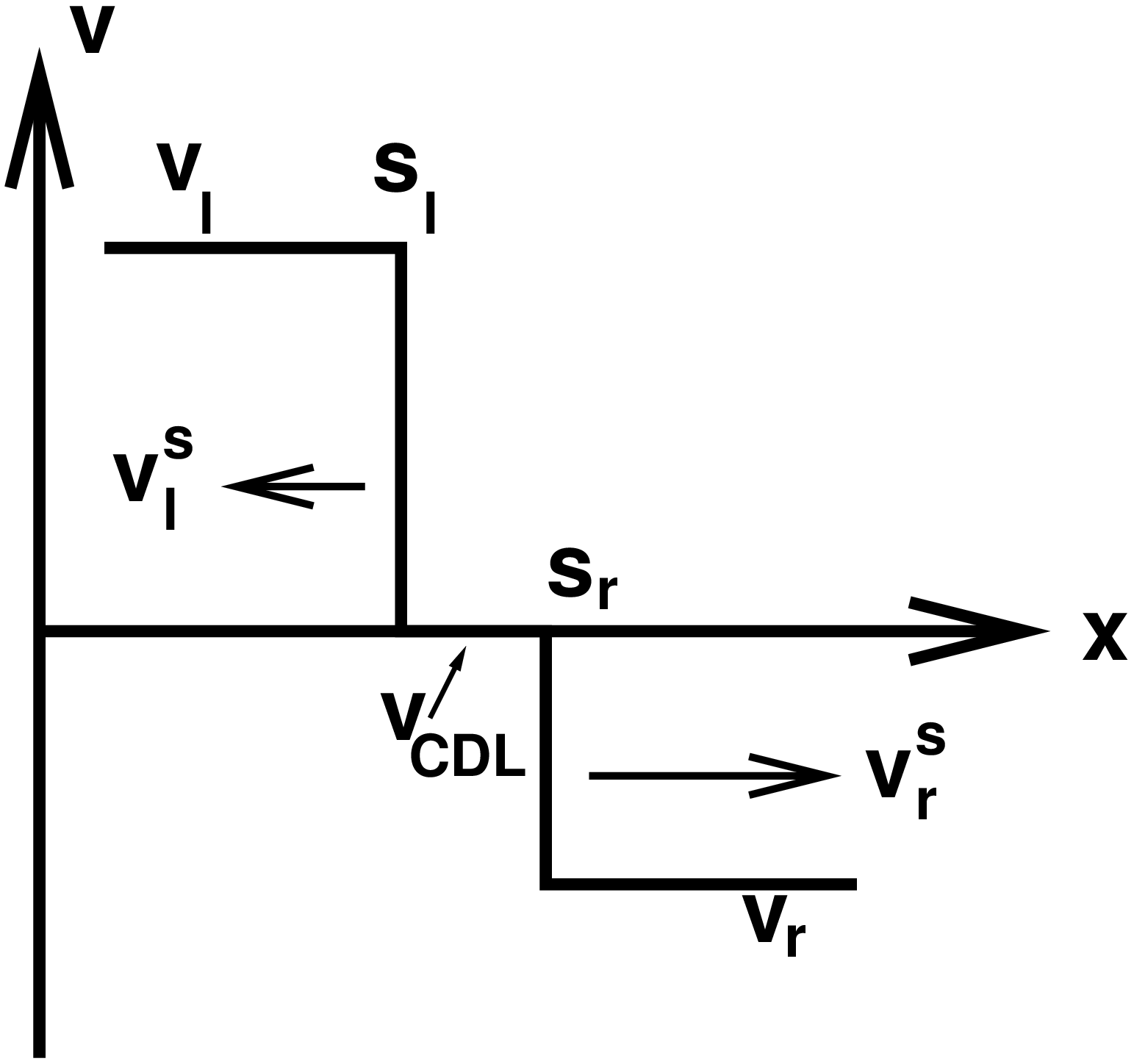}
          }
\caption{The self-similar 1D solution of isothermal colliding supersonic flows
  in density (top) and velocity (bottom). The interaction zone
  (labeled CDL) is bounded by two shocks, $s_{\mathrm{l}}$ and
  $s_{\mathrm{r}}$, having speeds $v^s_l$ and $v^s_r$ in the rest
  frame of the CDL. The density and velocity of the 1D interaction zone,
  we denote by $\rho_{\mathrm{1d}}$ and $v_{\mathrm{1d}}$, respectively. }
\label{fig:basic_structure_1d}
\end{figure}
Within the frame of Euler equations and in infinite space, the problem of
isothermal supersonically colliding flows can be solved analytically in 1D.
The solution, sketched in Fig.~\ref{fig:basic_structure_1d} and
Sect.~\ref{sec:anal-scaling_1d}, is self-similar and depends only on two free
parameters, the Mach-numbers of the left and right upwind flow.  In 2D the
situation is more complicated: the solution is unstable \citep{vishniac:94,
  blondin-marks:96}, the shocks confining the CDL are non-stationary and
oblique, the interior of the CDL is supersonically turbulent.

Nevertheless, in infinite space it seems reasonable to {\it assume}
that the solution, on average, may still evolve in a self-similar
manner.  We base this assumption on the following two observations.
First, the isothermal Euler equations are scale-free in infinite
space. Second, the free parameters of the problem
($\rho_{\mathrm{u}}$, $M_{\mathrm{u}}$, and $a$) do not introduce any
fixed length or time scale. Under these conditions, it is possible
that the solution also does not depend on length or time separately,
but only on their ratio.  If so, all length scales should evolve
equally with time, which implies, in particular, that the solution
then should not depend on the extension of the CDL. We stress,
however, that {\it we have no proof of the above assumption of
  self-similarity.}

In the remainder of this section, we elaborate a bit further on the
implications of the assumed self-similarity. In Sect.~\ref{sec:num_results} we
will see that the relations derived here give a good approximation of the
numerical results, but we stress already here three important points. The
numerical simulations are carried out in finite space (not infinite);
numerical dissipation might play a role; and the simulations are stopped for
the most part while the CDL is still small, about half the size of the
y-extent of the computational domain.  Important aspects that can only be
obtained from the numerical solution include quantities related to the driving
of the turbulence, the values of proportionality constants, and the interior
structure of the CDL. We neglect this last aspect, however, in the current
paper to focus on mean quantities instead.
\subsection{Self-similar 1D solution}
\label{sec:anal-scaling_1d}
Denoting the density and velocity of the CDL by $\rho_{\mathrm{1d}}$
and $v_{\mathrm{1d}}$, and those of the left and right upwind flows
by $\rho_{\mathrm{i}}$ and $v_{\mathrm{i}}$ ($i=l,r$), the 
solution in the rest frame of the CDL is given by
\begin{eqnarray}
\label{eq:self_sim1} 
\rho_{\mathrm{1d}} / \rho_{\mathrm{i}} & = & M_{\mathrm{i}}^\mathrm{2} + 1 
                         \approx M_{\mathrm{i}}^2,\\
\label{eq:self_sim2}
v_{\mathrm{1d}} & = & 0,       \\              
\label{eq:self_sim3}
|v^{s}_{\mathrm{i}}| & = & aM_{\mathrm{i}} / (M_{\mathrm{i}}^{2} - 1)
                   \approx a/ M_{\mathrm{i}} << a.
\end{eqnarray}
Here, $v^{s}_{\mathrm{i}}$ is the velocity of the confining shocks and $a$ is
again the isothermal sound speed. The approximations hold for large
Mach-numbers. The self-similar character is apparent: the solution is not a
function of $x$ and $t$ but only a function of $x/t$ through
$v^{s}_{\mathrm{i}}$.

A relation between characteristic length and time scales of the solution, the
self-similarity variable $\kappa_{\mathrm{1d}}$, can be obtained as follows.
As a length scale, we take the spatial extension $\ell_{\mathrm{1d}}$ of the
CDL, and as a time scale the time $\tau$ needed to accumulate the
corresponding column density ${ N }_{\mathrm{1d}}$. From the relations
\begin{equation}
 N_{\mathrm{1d}} = \rho_{\mathrm{1d}} \ell_{\mathrm{1d}}.
\label{eq:mass_column}
\end{equation}
and 
\begin{equation}
 N_{\mathrm{1d}} = \tau \left( \rho_l v_l + \rho_r v_r\right)
\label{eq:mass_cons}
\end{equation}
and using $\rho_\mathrm{l}/\rho_\mathrm{r} =
M^{2}_\mathrm{r}/M^{2}_\mathrm{l}$ (see Eq.~\ref{eq:self_sim1}), we 
obtain
\begin{equation}
\kappa_{\mathrm{1d}} \equiv \frac{\ell_{\mathrm{1d}}}{\tau}
  = a \frac{M_l + M_r}{M_l \cdot M_r}.
\label{eq:1d_kappa}
\end{equation}
Thus for strong shocks $\kappa_{\mathrm{1d}}$ is nothing else
than $|v^{s}_{\mathrm{l}}| + |v^{s}_{\mathrm{r}}|$. Specializing to
symmetric settings ($\mathrm{l} = \mathrm{r}$) yields
$\rho_{\mathrm{1d}} / \rho_{\mathrm{u}} = M_{\mathrm{u}}^{2}$ and
$\kappa_{\mathrm{1d}} =2a/M_{\mathrm{u}}$.
\subsection{Scaling properties of the 2D symmetric solution}
\label{sec:anal-scaling_2d}
In the following, we derive scaling relations for the 2D
  solution, assuming self-similarity.  We confront these relations
with corresponding numerical results in Sect.~\ref{sec:num_results}.
\subsubsection{Density, Mach-number, self-similarity variable}
In the following, all velocities are again given in the rest frame of
the CDL and we {\it assume} that a self-similar solution exists.  A
natural choice for the (constant) self-similarity variable
then is again $\kappa_{\mathrm{2d}} \equiv \ell_{\mathrm{cdl}}/\tau$.
Using the definitions of Sect.~\ref{sec:num_settings} we must have, as
in the 1D case,
\begin{eqnarray}
\label{eq:col-dens-2d-1}
 N  & = & \rho_{\mathrm{m}} \ell_{\mathrm{cdl}},\\
\label{eq:col-dens-2d-2}
 N  & = & 2 \tau \rho_{\mathrm{u}} v_{\mathrm{u}}.
\end{eqnarray}
Dividing the two equations through each other yields $\kappa_{\mathrm{2d}} = 2
\rho_{\mathrm{u}} v_{\mathrm{u}} / \rho_{\mathrm{m}}$. As
$\kappa_{\mathrm{2d}}$ is a constant, the CDL mean density $\rho_{\mathrm{m}}$
must be constant in time.  The root-mean-square velocity
$v_{\mathrm{rms}}^{2}$ then has to be constant in time as well, at least if
the CDL density and velocity, $\rho$ and $v$, are uncorrelated (in which case
we can replace the average over the product $\rho v^{2}$ by the product of the
averages of $\rho$ and $v^{2}$) and if kinetic pressure dominates over thermal
pressure. This can be seen from equating the total upwind pressure with the
total pressure within the CDL,
\begin{equation}
\rho_{\mathrm{u}} (a^{2} + v_{\mathrm{u}}^{2}) = \rho_{\mathrm{m}} (a^2 + v_{\mathrm{rms}}^2).
\label{eq:p1_cdl}
\end{equation}
The simplest ansatz for $\rho_{\mathrm{m}}$ and $v_{\mathrm{rms}}$
is that they only depend on the upstream Mach-number,
\begin{eqnarray}
\label{eq:ansatz_rho}
\rho_{\mathrm{m}}/\rho_{\mathrm{u}} & = &  \eta_{\mathrm{1}} 
                                             M_{\mathrm{u}}^{\beta_{\mathrm{1}}}, \\
\label{eq:ansatz_v}
v_{\mathrm{rms}}/a                    & = & \eta_{\mathrm{2}} 
                                            M_{\mathrm{u}}^{\beta_{\mathrm{2}}}.
\end{eqnarray}
Using the ansatz for $\rho_{\mathrm{m}}$ we obtain a first expression for
$\kappa_{\mathrm{2d}}$ from Eqs.~\ref{eq:col-dens-2d-1}
and~\ref{eq:col-dens-2d-2},
\begin{equation}
\kappa_{\mathrm{2d}} = 2 a \eta_{\mathrm{1}}^{-1} M_{\mathrm{u}}^{1-\beta_{\mathrm{1}}}
                 \propto a M_{\mathrm{u}}^{1-\beta_{\mathrm{1}}}.
\label{eq:2d_kappa_coldens}
\end{equation}
A second expression for $\kappa_{\mathrm{2d}}$, we obtain from Eq.~\ref{eq:p1_cdl}
\begin{equation}
\rho_{\mathrm{u}}a^{2}(1+M_{\mathrm{u}}^{2}) = \rho_{\mathrm{m}} (a^2 + v_{\mathrm{rms}}^2) = 
                  \frac{a^2  N }{\ell_{\mathrm{cdl}}} 
                  ( 1 + \eta_{\mathrm{2}}^{2} M_{\mathrm{u}}^{2 \beta_{\mathrm{2}}} ).
\label{eq:p_cdl}
\end{equation}
Again using Eq.~\ref{eq:col-dens-2d-2} to replace $ N $, one
obtains
\begin{equation}
\kappa_{\mathrm{2d}} = 
   2 a M_{\mathrm{u}} \frac{ 1 + \eta_{\mathrm{2}}^{2} M_{\mathrm{u}}^{2 \beta_{\mathrm{2}}}}
              {1 +M_{\mathrm{u}}^{2}} 
  \approx 2 a \eta_{\mathrm{2}}^{2} M_{\mathrm{u}}^{2\beta_{\mathrm{2}} -1}
  \propto a M_{\mathrm{u}}^{2\beta_{\mathrm{2}} -1}.
\label{eq:2d_kappa_pressure}
\end{equation}
The approximation is good for high Mach-number flows, with
$\eta_{\mathrm{2}}^{2} M_{\mathrm{u}}^{2 \beta_{\mathrm{2}}} >> 1$,
and for $\beta_{\mathrm{2}} > 0$, which is, however, to be expected for
supersonic turbulence. Comparing Eqs.~\ref{eq:2d_kappa_coldens}
and~\ref{eq:2d_kappa_pressure} gives
\begin{eqnarray}
\label{eq:beta12}
\beta_{\mathrm{2}} & = & 1 - \beta_{\mathrm{1}}/2, \\
\label{eq:eta12}
\eta_{\mathrm{1}}^{-1} & = & \eta_{\mathrm{2}}^{2}.
\end{eqnarray}
\subsubsection{Driving energy}
\label{sec:a_drive_eff}
From energy conservation, we have $ \dot{\cal E}_{\mathrm{diss}} = \dot{\cal
  E}_{\mathrm{drv}} - \dot{\cal E}_{\mathrm{kin}}$. Here $\dot{\cal
  E}_{\mathrm{drv}}$ is the energy flux density entering the CDL per
time and per unit length in the y-direction, and  $\dot{\cal E}_{\mathrm{diss}}$
denotes the energy density dissipated per time within an average
column of length $\ell_{\mathrm{cdl}}$ of the CDL. Finally, $\dot{\cal
  E}_{\mathrm{kin}}$ is the change per time of the kinetic energy
contained within such an average column.  We first turn to the driving
energy $\dot{\cal E}_{\mathrm{drv}}$ and come back to $\dot{\cal
  E}_{\mathrm{diss}}$ and $\dot{\cal E}_{\mathrm{kin}}$ in
Sect.~\ref{sec:energy_dissipation}.

Part of the total (left plus right) upwind kinetic energy flux density, ${\cal
  F}_{\mathrm{e_{\mathrm{kin}},u}} = \rho_{\mathrm{u}}v_{\mathrm{u}}^{3}$, is
thermalized at the shocks confining the CDL.  The remaining part, $\dot{\cal
  E}_{\mathrm{drv}}$, drives the turbulence in the CDL. We assume that
$\dot{\cal E}_{\mathrm{drv}}$ and ${\cal F}_{\mathrm{e_{\mathrm{kin}},u}}$ are
related by a function of the upwind Mach-number only,
\begin{equation}
\dot{\cal E}_{\mathrm{drv}} = f_{\mathrm{eff}}(M_{\mathrm{u}}){\cal F}_{\mathrm{e_{\mathrm{kin}},u}}.
\label{eq:def_feff}
\end{equation}
We call the function $f_{\mathrm{eff}}$ the driving efficiency.  An
expression for $f_{\mathrm{eff}}$ can be derived by using the jump
conditions for strong, oblique shocks,
\begin{eqnarray}
\rho_{\mathrm{d}}        & = & \rho_{\mathrm{u}} M_{\mathrm{\perp,u}}^{2} 
                           =   \rho_{\mathrm{u}} M_{\mathrm{u}}^{2} \sin^{2}\alpha, \nonumber \\
v_{\mathrm{\perp,d}}     & = & v_{\mathrm{\perp,u}} M_{\mathrm{\perp,u}}^{-2} 
                           =   \frac{a}{M_{\mathrm{u}} \sin \alpha}, \nonumber \\
v_{\mathrm{\parallel,d}} & = & v_{\mathrm{\parallel,u}}
                           = a M_{\mathrm{u}} \cos\alpha.
\label{eq:oblique_jump}
\end{eqnarray}
The subscript d denotes downstream quantities, right after shock
passage; the subscripts ${\mathrm{\perp}}$ and ${\mathrm{\parallel}}$
denote flow components perpendicular and parallel to the shock,
respectively; and $\alpha$ is given in Fig.~\ref{fig:sketch2d}.  Using
Eq.~\ref{eq:oblique_jump} we obtain
\begin{eqnarray}
\dot{\cal E}_{\mathrm{drv}} & = & \frac{1}{\mathrm{Y}}\int_{s_{\mathrm{l,r}}} ds 
                                \frac{\rho_{\mathrm{d}} v_{\mathrm{d}}^{2}}{2} 
                                v_{\mathrm{\perp,d}} \nonumber \\
                          & = & \frac{\rho_{\mathrm{u}}v_{\mathrm{u}}^{3}}{2Y} 
                                \int_{\mathrm{Y}_{\mathrm{l,r}}} dy (1 - \sin^{2}\alpha + 
                                \frac{1}{M_{\mathrm{u}}^{4}\sin^{2}\alpha}),
\label{eq:edrive_bowshock1}
\end{eqnarray}
where the integral over $s_{\mathrm{l,r}}$ and $\mathrm{Y}_{\mathrm{l,r}}$
runs over both shocks and where it was used that $\sin \alpha \; ds = dy$. The
last term on the right hand side of Eq.~\ref{eq:edrive_bowshock1} is omitted in
the following. This is justified, as the shocks we observe in our simulations
fulfill $\sin \alpha >> M_{\mathrm{u}}^{-2}$ for the most part (see
Sect.~\ref{sec:confshocks}). For $f_{\mathrm{eff}}(M_{\mathrm{u}})$ we thus
obtain
\begin{equation}
f_{\mathrm{eff}} = \frac{1}{2Y} \int_{\mathrm{Y_{\mathrm{l,r}}}} dy (1 - \sin^{2}\alpha)
                 \equiv 1 - \sin^{2}\alpha_{\mathrm{eff}}
\label{eq:a_feff}
\end{equation}
where we used the midpoint rule. The angle $\alpha_{\mathrm{eff}}$ can be
interpreted as an average bending angle. As the ansatz for the Mach-number
dependence of $f_{\mathrm{eff}}$ we thus take
\begin{equation}
f_{\mathrm{eff}} = 1 - \sin^{2}\alpha_{\mathrm{eff}} = 1 - \eta_{\mathrm{3}}M_{\mathrm{u}}^{\beta_{\mathrm{3}}}.
\label{eq:b_feff}
\end{equation}
\subsubsection{Energy dissipation}
\label{sec:energy_dissipation}
A first expression for the column-integrated dissipated energy per
time can be obtained from energy conservation, $\dot{\cal
  E}_{\mathrm{diss}} = \dot{\cal E}_{\mathrm{drv}} - \dot{\cal
  E}_{\mathrm{kin}}$. For $\dot{\cal E}_{\mathrm{drv}}$ we just
derived an expression, Eqs.~\ref{eq:def_feff} and ~\ref{eq:b_feff}.
For $\dot{\cal E}_{\mathrm{kin}}$ we obtain, within the frame of
self-similarity,
\begin{equation}
\dot{\cal E}_{\mathrm{kin}} = \frac{\rho_{\mathrm{m}} v_{\mathrm{rms}}^{2}}{2} \frac{d \ell_{\mathrm{cdl}}}{dt}
                        = \rho_{\mathrm{u}} a^{3}
                          \frac{\eta_{\mathrm{2}}}{2} M_{\mathrm{u}}^{3-\beta_{\mathrm{1}}},
\label{eq:ekin}
\end{equation}
where we used Eqs.~\ref{eq:ansatz_rho}, \ref{eq:ansatz_v},
and~\ref{eq:2d_kappa_pressure} to~\ref{eq:eta12}. Together we get
\begin{equation}
\dot{\cal E}_{\mathrm{diss}} = \rho_{\mathrm{u}} a^{3} M_{\mathrm{u}}^{3} \;
                           [1 - \eta_{\mathrm{3}}M_{\mathrm{u}}^{\mathrm{\beta_{\mathrm{3}}}}
                            -
                            0.5 \; \eta_{\mathrm{2}}^{2} M_{\mathrm{u}}^{\mathrm{-\beta_{\mathrm{1}}}}].
\label{eq:ediss_cons}
\end{equation}
The energy dissipated per time within an average column of length
$\ell_{\mathrm{cdl}}$ is thus independent of this length. If 
energy dissipation occurs only (as within the frame of Euler equations) or
at least dominantly in shocks, which implies that the average distance between
shocks increases and / or the average strength of the shocks decreases
as the CDL grows.

A second expression for $\dot{\cal E}_{\mathrm{diss}}$ can be obtained
from dimensional considerations. The energy dissipated per unit volume
per unit time must be proportional to $\rho_{\mathrm{diss}}
v_{\mathrm{diss}}^{3} \ell_{\mathrm{diss}}^{-1}$. Here,
$\rho_{\mathrm{diss}}$, $v_{\mathrm{diss}}$, and
$\ell_{\mathrm{diss}}$ are the characteristic density, velocity, and
length scale of the dissipation. The energy
dissipation within an average column of length $\ell_{\mathrm{cdl}}$
can thus be written as $\dot{\cal E}_{\mathrm{diss}} \propto
\rho_{\mathrm{diss}} v_{\mathrm{diss}}^{3} \ell_{\mathrm{diss}}^{-1}
\ell_{\mathrm{cdl}}$.  As all length scales must evolve equally with
time within the frame of self-similarity,
$\ell_{\mathrm{cdl}}/\ell_{\mathrm{diss}}$ must be constant, thus
\begin{equation}
\dot{\cal E}_{\mathrm{diss}} \propto \rho_{\mathrm{diss}} v_{\mathrm{diss}}^{3}.
\label{eq:ediss2}
\end{equation}

Comparison of Eqs.~\ref{eq:ediss_cons} and~\ref{eq:ediss2} suggests
$v_{\mathrm{diss}} \propto a M_{\mathrm{u}}$ and a more complicated
Mach-number dependence for $\rho_{\mathrm{diss}}$. As $v_{\mathrm{rms}}$ is
the only velocity scale we have, it seems natural to assume that
$v_{\mathrm{diss}} \propto v_{\mathrm{rms}}$. It then follows that
$v_{\mathrm{rms}} \propto a M_{\mathrm{u}}$ or $\beta_{\mathrm{2}}=1$ (and
$\beta_{\mathrm{1}}=0$). We note that~\citet{gammie-ostriker:96} even found
$v_{\mathrm{diss}} = v_{\mathrm{rms}}$ for a 1D case.
\subsection{Summary of expected scaling relations}
\label{sec:expectedrelations}
If a self-similar solution exists, we expect the following
dependencies:
\begin{eqnarray}
\label{eq:exp_rho}
\rho_{\mathrm{m}}        & = & \eta_{\mathrm{1}} \rho_{\mathrm{u}} M_{\mathrm{u}}^{\beta_{\mathrm{1}}} 
                           =   \eta_{\mathrm{1}} \rho_{\mathrm{u}},\\
\label{eq:exp_mach}
M_{\mathrm{rms}}         & = & \eta_{\mathrm{2}} M_{\mathrm{u}}^{\beta_{\mathrm{2}}}
                           =   \eta_{\mathrm{1}}^{-1/2} M_{\mathrm{u}},\\
\label{eq:exp_kappa}
\kappa_{\mathrm{2d}}     & = & \ell_{\mathrm{cdl}}/\tau 
                           =   2 \eta_{\mathrm{1}}^{-1} a M_{\mathrm{u}},\\
\label{eq:exp_edrv}
\dot{\cal E}_{\mathrm{drv}}  & = & \rho_{\mathrm{u}} a^{3} M_{\mathrm{u}}^{3} 
                              (1 - \eta_{\mathrm{3}} M_{\mathrm{u}}^{\beta_{\mathrm{3}}}),\\
\label{eq:exp_ekin}
\dot{\cal E}_{\mathrm{kin}}  & = & \rho_{\mathrm{u}} a^{3} M_{\mathrm{u}}^{3} 
                               \; 0.5 \; \eta_{\mathrm{2}}^{2},\\
\label{eq:exp_ediss}
\dot{\cal E}_{\mathrm{diss}} & = & \rho_{\mathrm{u}} a^{3} M_{\mathrm{u}}^{3} 
                              (1 -\eta_{\mathrm{3}} M_{\mathrm{u}}^{\beta_{\mathrm{3}}} 
                                 - 0.5 \; \eta_{\mathrm{2}}^{2}).
\end{eqnarray}
Note the differences to the 1D solution: Eq.~\ref{eq:exp_rho} predicts
the CDL mean density to be independent of $M_{\mathrm{u}}$ and
$\kappa_{\mathrm{2d}} \propto a M_{\mathrm{u}}$, in contrast to
$\rho_{\mathrm{1d}} \propto M_{\mathrm{u}}^{2}$ and
$\kappa_{\mathrm{1d}} \propto a / M_{\mathrm{u}}$.

In deriving the above relations, we made four basic assumptions:
a) we have simple Mach-number dependencies of $\rho_{\mathrm{m}}$,
$v_{\mathrm{rms}}$, and $f_{\mathrm{eff}}$, Eqs.~\ref{eq:ansatz_rho},
\ref{eq:ansatz_v}, and \ref{eq:b_feff}; b) the CDL density and
velocity are uncorrelated; c) we have high Mach-numbers in the sense
that $\eta_{\mathrm{2}}^{2} M_{\mathrm{u}}^{2 \beta_{\mathrm{2}}} >>
1$ or $M_{\mathrm{rms}}^{2} >> 1$; d) $v_{\mathrm{diss}} \propto
v_{\mathrm{rms}}$.

In Sect.~\ref{sec:num_results} we are going to check the validity of these
assumptions and confront Eqs.~\ref{eq:exp_rho} to~\ref{eq:exp_ediss} with
numerically obtained values. We expect good agreement as long as
$M_{\mathrm{rms}} >> 1$, thus dissipation in shocks likely dominates, and as
long as $\ell_{\mathrm{cdl}} << \mathrm{Y}$. The 'Euler character' of the
solution should prevail under these conditions.  We also determine those
quantities that cannot be derived analytically. These are, on the one hand,
the coefficients $\eta_{\mathrm{1}}$ and $\eta_{\mathrm{3}}$, as well as the
exponent $\beta_{\mathrm{3}}$. On the other hand, there are quantities for
which we have no analytical expression at all, like the wiggling of the
confining shocks, the associated distribution of the angle $\alpha$, or the
Mach-number dependence of the length of the confining shocks.
\section{Numerical results}
\label{sec:num_results}
We now present our numerical results.  After a brief phenomenological
description of the solution in Sect.~\ref{sec:pheno}, we give quantitative
results for initial conditions I=0 in Sect.~\ref{sec:symmetric_nocdl}. Results
for initial conditions I=1 and I=2 are given in
Sect.~\ref{sec:symmetric_withcdl}, and asymmetric settings are briefly
addressed in Sect.~\ref{sec:results_asym}. Discretization and domain studies
are the topic of Sect.~\ref{sec:griddomain}.
\subsection{Brief phenomenological description}
\label{sec:pheno}
\begin{figure}[tp]
\centerline{
\includegraphics[width=8.5cm]{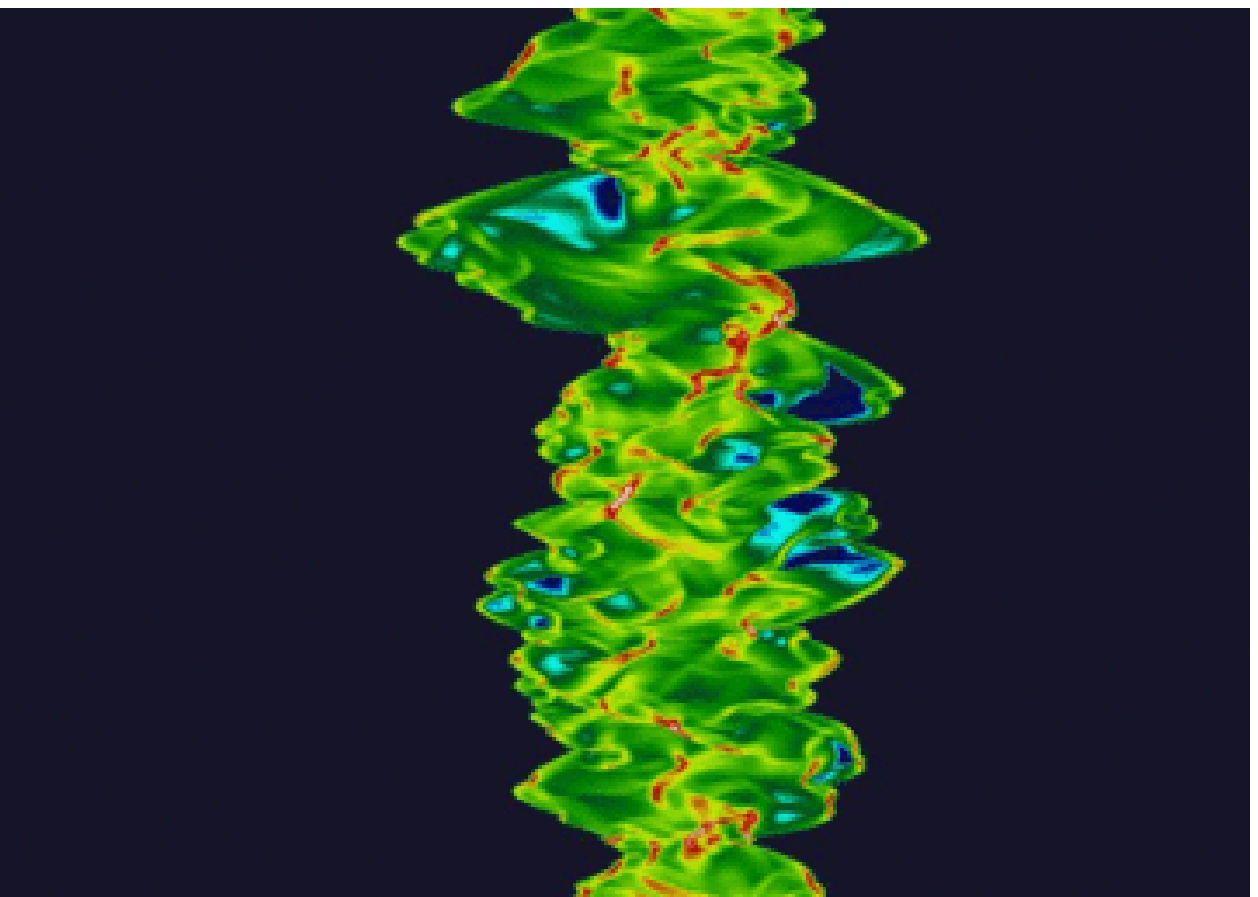}
           }
\vspace{.2cm}
\centerline{
\includegraphics[width=8.5cm]{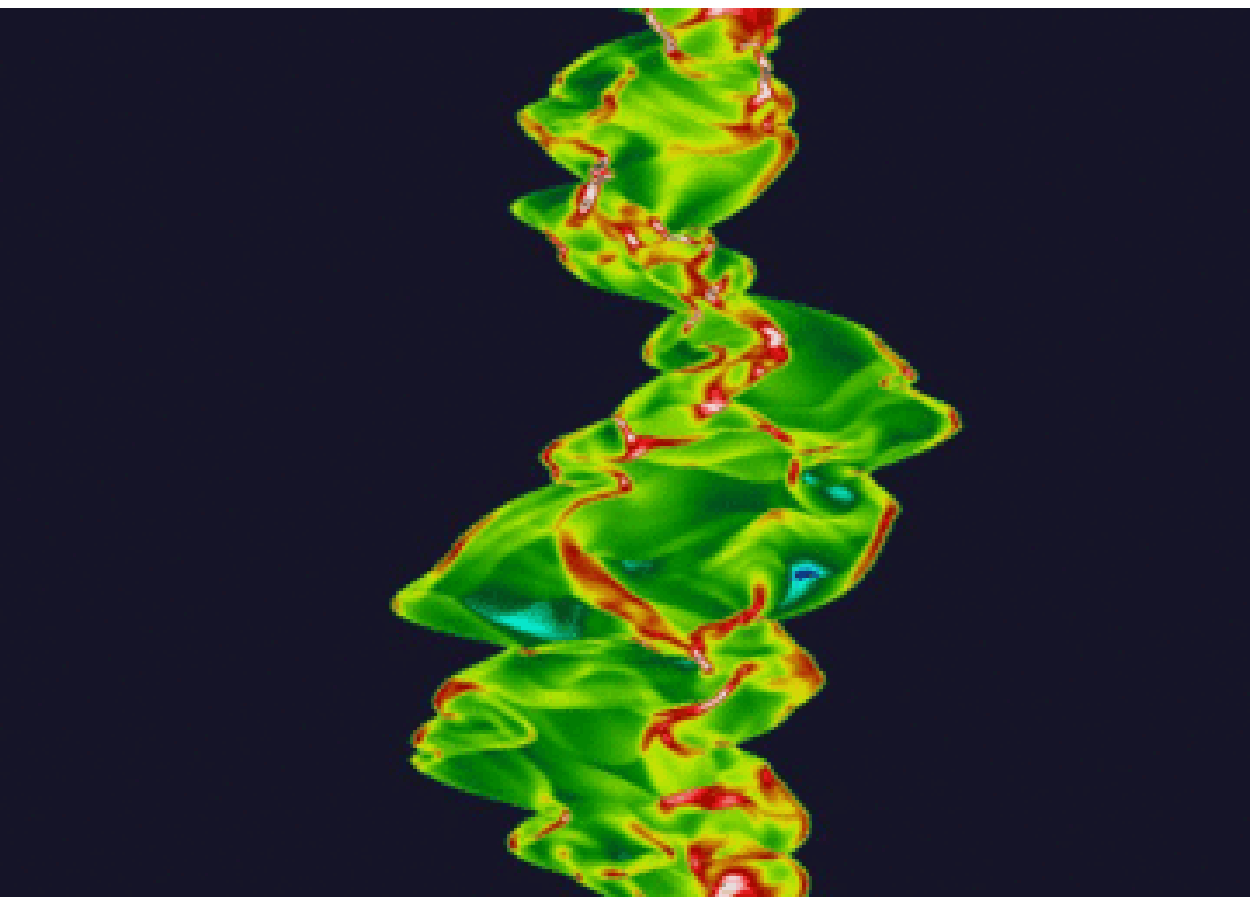}
           }
\vspace{.2cm}
\centerline{
\includegraphics[width=8.5cm]{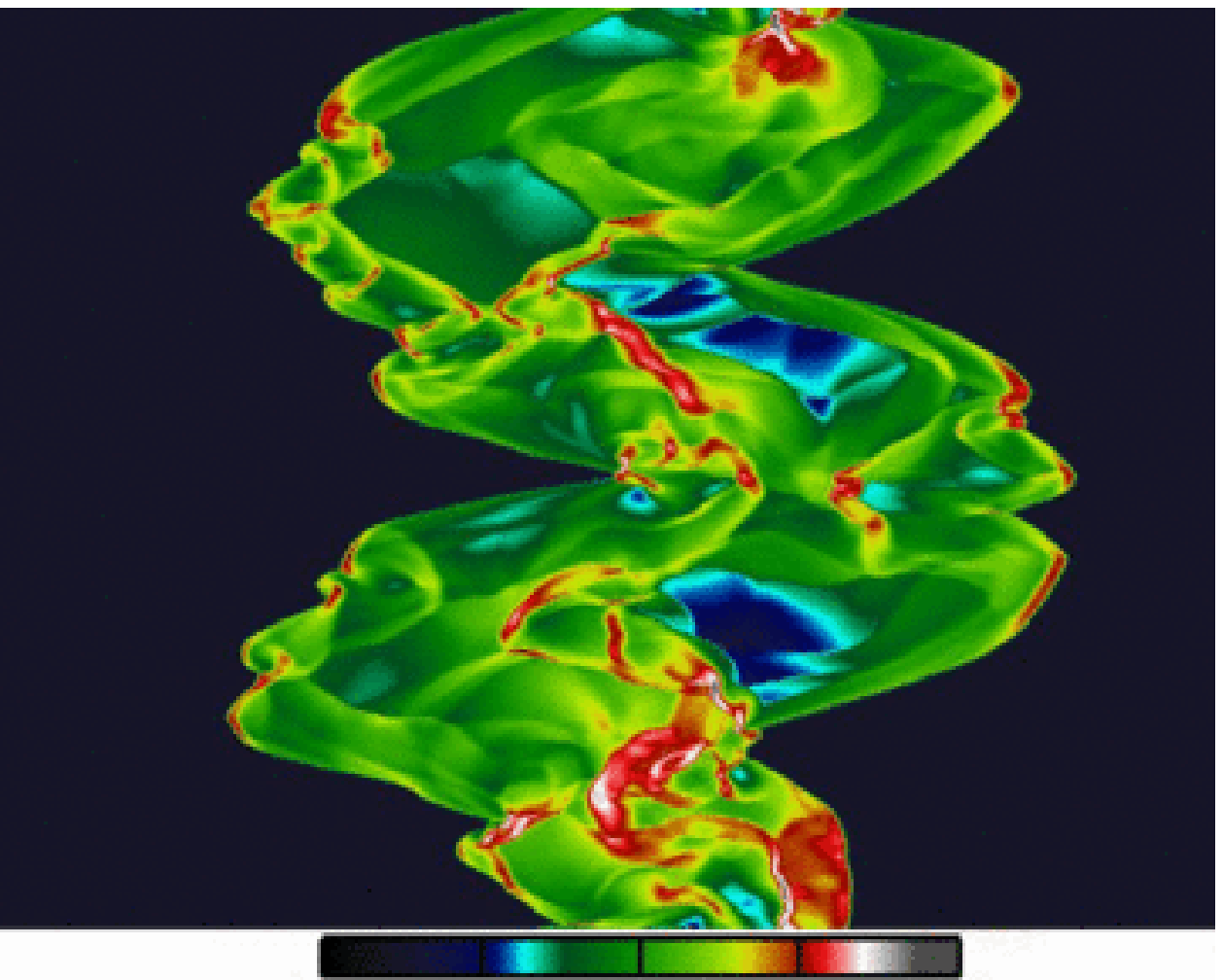}
           }
\caption{The interaction zone of run R22\_1.2.2, shown in density 
  (logarithmic scale, in units of $\rho_{\mathrm{u}}$, color bar from
  0 to 4), for three different times: $\ell( N ) \approx 34$
  (top), $\ell( N ) \approx 54$ (middle), $\ell( N ) \approx
  74$ (bottom). The spatial scale of patches, filaments, and 
  wiggling of the confining shocks increases with 
  $\ell( N )$.}
\label{fig:pheno_dens}
\end{figure}
We begin with a brief qualitative description of the CDL.  As an
example, the density structure of run R22\_1.2.2 is shown in
Fig.~\ref{fig:pheno_dens} for three different times.

A first characteristic is the local bending of the confining shocks.  The
spatial scale of these wiggles increases linearly with time, as the CDL
accumulates more and more matter and gets more and more extended.  The 
  inclination of the wiggles with respect to the direction of the upstream flows
decreases with increasing upstream Mach-number (see
Sect.~\ref{sec:confshocks}).  Occasionally, we observe a superimposed 'bending
mode' (e.g. bottom panel in Fig.~\ref{fig:pheno_dens}), which in
  appearance is somewhat similar to the bending modes of the NTSI described
  by~\citet{vishniac:94}.
  
  A second characteristic is the patchy appearance of the CDL.  The turbulent
  interior is organized in filaments and patches, regions within which a flow
  variable remains more or less constant.  The spatial extension of these
  patches increases as well as the CDL accumulates more and more matter. The
  flow variables clearly mirror the supersonic character of the turbulence:
  the contrast between high-density filaments and extended patches in
  Fig.~\ref{fig:pheno_dens} easily reaches two orders of magnitude, the
  root-mean-square velocity is well above sound, and the mean density is
  substantially reduced compared to the 1D case. Shocks within the CDL are
  ubiquitous.
\subsection{Settings without CDL at $t=0$}
\label{sec:symmetric_nocdl}
For symmetric settings, and if there is no CDL at time $t=0$, we expect to see
the self-similar relations we derived in Sect.~\ref{sec:anal-scaling_2d}.  We
express the time evolution of the solution we express in terms of
\begin{equation}
\label{eq:ellnmass}
\ell( N ) \equiv N /
N_{\mathrm{0}} = \frac{\rho_{\mathrm{m}} \ell_{\mathrm{cdl}}}
                      {\rho_{\mathrm{u}} \mathrm{Y}_{\mathrm{0}}}.
\end{equation}
This function monotonically increases at about the same rate as
the mean extension of the CDL, since $\rho_{\mathrm{m}} \approx
\eta_{\mathrm{1}} \rho_{\mathrm{u}}$ (Eq.~\ref{eq:exp_rho}). In fact,
$\rho_{\mathrm{m}} \approx 30 \rho_{\mathrm{u}}$
(Sect.~\ref{sec:means}) and thus $\ell( N ) = 60$ corresponds to
$\ell_{\mathrm{cdl}} \approx 2 \mathrm{Y}_{\mathrm{0}}$.  For the
symmetric case we consider in this paper, $\ell ( N ) $ is proportional
to the elapsed time. Using Eq.~\ref{eq:col-dens-2d-2} to express $N$,
we can write
\begin{equation}
\label{eq:ellntime}
\ell ( N ) \equiv N  / N_{\mathrm{0}} = \frac{2 \tau \rho_{\mathrm{u}} v_{\mathrm{u}}} 
                                             {\rho_{\mathrm{u}} \mathrm{Y}_{\mathrm{0}}}
 = \tau \frac{2 v_{\mathrm{u}}}{Y_{\mathrm{0}}},
\end{equation}
and $\ell ( N ) = 60$ then corresponds to a time $ \tau = 30 Y_{\mathrm{0}} /
v_{\mathrm{u}}$. Or, if we use $v_{\mathrm{u}} \approx 5 v_{\mathrm{rms}}$
(Sect.~\ref{sec:means}) and $\mathrm{Y}_{\mathrm{0}} \approx
\ell_{\mathrm{cdl}} / 2$ for $\ell ( N ) = 60$, we obtain $\tau
\approx 3 \ell_{\mathrm{cdl}} / v_{\mathrm{rms}}$.

Unless otherwise stated, averages and best fits in this section are
always taken over the interval $10 \le \ell( N ) \le 70 $ and over all
runs without CDL at time $t=0$. The interval was chosen such that
initialization effects have died away and that domain effects do not
matter yet (Sect.~\ref{sec:diff_y_ext}).

We mention here already that the two most extreme simulations in terms of
$M_{\mathrm{u}}$, R5\_0.2.4 and R87\_0.2.4, often differ somewhat from the
other simulations. In the case of R5\_0.2.4, we ascribe the deviation to the
only subsonic turbulence and the correlation of density and velocity
($M_{\mathrm{rms}} \approx 0.9$ and corr$(\rho,v) \approx -0.4$, see
Sect.~\ref{sec:means}). In the case of R87\_0.2.4, the shocks become sometimes
too strongly inclined with respect to the computational grid to be properly
resolved by our numerical grid (Sect.~\ref{sec:confshocks}).
\subsubsection{CDL mean quantities and correlations}
\label{sec:means}
\begin{figure}[tp]
\centerline{
\includegraphics[width=9.0cm]{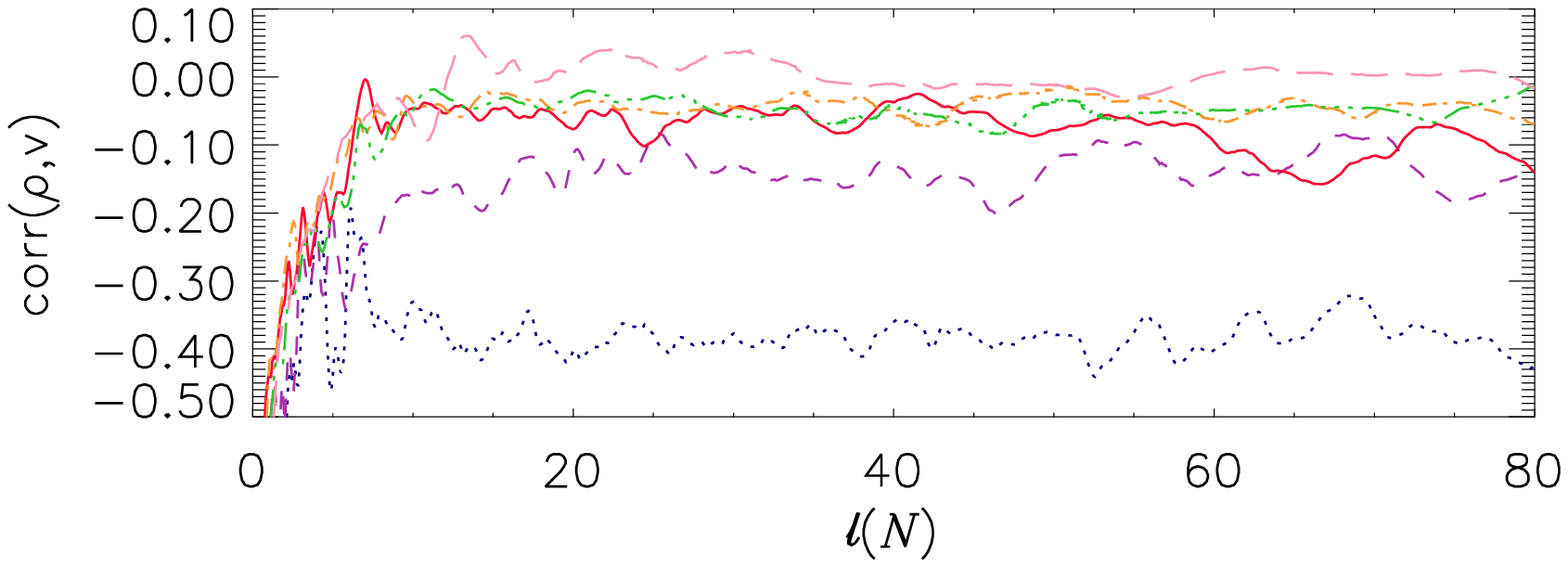}
}
\centerline{
\includegraphics[width=9.0cm]{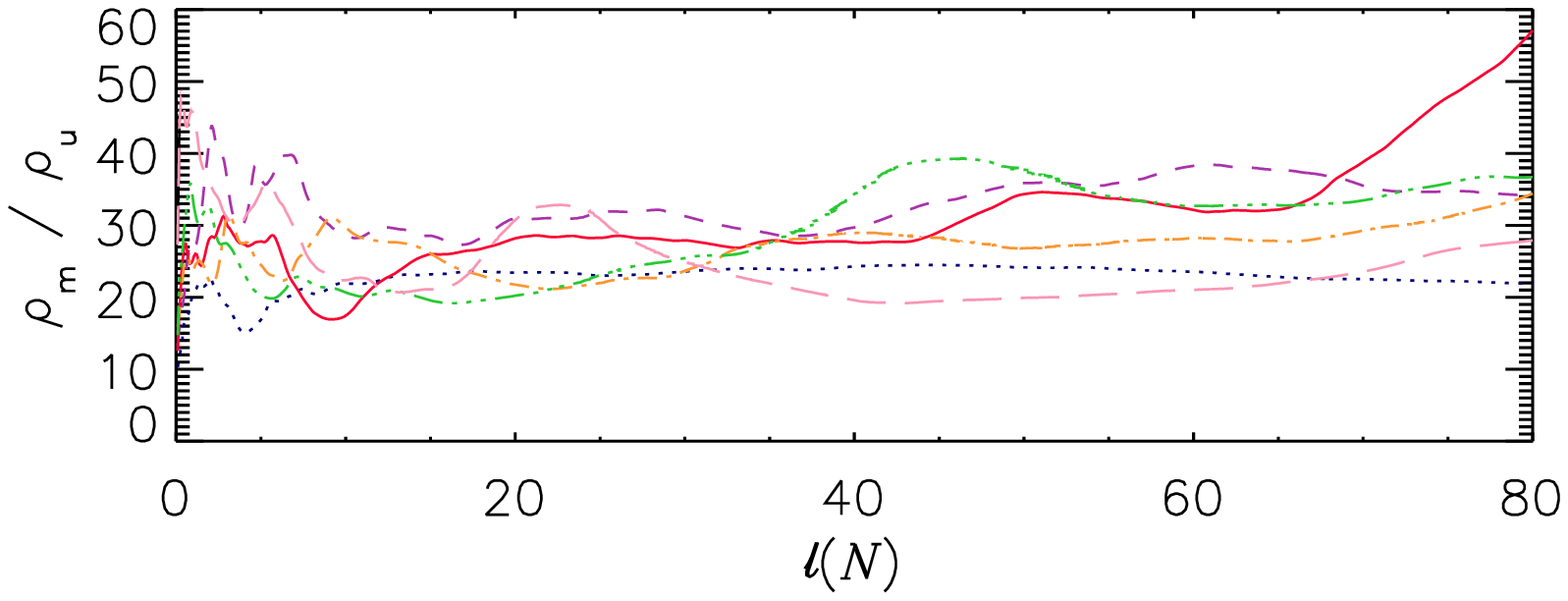}
}
\centerline{
\includegraphics[width=9.0cm]{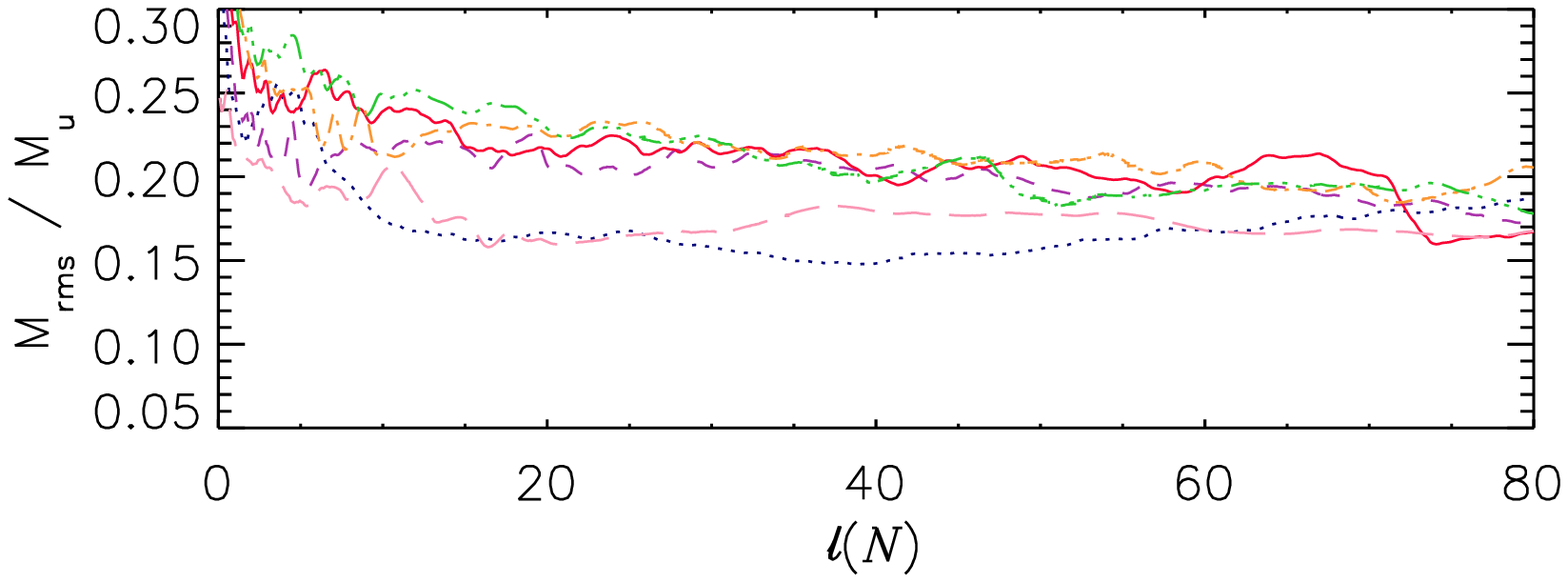}
}
\caption{Time evolution of corr$(\rho,v)$ (top), 
  $\rho_{\mathrm{m}}/\rho_{\mathrm{u}}$ (middle), and
  $M_{\mathrm{rms}}/M_\mathrm{u}$ (bottom) for runs R5\_0.2.4 (dotted,
  dark blue), R11\_0.2.4 (dashed, purple), R22\_0.2.4 (solid, red),
  R33\_0.2.4 (dash-dotted, orange), R43\_0.2.4 (dash-three-dots,
  green), and R87\_0.2.4 (long dashes, pink).  For these runs, $\ell(
  N ) = 60$ corresponds to $\ell_{\mathrm{cdl}} \approx
  \mathrm{Y}/2$.}
\label{fig:mean_tis}
\end{figure}

We first turn to the correlation of $\rho$ and $v$ and the CDL mean quantities
$\rho_{\mathrm{m}}$ and $M_{\mathrm{rms}}$, Eqs.~\ref{eq:exp_rho}
and~\ref{eq:exp_mach}. One of our basic assumptions in deriving these
self-similar relations, namely point b) that the CDL density and velocity are
uncorrelated, we find confirmed by our simulations.  For nearly all symmetric
simulations without initial CDL and for $ 10 \le \ell( N ) \le 70 $, we have
$0.1 \ge \mathrm{corr}(\rho,v) \ge -0.1$.  The only exceptions are the three
low Mach-number runs R11\_0.2.4, R11\_0.2.2, and R5\_0.2.4 with correlations
of about -0.2, -0.2, and -0.4 respectively. The top panel of
Fig.~\ref{fig:mean_tis} shows the time evolution of corr$(\rho,v)$ for five
selected runs that differ only in their upwind Mach-number, $5 \le
M_{\mathrm{u}} \le 90$.

In the middle and bottom panel of the the same figure,
$\rho_{\mathrm{m}}/\rho_{\mathrm{u}}$ and $M_{\mathrm{rms}}/M_{\mathrm{u}}$
are shown as a function of $\ell( N )$ for the same runs.  Two things are
apparent.  First, the ratios take similar values for all five runs, indicating
that indeed $\beta_{\mathrm{1}} \approx 0$ and $\beta_{\mathrm{2}} \approx 1$
for the exponents in Eqs.~\ref{eq:exp_rho} and~\ref{eq:exp_mach}.  Second,
the ratios are not constant with $\ell( N )$, indicating that the numerical
solution is indeed only approximately self-similar. We come back to this point
in Sect.~\ref{sec:discussion}.

To determine optimum exponents $\beta_{\mathrm{i}}$, $i=1,2$, we rewrite
Eqs.~\ref{eq:exp_rho} and~\ref{eq:exp_mach} as equations for
$\eta_{\mathrm{1}}$ and $\eta_{\mathrm{2}}$ and minimize the variance
$\sigma^{2}(\eta_{\mathrm{i}})$.  Considering all data points within $10 \le
\ell( N ) \le 70$ of all runs without a CDL at $t=0$, we find the smallest
variances for $\beta_{\mathrm{1}} = 0 $ and for $\beta_{\mathrm{2}} = 1$. The
corresponding means are $\mu(\eta_{\mathrm{1}}) \approx 28$ and
$\mu(\eta_{\mathrm{2}}) \approx 0.21$.  Although clearly identifiable, the
minima of $\sigma$ are relatively shallow. Changing $\beta_{\mathrm{1}}$ or
$\beta_{\mathrm{2}}$ by $\pm 0.1$, or excluding the very low Mach-number case
R5\_0.2.4 (for which $M_{\mathrm{rms}} \approx 0.9$) changes $\sigma$ by only
about 5\%.  By repeating the analysis but allowing for a linear dependence of
$\eta_{\mathrm{i}}$ on $\ell( N )$, we obtain the same optimum values for
$\beta_{\mathrm{1}}$ and $\beta_{\mathrm{2}}$ but with considerably smaller
variance. As $\ell( N )$ increases from 10 to 70, $\eta_{\mathrm{1}}$ rises
by about 25\% (from 25 to 31), while $\eta_{\mathrm{2}}$ decreases by about
15\% (from 0.22 to 0.19).

Part of our assumption a), namely the simple Mach-number dependencies of
$\rho_{\mathrm{m}}$ and $M_{\mathrm{rms}}$, thus seems justified.  With
$\eta_{\mathrm{2}} = 0.21$, assumption c), $\eta_{\mathrm{2}}^{2}
M_{\mathrm{u}}^{2} >> 1$, is also fulfilled for most of our simulations. An
exception is again run R5\_0.2.4, for which $\eta_{\mathrm{2}}^{2}
M_{\mathrm{u}}^{2} \approx 1$.

In summary, the simulation results, $\rho_{\mathrm{m}} \approx 28
\rho_{\mathrm{u}}$ and $M_{\mathrm{rms}} \approx 0.21 M_{\mathrm{u}}$,
essentially confirm the expected relations, Eqs.~\ref{eq:exp_rho}
and~\ref{eq:exp_mach}. $\eta_{\mathrm{1}}^{1/2} \eta_{\mathrm{2}}=1$,
predicted by Eq.~\ref{eq:eta12}, is fulfilled to within 10\% at any
given time. The mean density is (nearly) independent of
$M_{\mathrm{u}}$. As expected, the solution is only approximately
self-similar, $M_{\mathrm{rms}}$ decreases by about 15\% as $\ell( N)$
increases from 10 to 70.
\subsubsection{Confining shocks}
\label{sec:confshocks}
The turbulence within the CDL is driven by the upstream flows.  The confining
shocks of the CDL affect this driving in two ways. The less inclined the
shocks are on average with respect to the direction of the upstream flows
(smaller angle $\alpha_{\mathrm{eff}}$ in Eq.~\ref{eq:a_feff}), the more
kinetic energy survives shock passage and is available for driving the
turbulence. The smaller the spatial scale on which the angle $\alpha$ varies,
the smaller the scale on which the energy input changes. In the following, we
analyze how these shock properties depend on $M_{\mathrm{u}}$ and on
$\ell_{\mathrm{cdl}}$.

For this purpose, we specify the following basic quantities.  The discrete
x-position of the left and right shocks, $s_{\mathrm{l}}$ and
$s_{\mathrm{r}}$, defined for each discrete y-position $y_{j}$ as the two cell
boundaries where the Mach-number drops for the first time from its upwind
value $M_{u}$ to $0.8 M_{u}$. We determine the average extension of the CDL,
$\ell_{\mathrm{cdl}}$, as
\begin{equation}
\ell_{\mathrm{cdl}} = \frac{1}{J} \sum_{j=1}^{J} [ s_{\mathrm{r}}(y_{j}) - s_{\mathrm{l}}(y_{j}) ].
\label{eq:l_cdl_num}
\end{equation}
The length of the left and right shocks, $\ell_{\mathrm{sh,l}}$ and
$\ell_{\mathrm{sh,r}}$, are computed as
\begin{equation}
\ell_\mathrm{sh,i} = \sum_{j=1}^{J} [ (s_{\mathrm{i}}(y_{j}) - s_{\mathrm{i}}(y_{j-1}))^{2} + 
                                      (y_{j} - y_{j-1})^{2}]^{1/2},
\end{equation}
where $J$ is the number of cells in y-direction, and $i=l,r$. We define the
angle $\alpha_{\mathrm{l,r}}(y_{j})$ as the angle between the x-axis and the
tangent to the shock (see Fig.~\ref{fig:sketch2d}). Its numerical computation
is described in Appendix~\ref{app:alpha}. To obtain a number distribution, we
sort the values $\alpha_{\mathrm{l,r}}(y_{j}) \in [0,\pi/2]$ into 60 bins.
Finally, to obtain a measure for the scale on which the shocks are wiggled, we
look at the auto-correlation functions $\Gamma_{\mathrm{l,r}}$,
\begin{equation}
\Gamma_{\mathrm{i}}(y_{\mathrm{corr}}) = \frac{<[s_{\mathrm{i}}(y_{j}) - \bar{s}_{\mathrm{i}}]\cdot
                             [s_{\mathrm{i}}(y_{j}+y_{\mathrm{corr}}) - \bar{s}_{\mathrm{i}}]>}
                             {\sigma^{2}_{\mathrm{s}}},
\label{eq:def_autocorr}
\end{equation}
where $\sigma_{\mathrm{s}}^{2}$ is the variance of the shock position
$s_{\mathrm{i}}$, and $<.>$ denotes the mean over all discrete position
$y_{j}$. For each time, we determine $y_{\mathrm{corr_{\mathrm{0}}}}$
such that $\Gamma_{\mathrm{i}} (y_{\mathrm{corr_{\mathrm{0}}}}) =
0.5$. Averaging $y_{\mathrm{corr_{\mathrm{0}}}}$ over both shocks
gives a mean auto-correlation length $\ell_{\mathrm{corr}}$,
\begin{equation}
\ell_{\mathrm{corr}} = \frac{1}{2}
                       \left[ 
                       y_{\mathrm{corr_{\mathrm{0}}}}(s_{\mathrm{l}}) + 
                       y_{\mathrm{corr_{\mathrm{0}}}}(s_{\mathrm{r}})
                       \right].
\label{eq:lcorr}
\end{equation}
A larger auto-correlation length $\ell_{\mathrm{corr}}$ then indicates
that the shocks are wiggled on a larger spatial scale, but it does not
give the scale of the wiggles in absolute units (see below).

\begin{figure}[tp]
\centerline{\includegraphics[width=9.0cm]{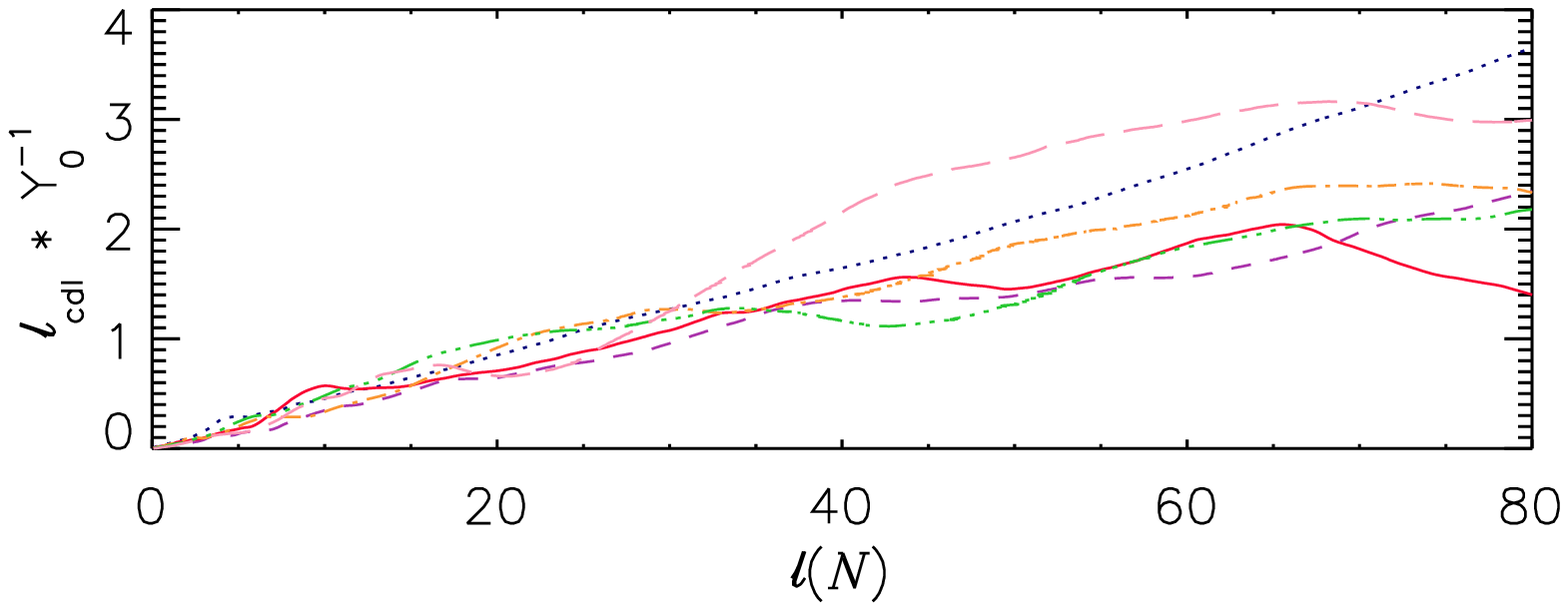}}
\centerline{\includegraphics[width=9.0cm]{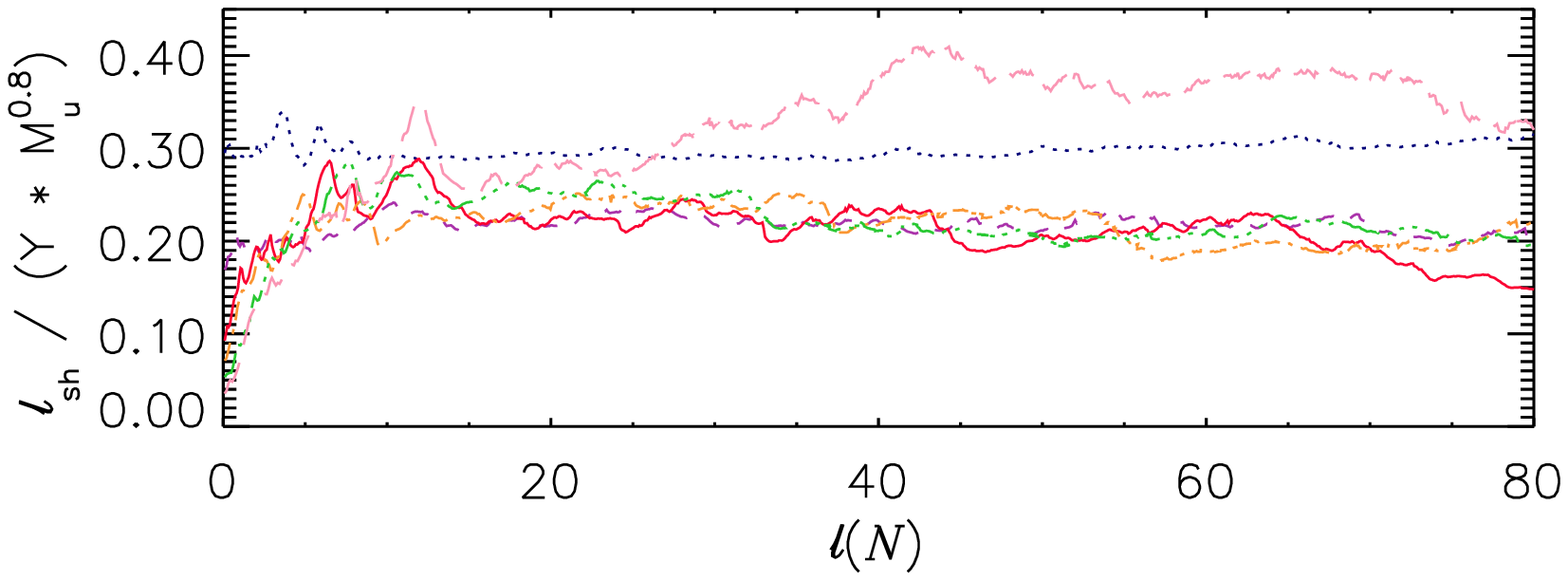}}
\centerline{\includegraphics[width=9.0cm]{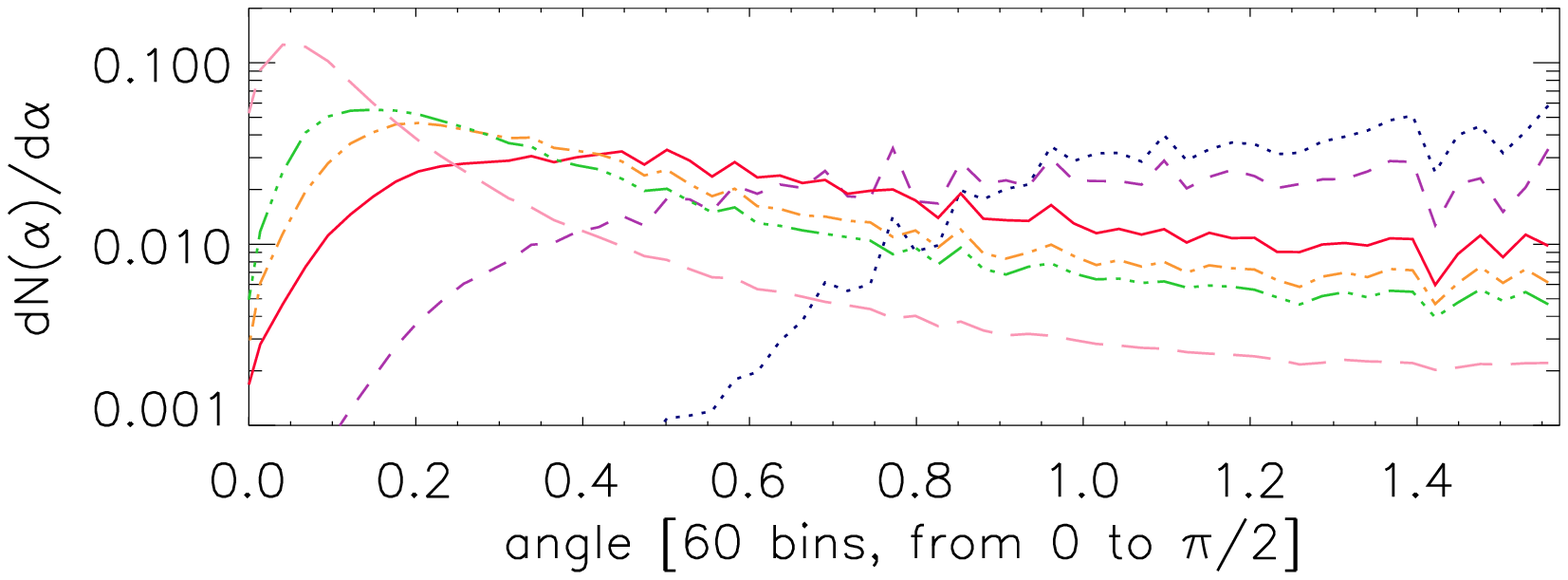}}
\centerline{\includegraphics[width=9.0cm]{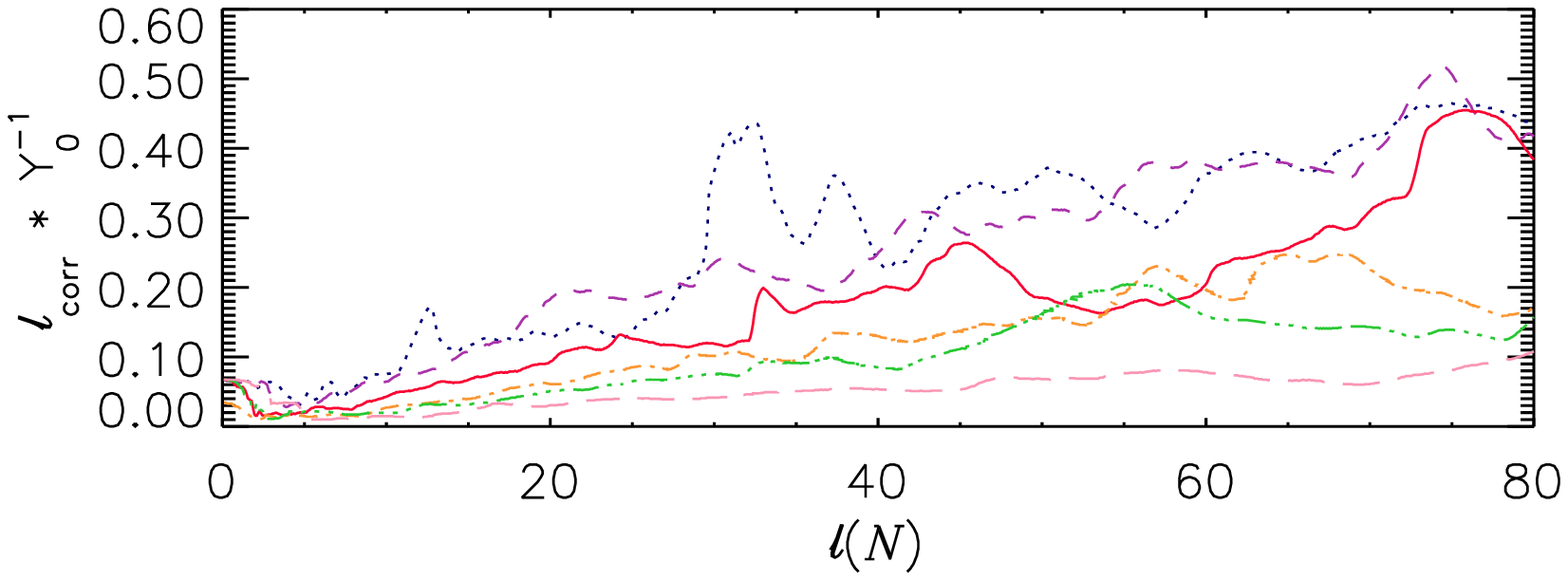}}
\centerline{\includegraphics[width=9.0cm]{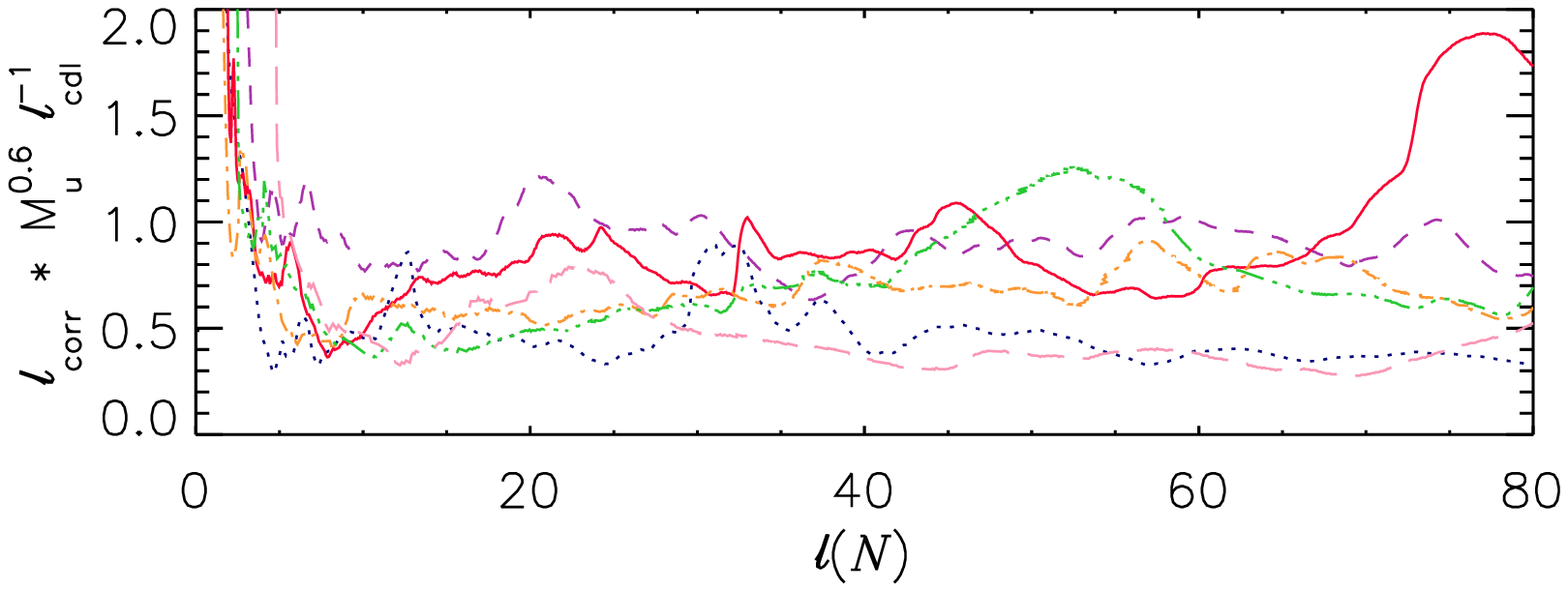}}
\caption{Quantities related to the confining shocks: average extension
  $\ell_{\mathrm{cdl}}$ of the CDL (first panel), total normalized
  shock length $l_{\mathrm{sh}}/ (\mathrm{Y} M_{\mathrm{u}}^{0.8})$
  (second panel), number distribution (60 bins) of obliqueness angle
  $\alpha$ averaged over $10 \le \ell( N ) \le 70$ (third panel),
  auto-correlation length $\ell_{\mathrm{corr}} /
  \mathrm{Y_{\mathrm{0}}}$ (fourth panel), and scaled auto-correlation
  length $\ell_{\mathrm{corr}} / (\ell_{\mathrm{cdl}}
  M_{\mathrm{u}}^{-0.6})$, (fifth panel).  Individual curves denote
  the same runs as in Fig.~\ref{fig:mean_tis}.}
\label{fig:shell_shock_tis}
\end{figure}

All four quantities, CDL extension, number distribution of angle
$\alpha$, shock length, and correlation length, are shown in
Fig.~\ref{fig:shell_shock_tis}.
\begin{figure}[tp]
\centerline{\includegraphics[width=9.0cm]{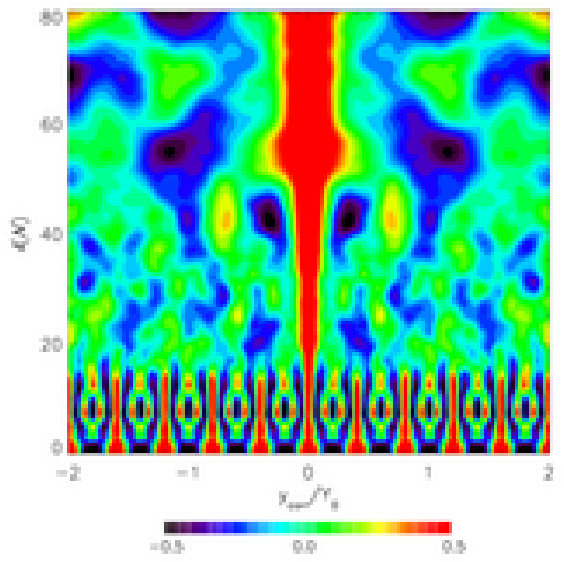}}
\centerline{\includegraphics[width=9.0cm]{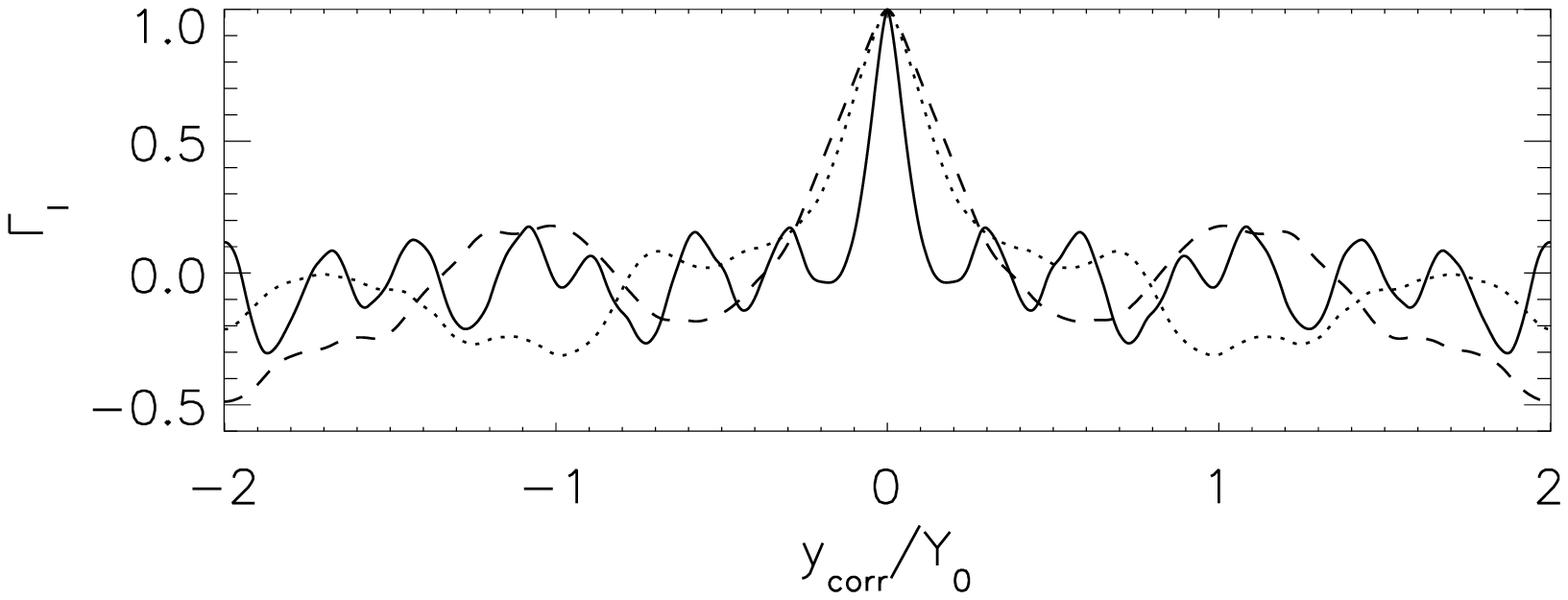}}
\caption{Variation of $\Gamma_{\mathrm{l}}$, color coded, as a
  function of $y_{\mathrm{corr}}$ for run R43\_0.2.4 (top panel). To
  allow for better display the color scale is limited to a range $-0.5
  \le \Gamma_{\mathrm{l}} \le +0.5$. Lower or higher values of
  $\Gamma_{\mathrm{l}}$ are uniformly colored in dark blue or red,
  respectively.  For the same run, $\Gamma_{\mathrm{l}}$ is shown as a
  function of $y_{\mathrm{corr}}$ for three selected times (bottom
  panel). $\ell ( N ) = 30$ (solid), $\ell ( N ) = 50$ (dotted), $\ell
  ( N ) = 70$ (dashed).}
\label{fig:gamma}
\end{figure}
The first panel of Fig.~\ref{fig:shell_shock_tis} shows the essentially linear
growth of the CDL with $\ell( N )$. The growth rate, however, slowly decreases
with increasing $\ell( N )$.  The slope of a linear fit in the range $40 <
\ell( N ) < 70$ is roughly 10\% flatter than the slope obtained in the range
$10 < \ell( N ) < 40$.  This fits with the slight increase in
$\rho_{\mathrm{m}}$, observable in the middle panel of
Fig.~\ref{fig:mean_tis}.  The second panel of Fig.~\ref{fig:shell_shock_tis}
shows that the average shock length $\ell_\mathrm{sh} = 0.5(\ell_\mathrm{sh,l}
+ \ell_\mathrm{sh,r})$ is fairly constant with respect to $\ell( N )$ but
increases with $M_{\mathrm{u}}$.  Assuming a dependence of the form
$\ell_\mathrm{sh} = \eta_{\mathrm{sh}} \mathrm{Y}
M_{\mathrm{u}}^{\beta_{\mathrm{sh}}}$, the variance $\sigma^{2}
(\eta_{\mathrm{sh}})$ becomes minimal for $\beta_{\mathrm{sh}}=0.8$.  As can
be seen, the two runs R5\_0.2.4 and R87\_0.2.4 again behave somewhat
differently. If we neglect these two runs, $\beta_{\mathrm{sh}}$ remains
unchanged but $\sigma$ is reduced by about 40\%.  The third panel of
Fig.~\ref{fig:shell_shock_tis} shows that larger upwind Mach-numbers lead to
less inclined shocks with respect to the direction of the upstream flows
(lower values of $\alpha$).  Shown is the number distribution of $\alpha$,
averaged over $10 \le \ell( N ) \le 70$. Individual runs show a slight shift
towards higher values of $\alpha$ as $\ell( N )$ increases.  This shift is,
however, small compared to the effect of $M_{\mathrm{u}}$.  The fourth panel
of Fig.~\ref{fig:shell_shock_tis} shows the auto-correlation length
$\ell_{\mathrm{corr}}$. It not only depends on $M_{\mathrm{u}}$ but is also
proportional to $\ell_{\mathrm{cdl}}$.  The best fit is found to be
$\ell_{\mathrm{corr}} \approx 0.7 \ell_{\mathrm{cdl}} M_{\mathrm{u}}^{-0.6}$.
The fifth panel of Fig.~\ref{fig:shell_shock_tis} shows $\ell_{\mathrm{corr}}$
scaled with this best fit. From these scaling properties of
$\ell_{\mathrm{corr}}$, we take that higher values of $M_{\mathrm{u}}$ lead to
smaller scale wiggling of the shocks with respect to $\ell_{\mathrm{cdl}}$.

The absolute value of $\ell_{\mathrm{corr}}$ clearly depends on the
choice of the threshold value in our definition, $\Gamma (
y_{\mathrm{corr}} ) = 0.5$. Figure~\ref{fig:gamma} illustrates the
variation of $\Gamma_{\mathrm{l}}$ as a function of
$y_{\mathrm{corr}}$ at the example of run R43\_0.2.4.
The top panel of Fig.~\ref{fig:gamma} shows that the initially present
sinusoidal wiggling of the confining shocks does not get lost
until about $\ell ( N ) = 15$, which is rather late compared to the
other runs.  Mode-like signatures again appear around $\ell ( N )
\gapprox 50$.  Our data give, however, no clear answer to how typical
and persistent such signatures are.  A basic problem is that their
wave length soon becomes comparable (within a factor of 2 or so) to
the domain size in the y-direction, which may affect the signatures.  From
the bottom panel of Fig.~\ref{fig:gamma}, on the other
hand, it can be taken that $\Gamma_{\mathrm{l}}$ essentially decreases linearly from 1
to about 0.2. The other simulations show a similar behavior.
Consequently, the above scaling properties of $\ell_{\mathrm{corr}}$
should also be obtained if smaller threshold values are used, down to
about $\Gamma ( y_{\mathrm{corr}} )= 0.2$.

Figs.~\ref{fig:mean_tis} and~\ref{fig:shell_shock_tis} also allow some insight
into why runs R5\_0.2.4 and R87\_0.2.4 sometimes fit not so well. The third
panel of Fig.~\ref{fig:shell_shock_tis} shows that our spatial resolution is
barely sufficient for run R87\_0.2.4, the largest upwind Mach-number we have
considered. The number distribution here peaks at around $\alpha \approx 0.1$.
In terms of discrete positions this means that the shock position changes by
about 15 cells in the x-direction as one moves from $y_{j}$ to $y_{j+1}$. Run
R5\_0.2.4, on the other hand, may deviate just because of its low Mach-number.
The turbulence within its CDL is subsonic, $M_{\mathrm{rms}} \approx 0.9$; and
with $\eta_{\mathrm{2}}^{2}M_{\mathrm{u}}^{2} \approx 1.1$ and
$\mathrm{corr}(\rho,v) \approx -0.4$ (Fig.~\ref{fig:mean_tis}, top panel), it
violates two of the basic assumptions made when deriving the self-similar
scaling laws in Sect.~\ref{sec:anal-scaling_2d}.

In summary, as $M_{\mathrm{u}}$ increases, the bounding shocks become
less inclined with respect to the direction of the upstream flows (smaller $\alpha$),
the fraction of upstream kinetic energy that survives the passage
through the bounding shocks increases, and the bounding shocks
themselves are wiggled on progressively smaller scales (smaller
$\ell_{\mathrm{corr} }/ \ell_{\mathrm{cdl}}$.
\subsubsection{Energy balance}
\label{sec:driving_efficiency}
\begin{figure}[tp]
\centerline{\includegraphics[width=9.0cm]{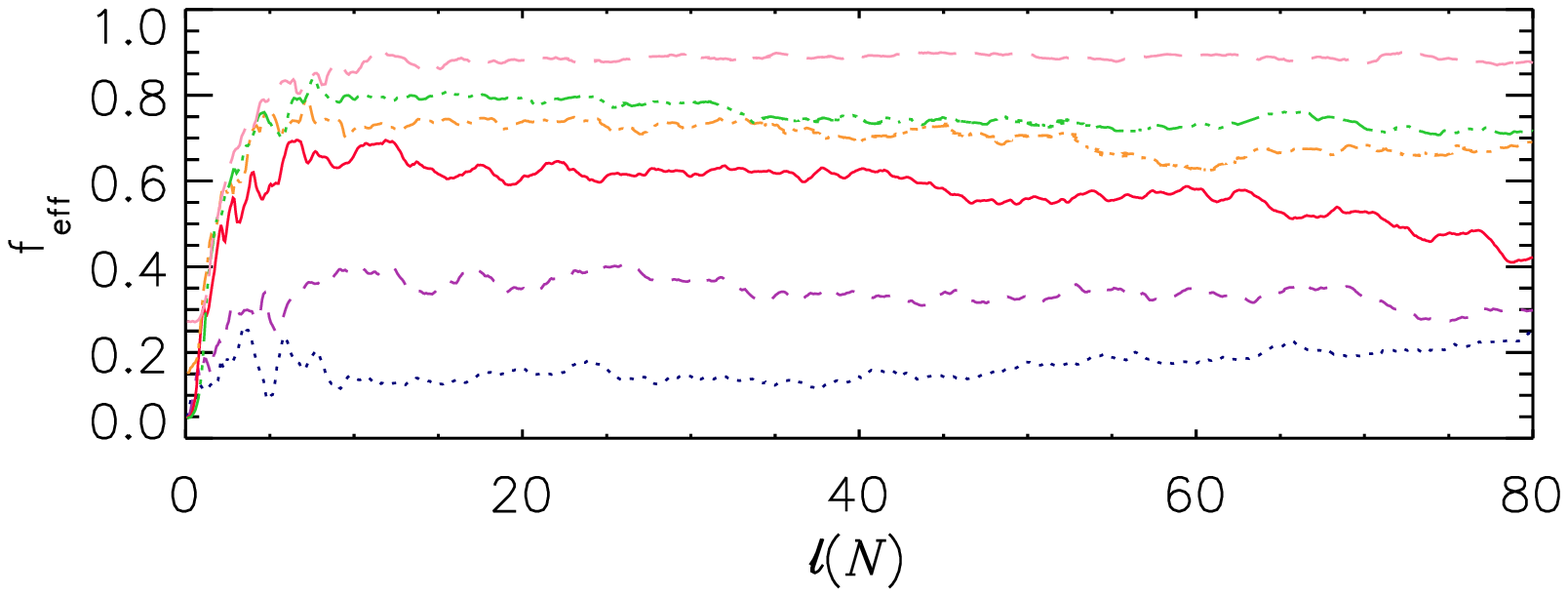}}
\centerline{\includegraphics[width=9.0cm]{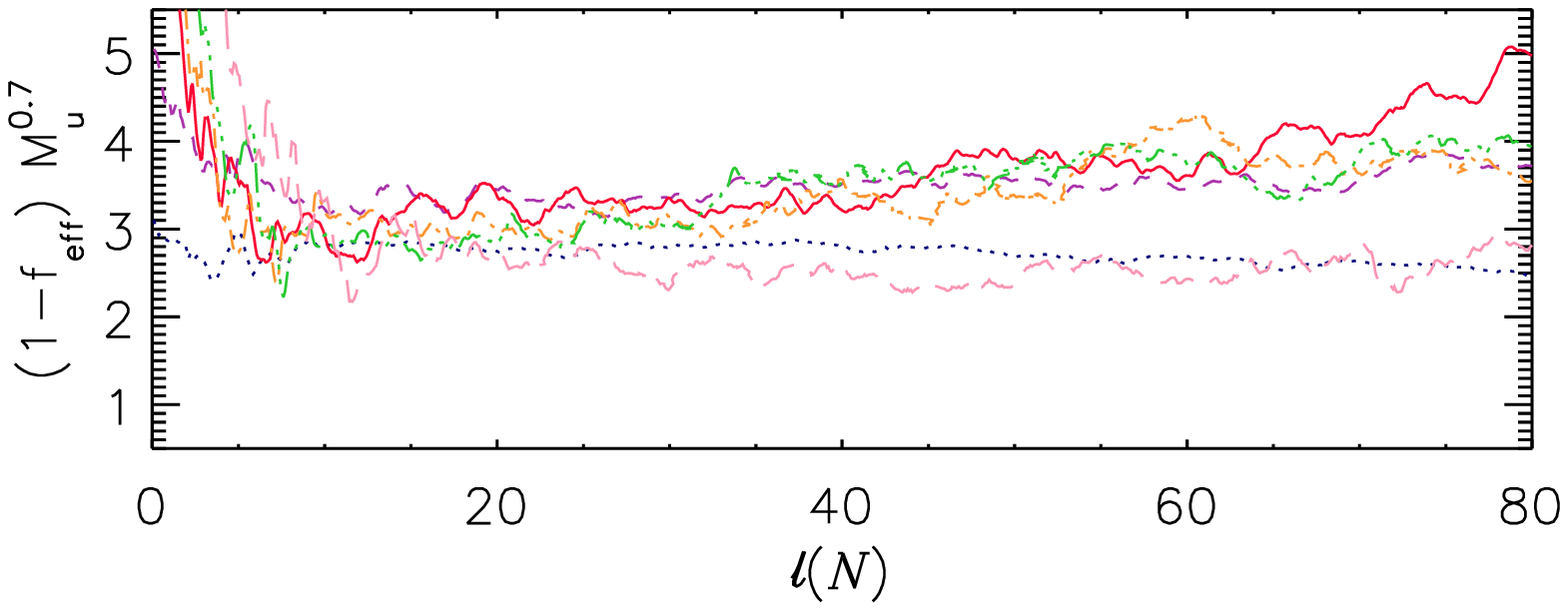}}
\caption{Driving efficiency (top panel) and best fit
  $\eta_{\mathrm{3}} = (1 - f_{\mathrm{eff}}) M_{\mathrm{u}}^{0.7}$
  (bottom panel). For details see text. }
\label{fig:driving-efficiency}
\end{figure}
Energy input into the CDL occurs only at its confining interfaces.  Energy
dissipation, on the other hand, occurs throughout the CDL volume.
Nevertheless, according to the analysis in Sect.~\ref{sec:anal-scaling_2d}
both $\dot{\cal E}_{\mathrm{drv}}$ and $\dot{\cal E}_{\mathrm{diss}}$ should
be independent of the CDL extension if dissipation is only due to shocks
  and if $\ell_{\mathrm{cdl}}$ is small compared to $\mathrm{Y}$.  The
average distance between shocks must then increase and / or the average
strength of the shocks must decrease as the CDL grows.

\begin{figure}[tp]
\centerline{\includegraphics[width=9.0cm]{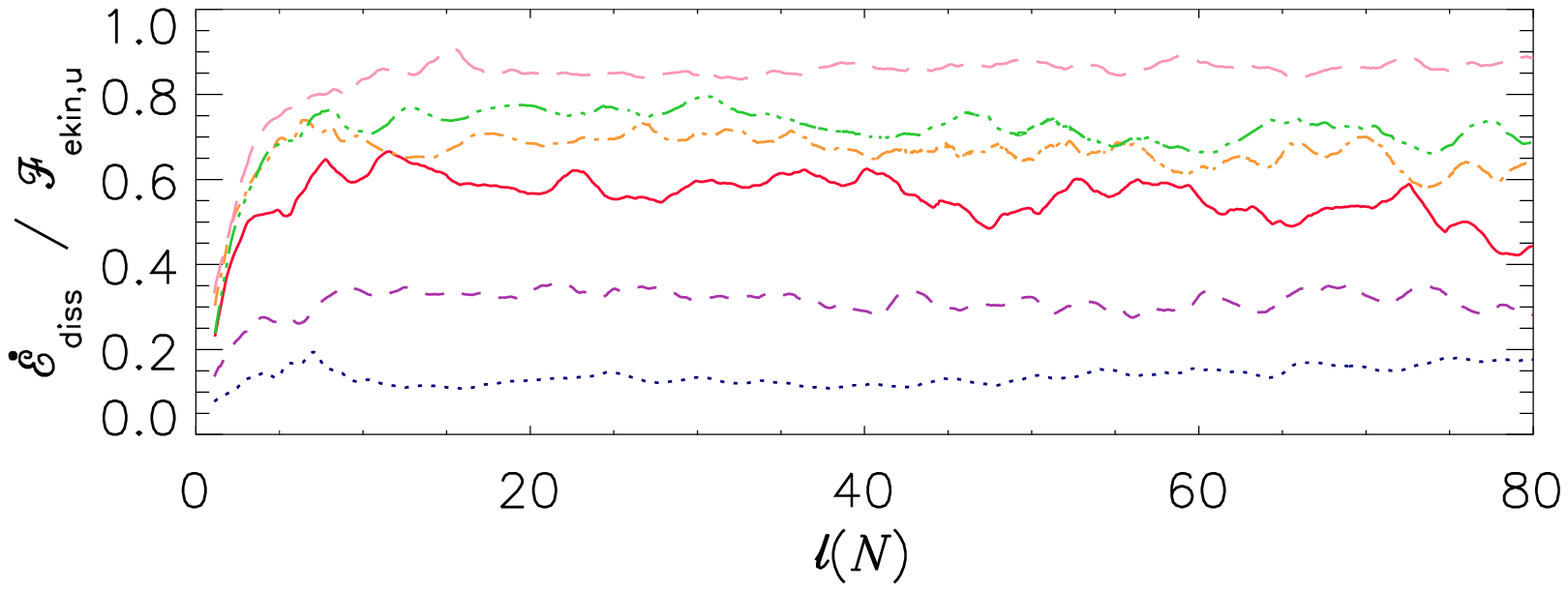}}
\centerline{\includegraphics[width=9.0cm]{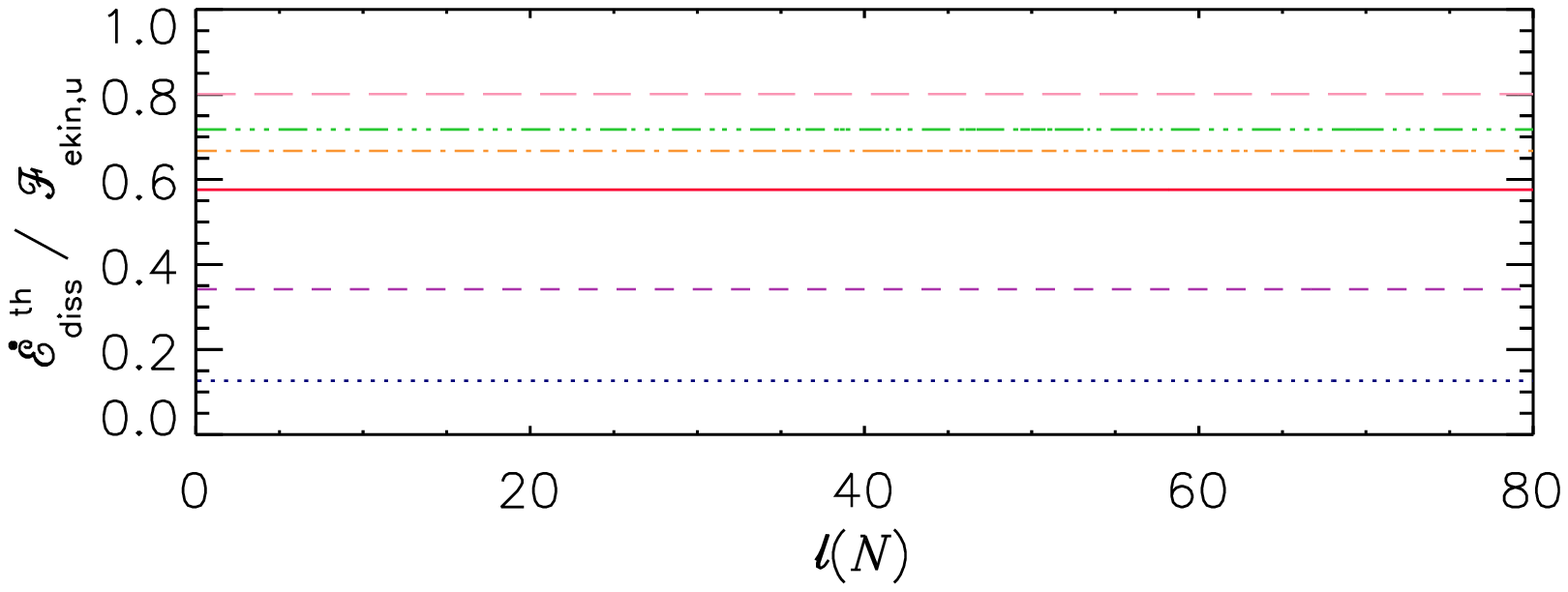}}
\centerline{\includegraphics[width=9.0cm]{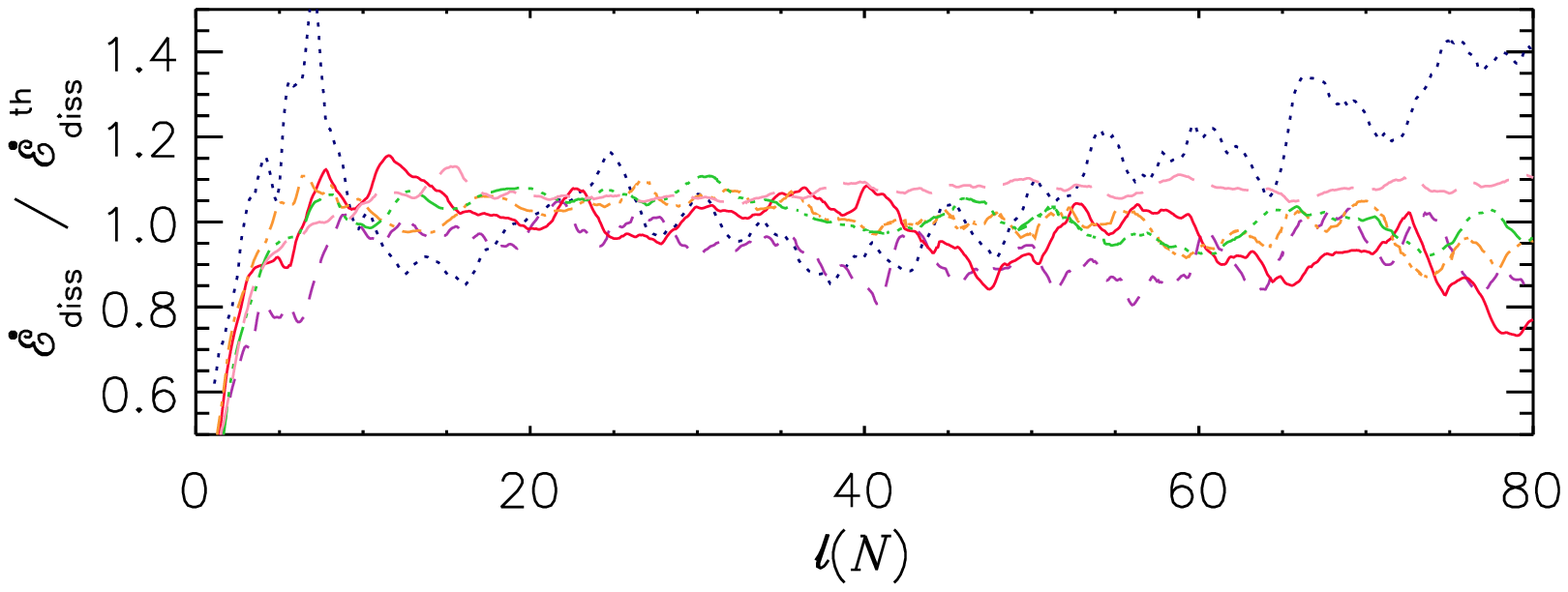}}
\caption{ Numerically obtained (top panel) and theoretically expected
  (middle panel) energy dissipation in units of the upstream kinetic energy
  flux density ${\cal F}_{\mathrm{e_{\mathrm{kin}},u}} = \rho_{\mathrm{u}}
  v_{\mathrm{u}}^{3}$. The constants in Eq.~\ref{eq:exp_ediss} were set to the
  best fit values, $\eta_{\mathrm{3}} = 3.3$, $\beta_{\mathrm{3}}= -0.7$, and
  $\eta_{\mathrm{2}} = 0.21$. We used $\eta_{\mathrm{3}} = 2.75$ for run
  R5\_0.2.4 (for details see text).  The bottom panel shows the ratio of the
  two quantities.  Individual curves denote the same runs as in
  Fig.~\ref{fig:mean_tis}.  For better display, $\dot{\cal E}_{\mathrm{diss}}$
  was smoothed using a running mean with time window $\Delta \ell ( N ) = \pm
  1$.}
\label{fig:energy-dissipation}
\end{figure}
\begin{figure}[htp]
\centerline{\includegraphics[width=9.0cm]{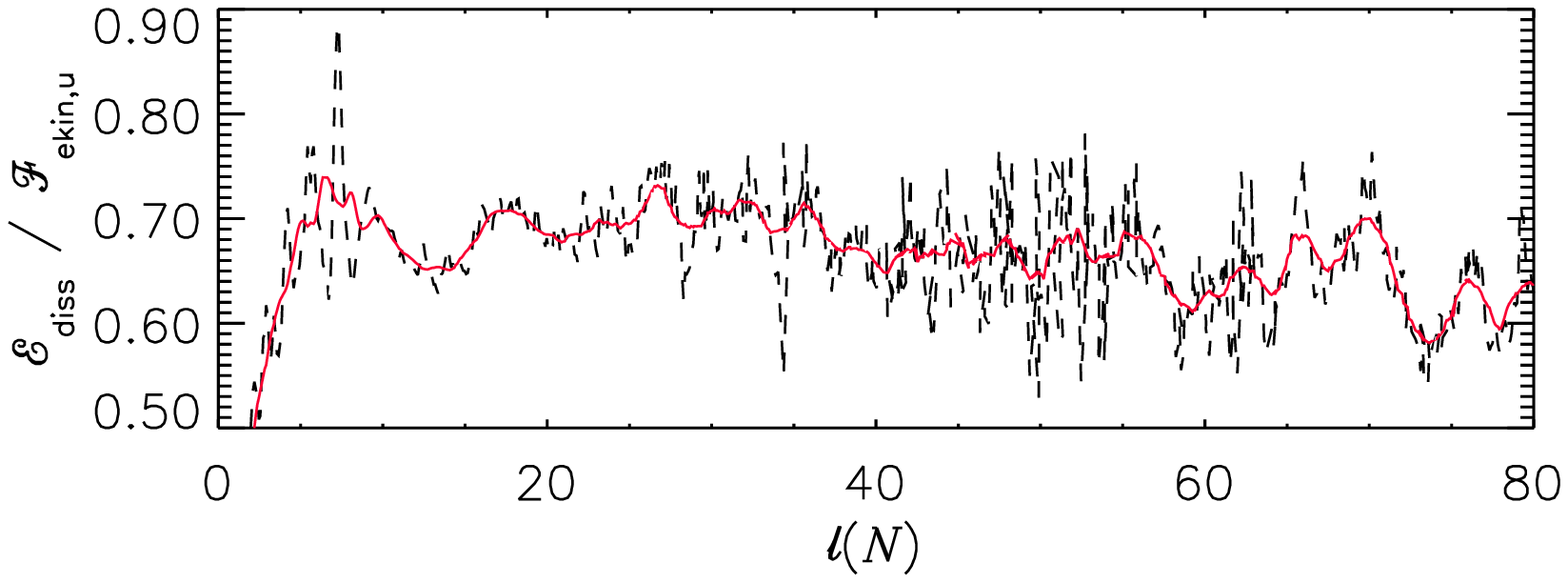}}
\caption{Effect of smoothing $\dot{\cal E}_{\mathrm{diss}}$ with a running mean and
  window $\Delta \ell ( N ) = \pm 1$,
  illustrated by run R33\_0.2.4. Shown is $\dot{\cal
    E}_{\mathrm{diss}}$ in units of  ${\cal F}_{\mathrm{e_{\mathrm{kin}},u}} =
\rho_{\mathrm{u}}v_{\mathrm{u}}^{3}$, before (dashed, black) and after (solid, red)
  smoothing, in units of erg cm$^{-3}$s$^{-1}$.}
\label{fig:smoothing}
\end{figure}
To determine $\dot{\cal E}_{\mathrm{drv}}$ we must compute the driving
efficiency $f_{\mathrm{eff}} = \dot{\cal E}_{\mathrm{drv}} / {\cal
  F}_{\mathrm{e_{\mathrm{kin}},u}}$.  The corresponding integral in
Eq.~\ref{eq:a_feff} is evaluated numerically, and the resulting driving
efficiency is shown in the top panel of Fig.~\ref{fig:driving-efficiency}.  As
can be seen, larger Mach-numbers lead to more efficient driving, and a smaller
part of the upstream kinetic energy is thermalized already at the confining
shocks. The driving efficiency $f_{\mathrm{eff}}$ increases by about a factor
of four between runs R5\_0.2.4 and R87\_0.2.4. Also noteworthy is that the
absolute value of the driving power $\dot{\cal E}_{\mathrm{drv}}$ differs by
more than 4 orders of magnitude between runs R5\_0.2.4 and R87\_0.2.4.  The
best fit for the assumed Mach-number dependence (minimization of
$\sigma(\eta_{\mathrm{3}})$ in Eq.~\ref{eq:b_feff}) yields $\beta_{\mathrm{3}}
= -0.7$. The corresponding values of $\eta_{\mathrm{3}} = (1 -
f_{\mathrm{eff}})M_{\mathrm{u}}^{0.7}$ are shown in the bottom panel of
Fig.~\ref{fig:driving-efficiency}. From the figure we take that the second
part of our assumption a), the simple Mach-number dependence of
$f_{\mathrm{eff}}$, seems justified. The figure also shows that
$f_{\mathrm{eff}}$, and thus the driving power $\dot{\cal E}_{\mathrm{drv}}$,
is not strictly independent of $\ell_{\mathrm{cdl}}$ but decreases with
increasing $\ell( N )$.  Repeating the best fit analysis but allowing for a
linear dependence of $\eta_{\mathrm{3}}$ on $\ell( N )$ again leads to
$\beta_{\mathrm{3}} = -0.7$, while $\eta_{\mathrm{3}}$ changes from 3.1 to 3.6
as $\ell( N )$ goes from 10 to 70. The average value of $\eta_{\mathrm{3}}$ is
3.3. Omission of the extreme runs R5\_0.2.4 and R87\_0.2.4 does not change the
result.

We determine the dissipated energy as $\dot{\cal E}_{\mathrm{diss}}= \dot{\cal
  E}_{\mathrm{drv}} - \dot{\cal E}_{\mathrm{kin}} $
(Sect.~\ref{sec:a_drive_eff}), where $\dot{\cal E}_{\mathrm{kin}}$ is the
change per time of the kinetic energy within an average column of the CDL, and
$\dot{\cal E}_{\mathrm{kin}}$ is directly from our simulation data.
Figure~\ref{fig:energy-dissipation} shows the numerically obtained value
$\dot{\cal E}_{\mathrm{diss}}$ (top panel) and the theoretically expected
value (Eq.~\ref{eq:exp_ediss}) $\dot{\cal E}_{\mathrm{diss}}^{\mathrm{th}}$
(middle panel), both in units of ${\cal F}_{\mathrm{e_{\mathrm{kin}},u}} =
\rho_{\mathrm{u}} v_{\mathrm{u}}^{3}$, as well as the ratio of the two (bottom
panel).  For better display, the theoretical value, which must not depend on
$\ell ( N )$, is shown as a (constant) function of $\ell ( N )$.  For the
constants in Eq.~\ref{eq:exp_ediss} we used the numerically obtained average
values, $\eta_{\mathrm{3}} = 3.3$, $\beta_{\mathrm{3}}= -0.7$, and
$\eta_{\mathrm{2}} = 0.21$.  We used $\eta_{\mathrm{3}} = 2.75$ only for
R5\_0.2.4, in accordance with the bottom panel of
Fig.~\ref{fig:driving-efficiency}.  The numerically obtained value was
smoothed for better display using a running mean with window size $\Delta
\ell( N ) = \pm 1$.  The effect of the smoothing is illustrated in
Fig.~\ref{fig:smoothing} with the example of run R11\_0.2.4.

Looking at the data of $\dot{\cal E}_{\mathrm{diss}}$ and $\dot{\cal
  E}_{\mathrm{drv}}$, three points may be stressed.  First, $\dot{\cal
  E}_{\mathrm{diss}}$ (Fig.~\ref{fig:energy-dissipation}, top panel)
mirrors $\dot{\cal E}_{\mathrm{drv}} = {\cal
  F}_{\mathrm{e_{\mathrm{kin}},u}} f_{\mathrm{eff}}$
(Fig.~\ref{fig:driving-efficiency}, top panel), and the values
  usually differ by less than 10\%.  This is not surprising. It
implies, however, that for larger upstream Mach-numbers, a larger 
fraction of the upstream kinetic energy is thermalized only within the
volume of CDL and not already at its confining shocks.  For
$M_{\mathrm{u}} \gapprox 20$, the energy dissipated within the CDL
exceeds 50\% of the upstream kinetic energy
(Fig.~\ref{fig:energy-dissipation}, top panel).

Second, the bottom panel of Fig.~\ref{fig:energy-dissipation} shows that
$\dot{\cal E}_{\mathrm{diss}}^{\mathrm{th}}$ and $\dot{\cal
  E}_{\mathrm{diss}}$ agree to within 10\% most of the time.  Given
the wide range covered (5 orders of magnitude in $\dot{\cal
  E}_{\mathrm{diss}}$, a factor of 20 in $M_{\mathrm{u}}$, and an
increase by a factor of 7 in $\ell ( N )$), we conclude that the
self-similar solution gives a good estimate.
  
Third, from the same figure it can be seen that $\dot{\cal
  E}_{\mathrm{diss}}$ generally decreases, except for run R5\_0.2.4.
Excluding R5\_0.2.4, a linear fit to $\dot{\cal E}_{\mathrm{diss}} /
\dot{\cal E}_{\mathrm{diss}}^{\mathrm{th}}$ yields a decrease of 10\%
as $\ell ( N )$ increases from 10 to 70. A similar fit to $\dot{\cal
E}_{\mathrm{drv}} / \dot{\cal E}_{\mathrm{drv}}^{\mathrm{th}}$ with
$\dot{\cal E}_{\mathrm{drv}}^{\mathrm{th}} = \rho_{\mathrm{u}}
v_{\mathrm{u}}^{3} ( 1 - 3.3 M_{\mathrm{u}}^{-0.7})$ yields an even
slightly larger decrease of 13\%.  The net dissipation, $\dot{\cal
E}_{\mathrm{diss}} / \dot{\cal E}_{\mathrm{drv}}$, in fact increases
by 3\%.  Thus, as the CDL size increases, the absolute dissipation
within an average column decreases while the net dissipation
increases.

In summary, the predicted scaling laws, Eqs.~\ref{eq:exp_edrv}
to~\ref{eq:exp_ediss}, are -- within the range of applicability -- essentially
confirmed by the simulations.  The fraction of upstream kinetic energy
dissipated only within the CDL, and not at the confining shocks, thus
increases with $M_{\mathrm{u}}$.  Best-fit analysis for the numerical
constants yields $f_{\mathrm{eff}} = 1 - 3.3 \; M_{\mathrm{u}}^{-0.7}$.  Both
$\dot{\cal E}_{\mathrm{drv}}$ and $\dot{\cal E}_{\mathrm{diss}}$ decrease
slightly with increasing $\ell_{\mathrm{cdl}}$. The net dissipation rate
$\dot{\cal E}_{\mathrm{diss}} / \dot{\cal E}_{\mathrm{drv}}$ increases, but
only slightly (3\% increase as $\ell ( N )$ goes from 10 to 70.)
\begin{figure}[tp]
\centerline{\includegraphics[width=9.0cm]{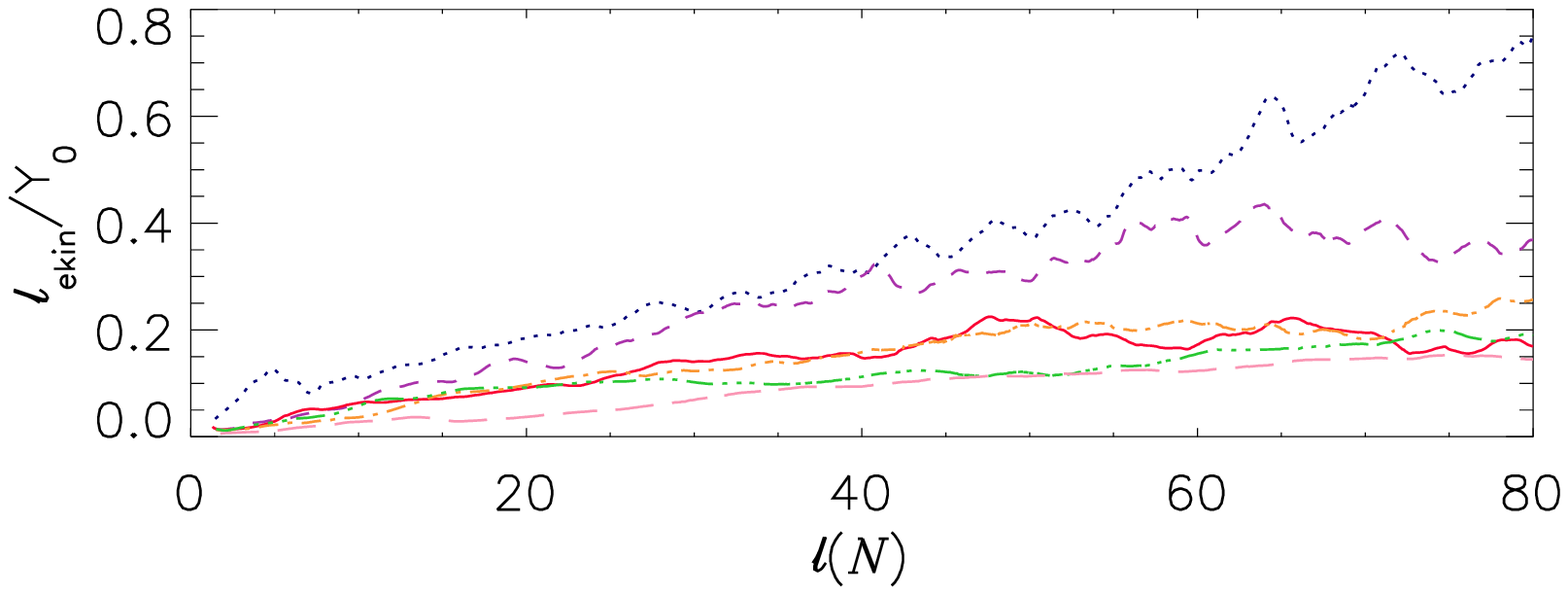}}
\centerline{\includegraphics[width=9.0cm]{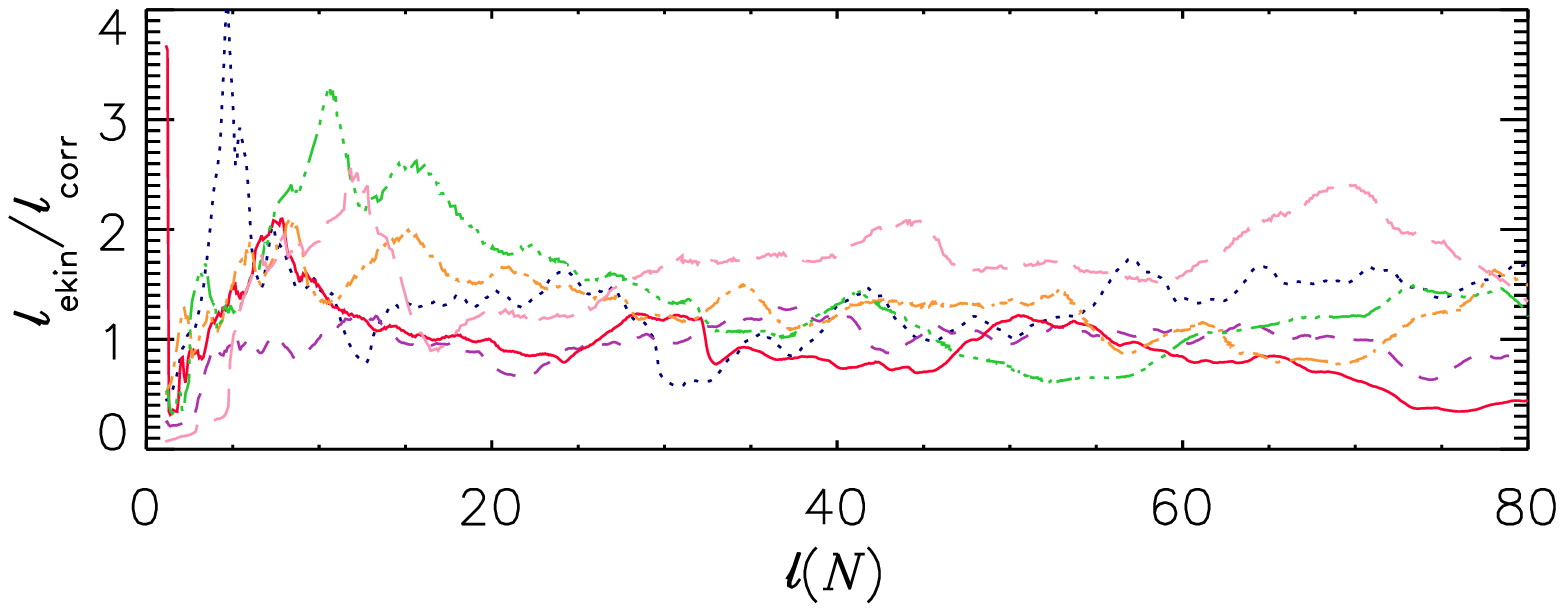}}
\centerline{\includegraphics[width=9.0cm]{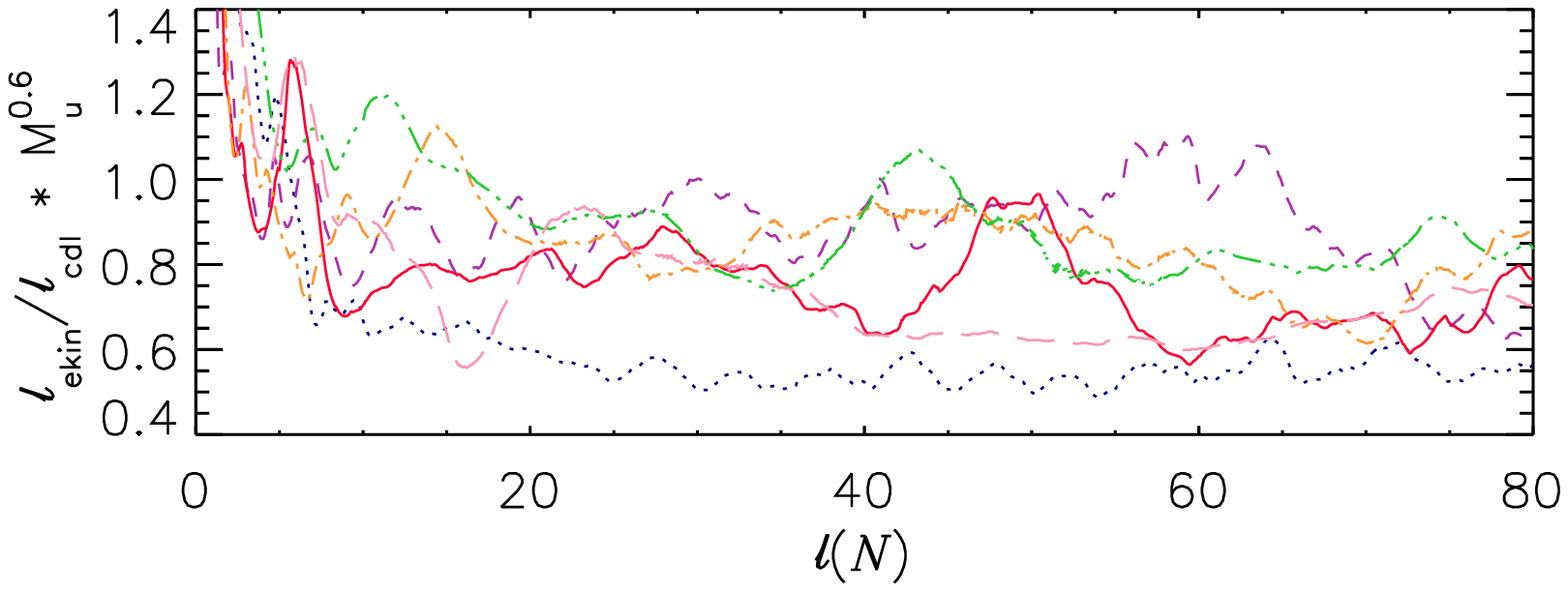}}
\caption{Characteristic length $\ell_{\mathrm{e_{\mathrm{kin}}}}$ of
  the turbulence (top), in units of $\ell_{\mathrm{corr}}$ (middle),
  and scaled with best-fit $\ell_{\mathrm{cdl}} M_{\mathrm{u}}^{0.6}$
  (bottom) as functions of $\ell( N )$. Individual curves denote the
  same runs as in Fig.~\ref{fig:mean_tis}. For better display,
  $\ell_{\mathrm{ekin}}$ was smoothed by a running mean with window
  $\Delta \ell ( N ) = \pm 1$.}
\label{fig:turbulence-length}
\end{figure}
\begin{figure}[tp]
  \centerline{\includegraphics[width=8.0cm]{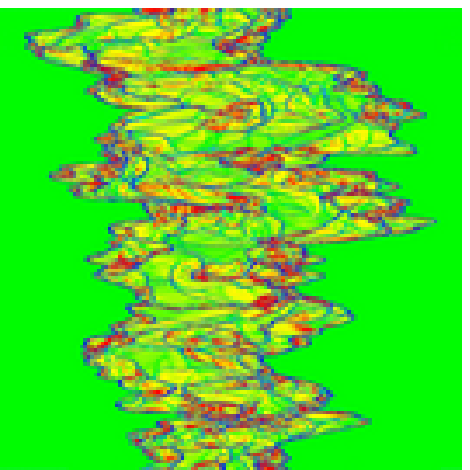}}
\vspace{.2cm}
  \centerline{\includegraphics[width=8.0cm]{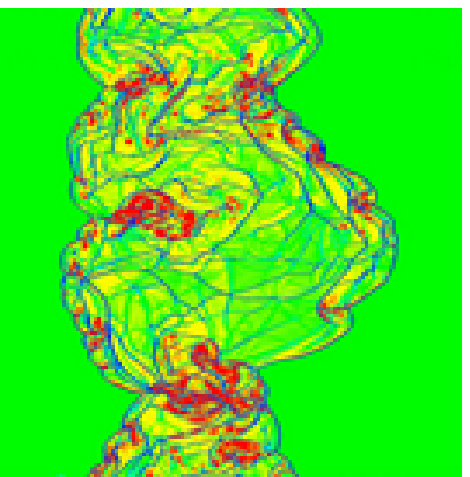}}
\caption{Plots of $\mathrm{div}(\vec{v})$ for two runs that are
  identical except for their upstream Mach-number. Larger upstream
  Mach-numbers lead, on average, to finer structure within the CDL and
  smaller scale wiggling of the confining shocks. Shown are runs
  R33\_0.2.4 (top) and R11\_0.2.4 (bottom), both at a time when
  $\ell_{\mathrm{cdl}} \approx 2 \, \mathrm{Y}_{\mathrm{0}} =
  \mathrm{Y}/2$. Blue (dark lines) indicates convergence, red (dark
  patches) divergence.}
\label{fig:div-ti1.5-tihf}
\end{figure}
\subsubsection{Length scales  of the turbulence}
\label{sec:driving-wave length}
In Sect.~\ref{sec:confshocks} we looked at the scaling properties of
the confining shocks and pointed out that shorter auto-correlation
lengths $\ell_{\mathrm{corr}}$ imply smaller-scale wiggling, thus
smaller scale changes of the kinetic energy entering the CDL. In the
following, we show that the interface based quantity
$\ell_{\mathrm{corr}}$ is proportional to the length scale derived
from the volume properties of the turbulence.  We take this as
evidence of the tight coupling between volume and interface
properties, between the turbulence and its driving.

On dimensional grounds, we can define two length scales based on
volume properties of the turbulence,
\begin{eqnarray}
\label{eq:lambda_ekin}
\ell_{\mathrm{e_{\mathrm{kin}}}} & \equiv & 
              \frac{ N ^{-1/2} {\cal E}_{\mathrm{kin}}^{3/2}}{\dot{\cal E}_{\mathrm{diss}}}, \\
\ell_{\mathrm{v_{\mathrm{rms}}}}   & \equiv & 
              \frac{ N v_{\mathrm{rms}}^{3}}{\dot{\cal E}_{\mathrm{diss}}},
\label{eq:lambda_vrms}
\end{eqnarray}
where ${\cal E}_{\mathrm{kin}} = 
\frac{\ell_{\mathrm{cdl}}}{\mathrm{2V}}\int_{\mathrm{V}}
\rho v^{2}$ is the average column integrated kinetic energy
density. Here $V$ is again the 2D volume of the CDL, introduced
in Sect.~\ref{sec:num_settings}. The two scales are equal up to
a numerical constant if the density and velocity are uncorrelated,
in which case we can replace the average over the product $\rho
v^{2}$ by the product of the averages of $\rho$ and $v^{2}$, ${\cal
E}_{\mathrm{kin}} = \ell_{\mathrm{cdl}} \rho_{\mathrm{m}} 
v_{\mathrm{rms}}^{2} = N v_{\mathrm{rms}}^{2}$.
As this is the case in most of our simulations we look at only one of
the above quantities in the following,
$\ell_{\mathrm{e_{\mathrm{kin}}}}$, shown in
the top panel of Fig.~\ref{fig:turbulence-length}. For better display, as
$\ell_{\mathrm{e_{\mathrm{kin}}}}$ inherits the large time variability
of $\dot{\cal E}_{\mathrm{diss}}$, it is smoothed in the same way as
$\dot{\cal E}_{\mathrm{diss}}$ in the bottom panel of
Fig.~\ref{fig:energy-dissipation}.
  
Assuming a relation of the form $\ell_{\mathrm{e_{\mathrm{kin}}}} =
\alpha_{\mathrm{e_{\mathrm{kin}}}} \ell_{\mathrm{corr}} $, we obtain optimal
fits (minimum of $\sigma^{2}(\alpha_{\mathrm{e_{\mathrm{kin}}}})$) for
$\alpha_{\mathrm{e_{\mathrm{kin}}}} \approx 1.3 $. The fits become only
slightly better if a weak linear dependence of
$\alpha_{\mathrm{e_{\mathrm{kin}}}}$ on $\ell( N )$ is allowed (13\% change as
$\ell( N )$ goes from 10 to 70).  $\ell_{\mathrm{e_{\mathrm{kin}}}} /
\ell_{\mathrm{corr}}$ is shown in the middle panel of
Fig.~\ref{fig:turbulence-length}.  Looking directly at the dependence of
$\ell_{\mathrm{e_{\mathrm{kin}}}}$ on $\ell_{\mathrm{cdl}}$ and
$M_{\mathrm{u}}$, we find $\ell_{\mathrm{e_{\mathrm{kin}}}} \propto
\ell_{\mathrm{cdl}} M_{\mathrm{u}}^{-0.6}$.  This is the same dependence we
found for $\ell_{\mathrm{corr}}$ in Sect.~\ref{sec:confshocks},
$\ell_{\mathrm{e_{\mathrm{kin}}}}$ scaled with this best fit is shown in the
bottom panel of Fig.~\ref{fig:turbulence-length}.
  
With increasing upstream Mach-number the characteristic length scale
$\ell_{\mathrm{e_{\mathrm{kin}}}}$ thus decreases with respect to the CDL
extension. This is consistent with our observation that for the same
$\ell_{\mathrm{cdl}}$ the interior of the CDL shows finer structuring
(patches, filaments) for higher values of $M_{\mathrm{u}}$.
Figure~\ref{fig:div-ti1.5-tihf} illustrates this observation with the
example of runs R11\_0.2.4 and R33\_0.2.4. Shown in the figure is
$\mathrm{div}(\vec{v})$, as the flow patterns, especially shocks, are
better visible in this quantity than in density.
  
In summary, our simulations show that the inherent length scale of the
turbulence is proportional to the auto-correlation length of the
confining shocks, independent of $M_{\mathrm{u}}$ and
$\ell_{\mathrm{cdl}}$.  With increasing $M_{\mathrm{u}}$, both length
scales decrease relative to the CDL extension,
$\ell_{\mathrm{e_{\mathrm{kin}}}} / \ell_{\mathrm{cdl}} \propto
M_{\mathrm{u}}^{-0.6}$. The appearance of the CDL, the size of its 
patches and filaments, behaves similarly.
\subsection{Settings with CDL at $t=0$}
\label{sec:symmetric_withcdl}
\begin{figure}[tp]
\centerline{\includegraphics[width=9.0cm]{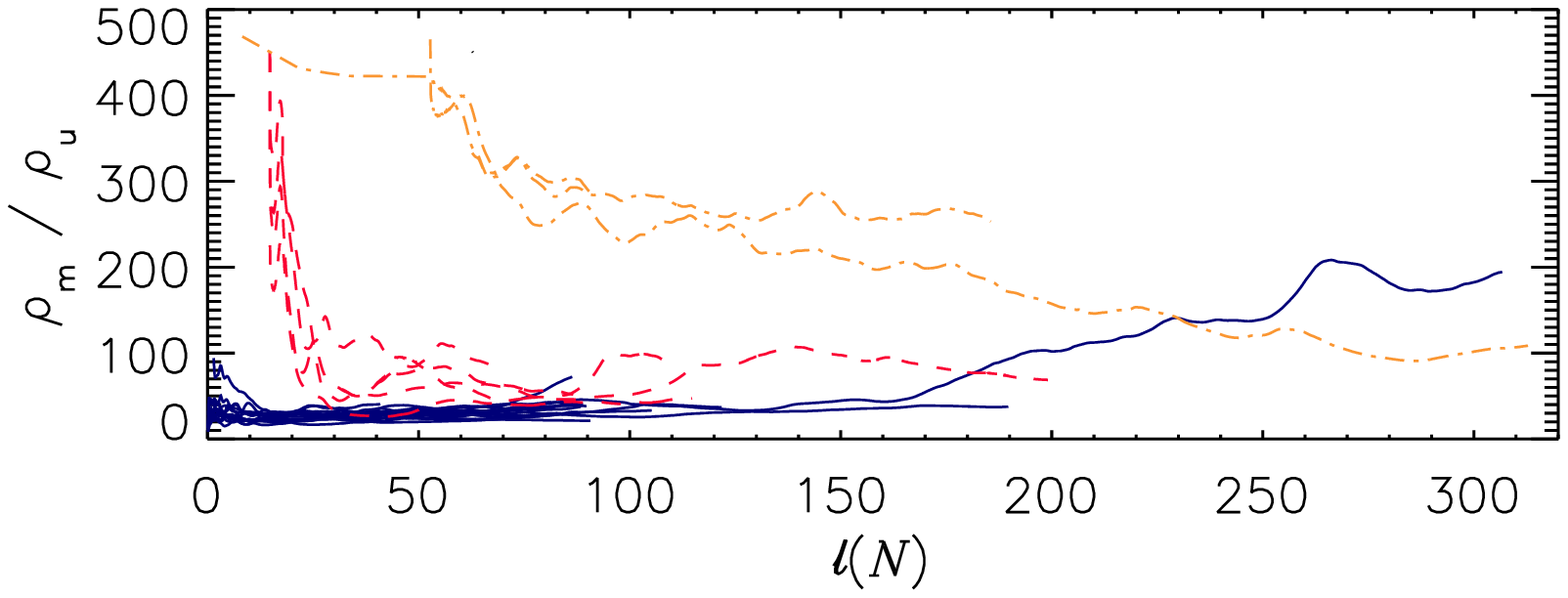}}
\centerline{\includegraphics[width=9.0cm]{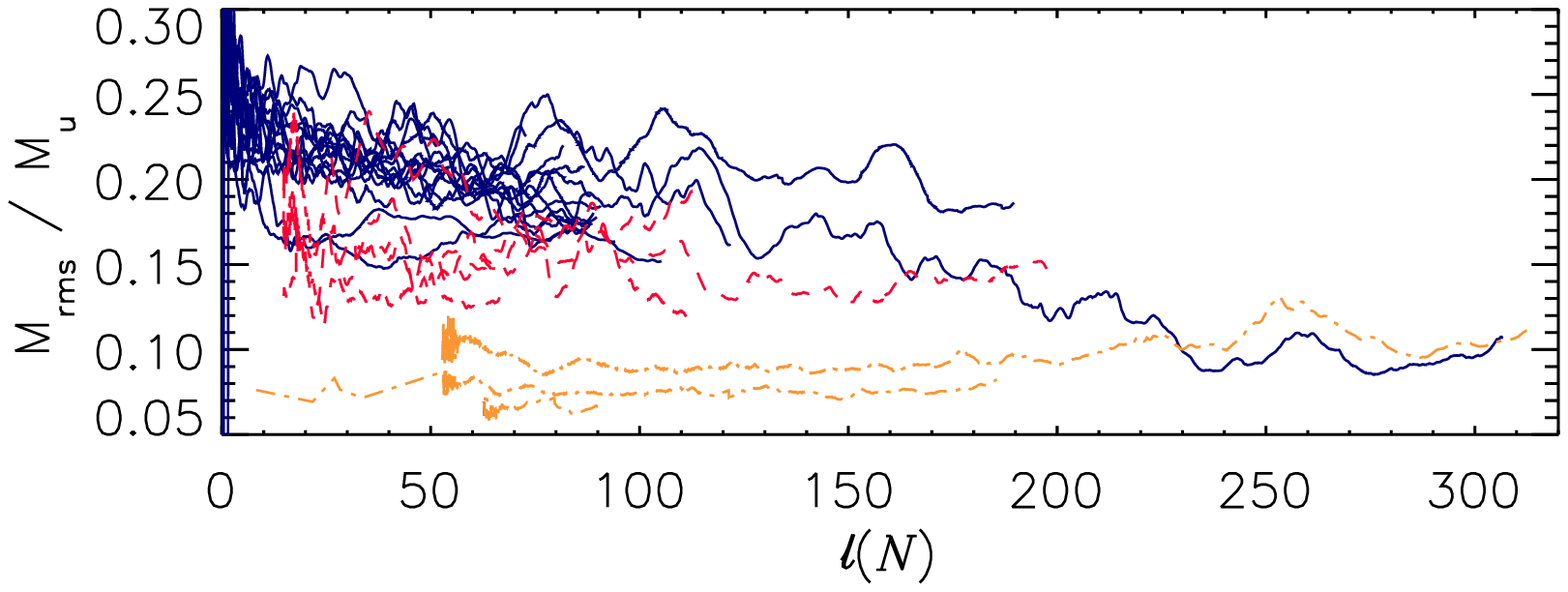}}
\centerline{\includegraphics[width=9.0cm]{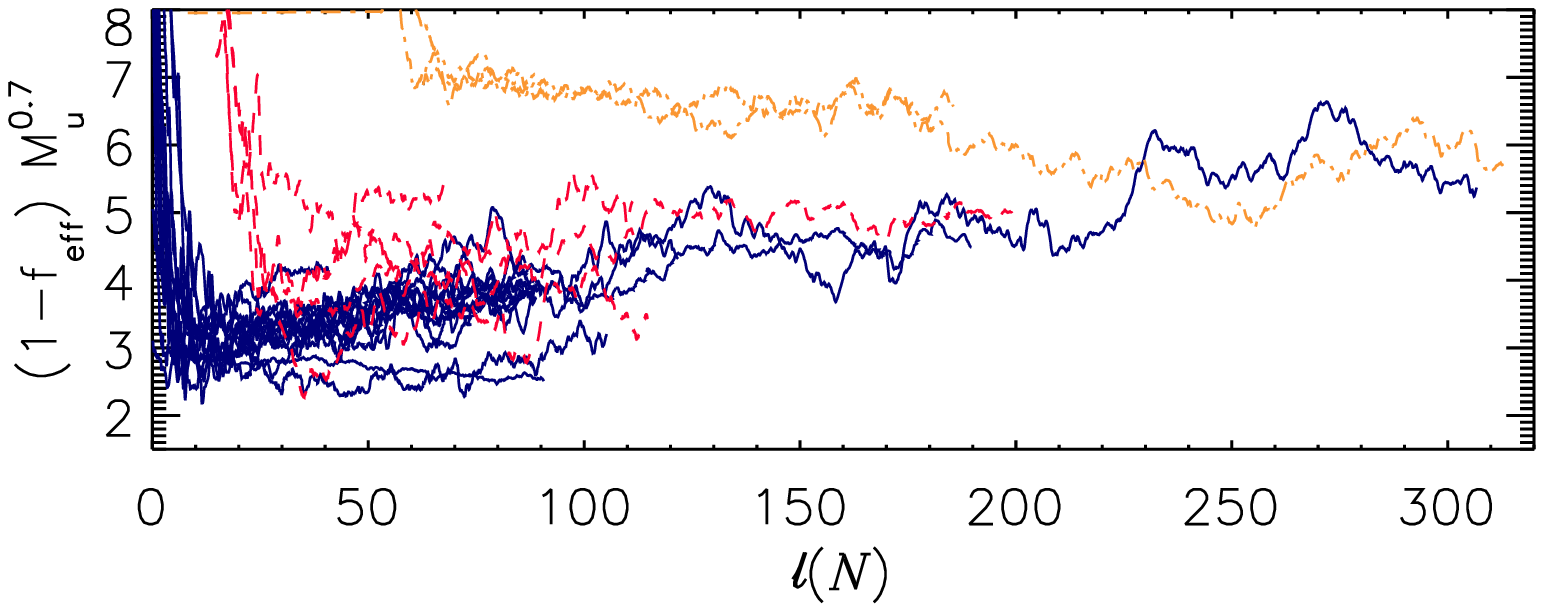}}
\centerline{\includegraphics[width=9.0cm]{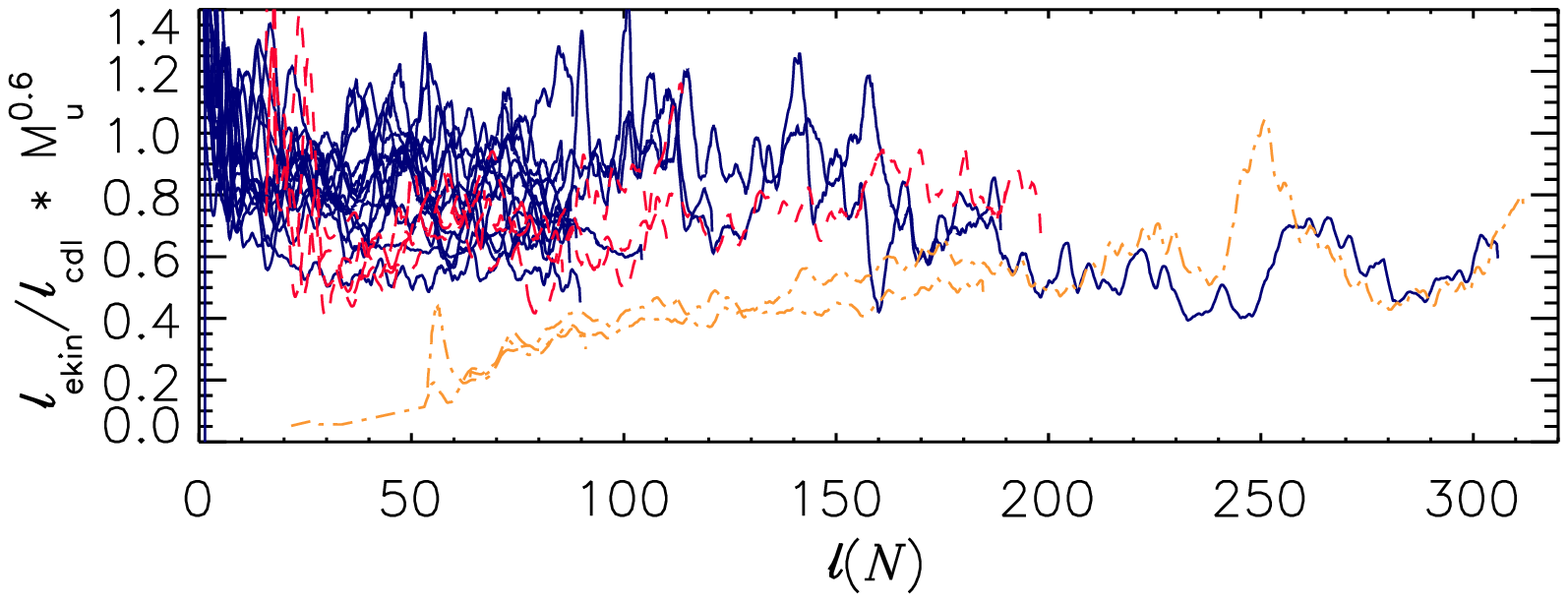}}
\caption{Comparing runs with and without an initial CDL. Shown are 
  $\rho_{\mathrm{m}} / \rho_{\mathrm{u}}$ (first),
  $M_{\mathrm{rms}}/M_\mathrm{u}$ (second), the scaled driving
  efficiency $(1-f_{\mathrm{eff}}) M_{\mathrm{u}}^{0.7}$ (third), and
  the scaled characteristic length of the turbulence
  $\ell_{\mathrm{e_{\mathrm{kin}}}}/\ell_{\mathrm{cdl}} \cdot
  M_{\mathrm{u}}^{0.6}$ for all symmetric runs. Line styles and colors
  denote initial conditions, 0 (solid line, blue), 1 (dashed line, red),
  and 2 (dash-dotted line, orange).}
\label{fig:diff_ini}
\end{figure}
\begin{figure}[tp]
\centerline{\includegraphics[width=9.0cm]{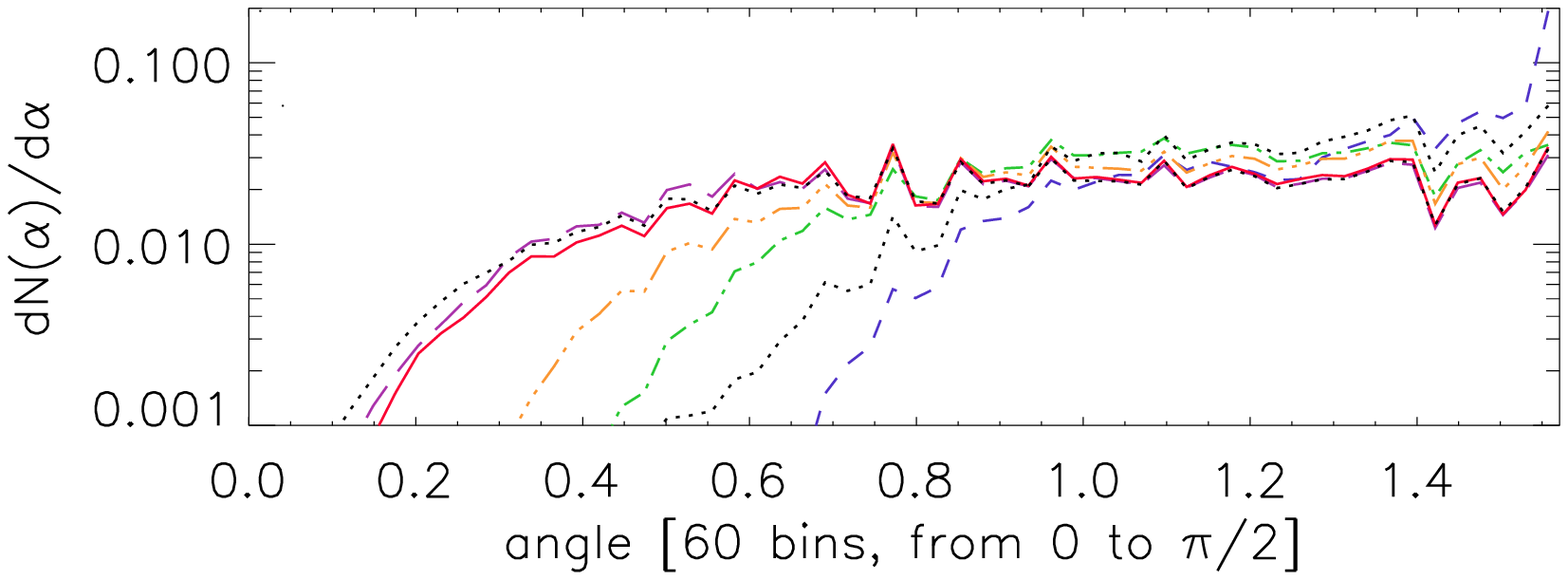}}
\caption{Time evolution of angle distribution for run R22\_2.2.2.
Shown is the average angle distribution for 
$ 10 < \ell( N ) <  70$ (dashed, blue), 
$ 70 < \ell( N ) < 130$ (dash-dotted, green), 
$130 < \ell( N ) < 190$ (dash-three-dots, orange), 
$190 < \ell( N ) < 250$ (long dashes, purple), 
$250 < \ell( N ) < 310$ (solid, red). 
Also shown are the distributions for run R5\_0.2.4 (black dots, right line) and
for run R11\_0.2.4 (black dots, left line), both averaged 
over  $10 < \ell( N ) < 70$.
}
\label{fig:ddd_time_evol_ang_dist}
\end{figure}
We performed additional runs to study the influence of an initially
present CDL. Figure~\ref{fig:diff_ini} illustrates the results for
some selected quantities.  Shown are all the runs we
performed with initial condition I=0 (no CDL at
$t=0$), I=1 (moderate CDL at $t=0$), and I=2 (massive CDL at $t=0$).

Comparison of the I=1 and I=0 curves in Fig.~\ref{fig:diff_ini} shows that an
initially present CDL of moderate column density ($ N =14 \, N_{\mathrm{0}}$)
soon develops characteristics similar to those found in simulations without
initial CDL. A quasi-steady state is reached for $\ell( N ) \gapprox 40$.  The
I=1 and I=0 curves then agree to within about a factor of two for volume
quantities like $\rho_{\mathrm{m}}$ and $M_{\mathrm{rms}}$ (first two panels
in Fig.~\ref{fig:diff_ini}).  Agreement seems slightly better for interface
related quantities. For $(1-f_{\mathrm{eff}})\,M^{0.7}$, shown in the third
panel of Fig.~\ref{fig:diff_ini}, the I=1 and I=0 curves lie more or less on
top of each other. The same is true for $\ell_{\mathrm{e_{\mathrm{kin}}}}/
\ell_{\mathrm{cdl}} \,M_{\mathrm{u}}^{0.6}$, shown in the bottom panel of
Fig.~\ref{fig:diff_ini}.

The situation is slightly different for runs with an initially rather massive
CDL (I=2, with initially $ N =56\, N_{\mathrm{0}}$). Also in these simulations
the CDL gets more and more turbulent. For all the quantities shown in
Fig.~\ref{fig:diff_ini}, the I=2 curves approach the I=1 and I=0 curves.
However, it takes these runs much longer to saturate.  Only for $\ell( N ) >
240$ the curves finally seem to saturate, at similar values as the I=0 and I=1
curves. That saturation does indeed occur around that time is also supported
by Fig.~\ref{fig:ddd_time_evol_ang_dist}. As can be seen, the average angle
distribution of the confining shocks for run R22\_2.2.2 first shifts to higher
and higher values as $\ell( N )$ increases.  It then stagnates for the last
two averaging periods, $ 190 < \ell( N ) < 250$ and $250 < \ell( N ) < 310$.

In summary, we conclude that our symmetric simulations all end up in a
similar quasi-steady final state. An initially present CDL only delays
the development. The incoming flows also manage to generate (and sustain) a
similar level of turbulence also within an initially massive CDL.
\begin{figure}[tp]
\centerline{\includegraphics[width=9.0cm]{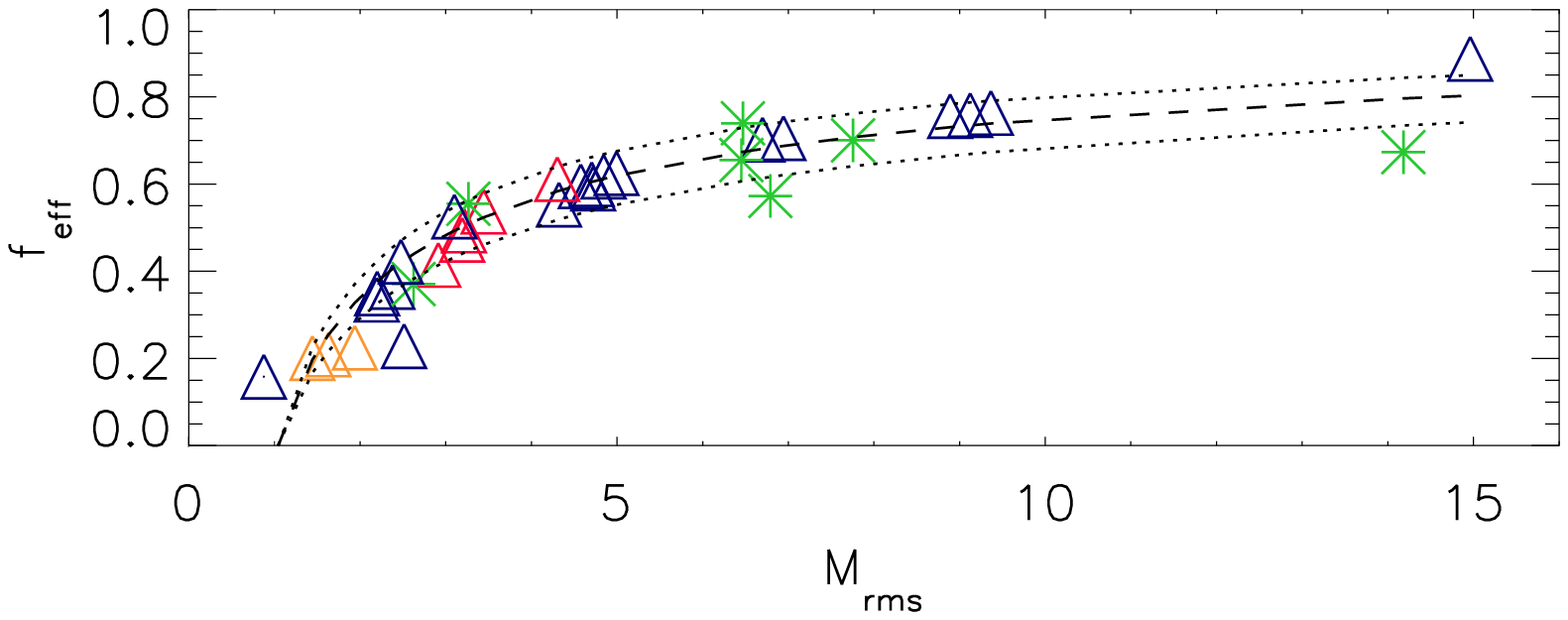}}
\caption{Average $f_{\mathrm{eff}}$ as a function of $M_{\mathrm{rms}}$ for
  all our symmetric simulations (triangles). In addition, we included 
  data from our asymmetric runs (asterisks), for which $1.6 M_{\mathrm{r}} \le
  M_{\mathrm{l}} \le 64 M_{\mathrm{r}}$ and which initially have no CDL. 
  Averages were taken over $10 \le \ell( N ) \le 70$ for simulations 
  without initial CDL (blue triangles and green asterisks), over $40 \le
  \ell( N ) \le 70$ for runs with a moderate initial CDL
  (red triangles), and over $70 \le \ell( N ) \le 140$ for runs with a
  massive initial CDL (orange triangles). Lines show $f_{\mathrm{eff}} = 1 -
  M_{\mathrm{rms}}^{\xi}$ with $\xi=-0.6$ (dashed) and $\xi=-0.6
  \pm 0.1$ (dotted).}
\label{fig:feff_vs_mrms}
\end{figure}
\subsection{Asymmetric cases}
\label{sec:results_asym}
We also computed a few asymmetric cases, where the two upwind Mach-numbers are
different, $M_{\mathrm{l}} \ne M_{\mathrm{r}}$. For the same reason as given
in Sect.~\ref{sec:anal-scaling}, we expect the solution to only depend on
$M_{\mathrm{l}}$ and $M_{\mathrm{r}}$.  These dependencies are more
complicated than those assumed in Sect.~\ref{sec:anal-scaling} as we now have
two different upwind Mach-numbers. The simple dependencies of
Sect.~\ref{sec:anal-scaling} should, however, be recovered in the limit
$M_{\mathrm{l}} \rightarrow M_{\mathrm{r}}$.

The basic physical reason for the more complicated dependencies on the
upwind Mach-numbers lies in the strong back coupling between the
turbulence within the CDL and the driving of this turbulence by the
upwind flows. Our asymmetric simulations demonstrate clearly (much
more clearly than the symmetric simulations) that the turbulence
crucially affects the driving: although $M_{\mathrm{l}}$ and
$M_{\mathrm{r}}$ are strongly different, the corresponding driving
efficiencies are about equal, $f_{\mathrm{eff,l}} \approx
f_{\mathrm{eff,r}}$. Thus the efficiency does not depend primarily on
the upwind flow. In fact, Fig.~\ref{fig:feff_vs_mrms} shows that for
both symmetric and asymmetric runs $f_{\mathrm{eff}}$ (averaged now
over both shocks) can be described well by
\begin{equation}
f_{\mathrm{eff}} = 1 - M_{\mathrm{rms}}^{-0.6}.
\end{equation}
The angle distribution of the two shocks behaves accordingly in that
it is similar for both shocks and determined by $M_{\mathrm{rms}}$
rather than by either $M_{\mathrm{l}}$ or $M_{\mathrm{r}}$. 

A more detailed analysis of the asymmetric case, including an
approximate analytical solution, will be presented in a
subsequent paper.
\begin{figure}[tp]
\centerline{\includegraphics[width=9.0cm]{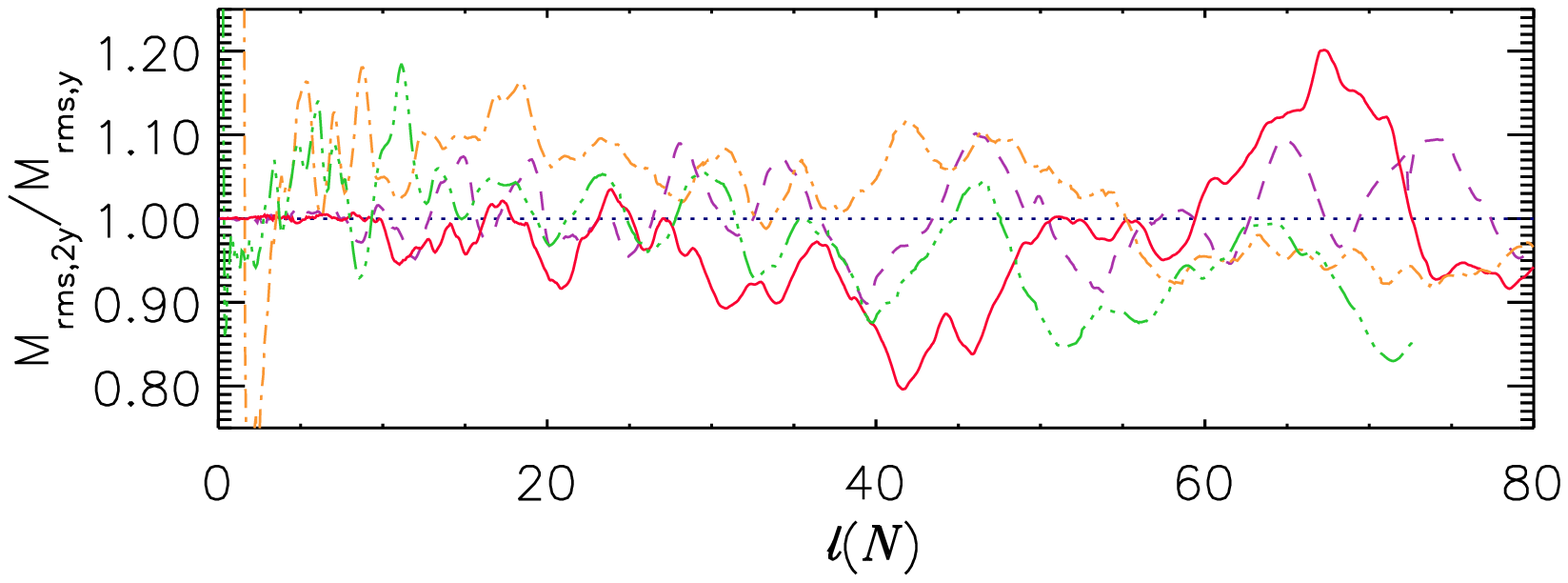}}
\centerline{\includegraphics[width=9.0cm]{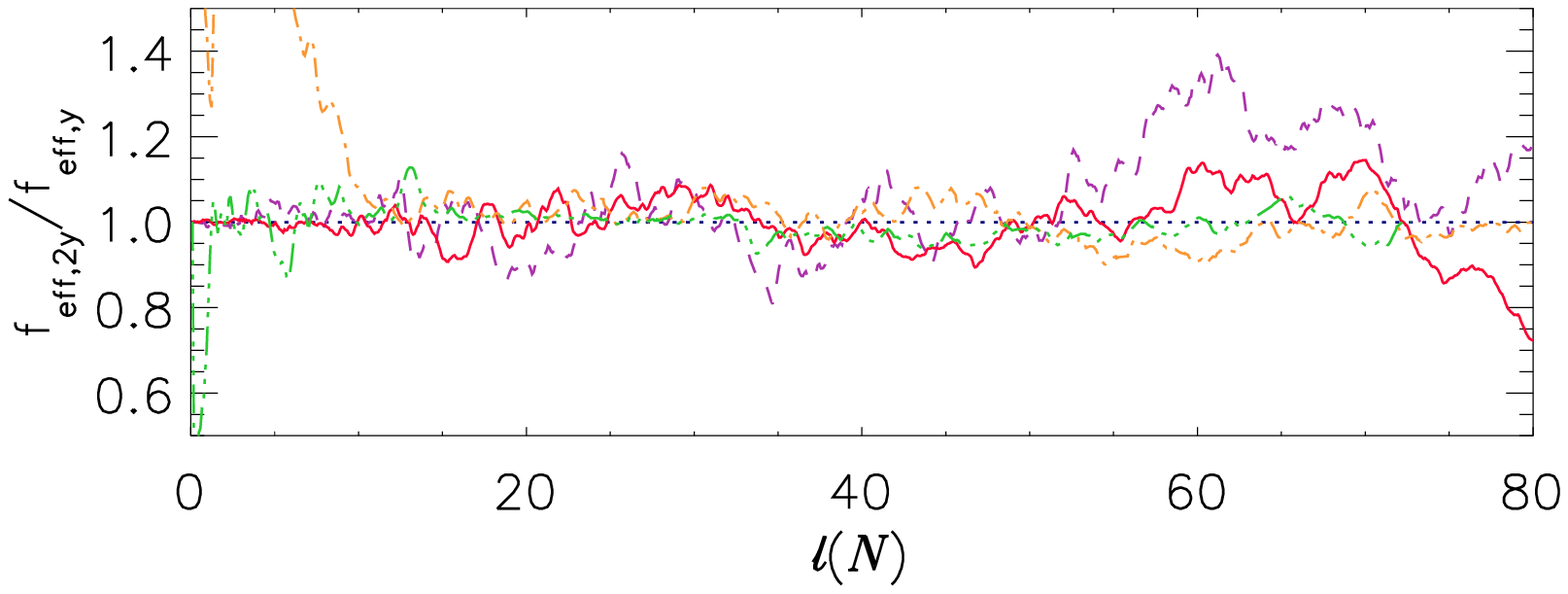}}
\centerline{\includegraphics[width=9.0cm]{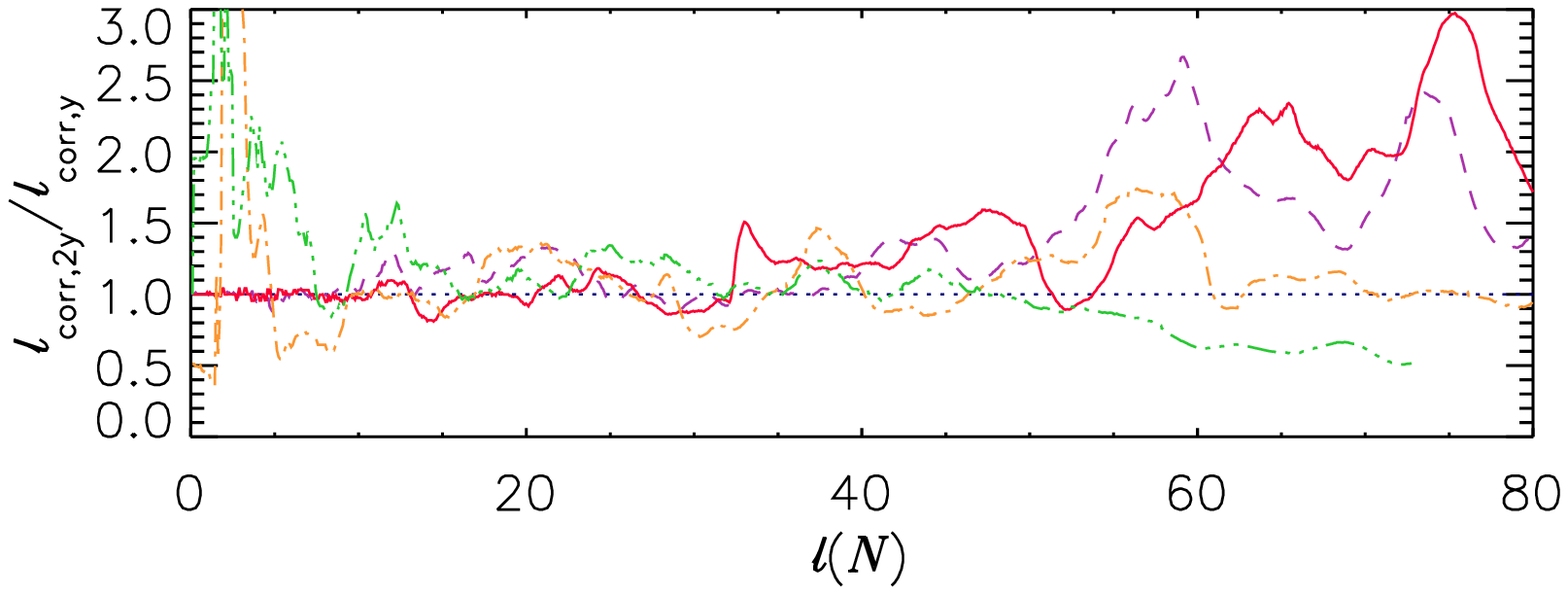}}
\caption{Comparing runs that differ only in the y-extent of the
  domain ($\mathrm{Y} = 2 \mathrm{Y}_{\mathrm{0}}$ and $\mathrm{Y} = 4
  \mathrm{Y}_{\mathrm{0}}$). Shown are $ M_{\mathrm{rms,2Y}} /
  M_{\mathrm{rms,Y}}$ ({\bf top}), $ f_{\mathrm{eff,2Y}} /
  f_{\mathrm{eff,Y}}$ ({\bf middle}), and $ \ell_{\mathrm{corr,2Y}} /
  \ell_{\mathrm{corr,Y}}$ ({\bf bottom}).  Individual curves denote
  runs R11\_0.2.* (dashed, purple), R22\_0.2.* (solid, red),
  R33\_0.2.* (dash-dotted, orange), R43\_0.2.* (dash-three-dots,
  green).}
\label{fig:m_feff_lcorr_tis_y_2y}
\end{figure}
\begin{figure}[tp]
\centerline{\includegraphics[width=9.0cm]{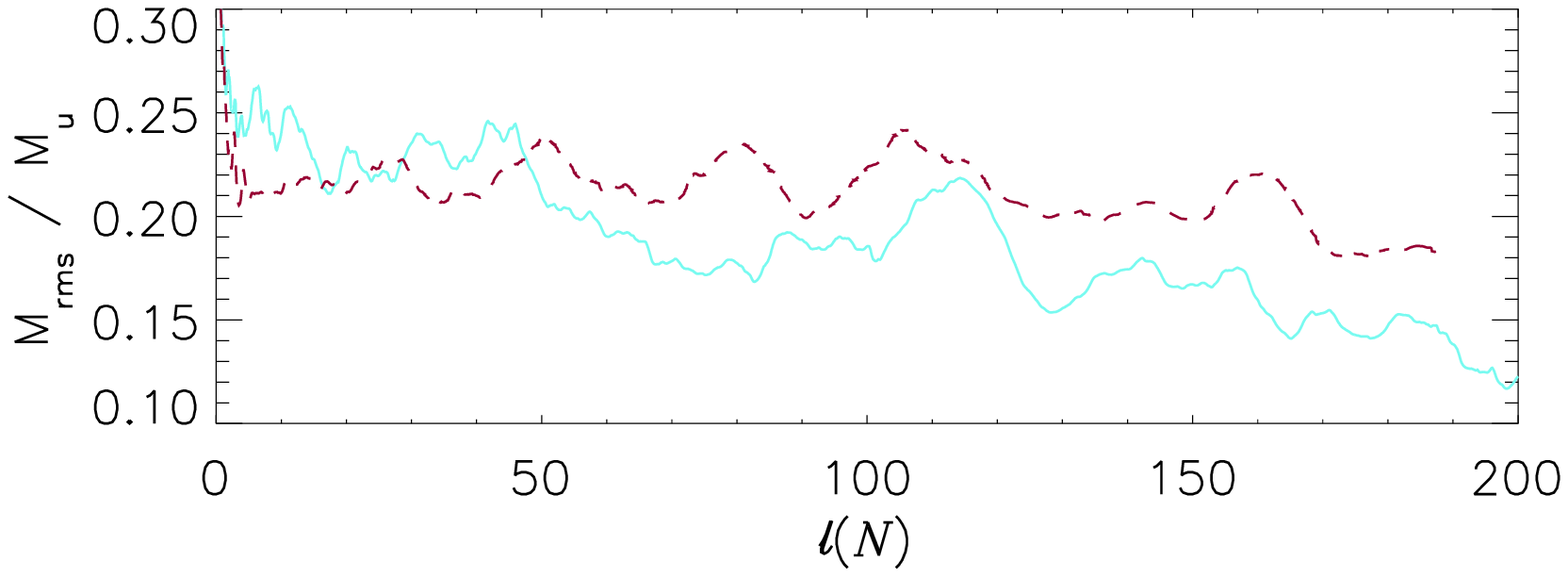}}
\centerline{\includegraphics[width=9.0cm]{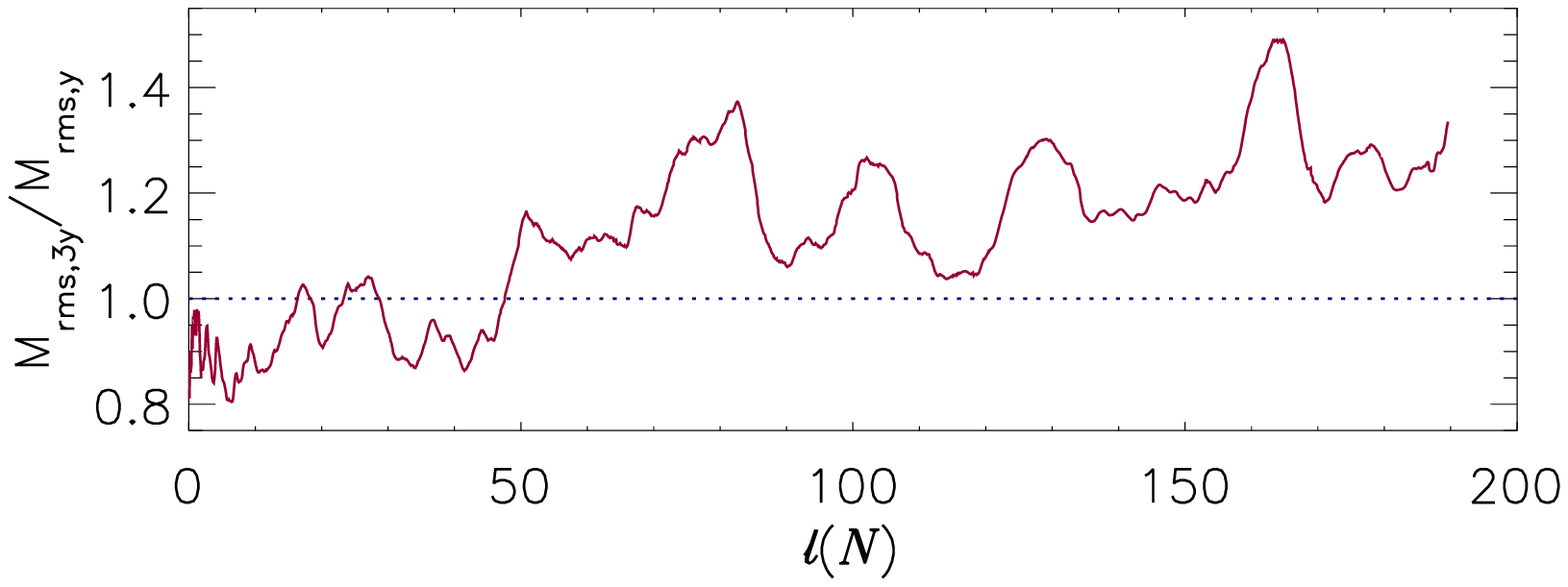}}
\caption{Comparison of runs R22\_0.2.2 and R22\_0.2.6, illustrating
  the effect of a three-times larger y-extent of the computational
  domain on long time scales. Shown are
  $M_{\mathrm{rms}}/M_{\mathrm{u}}$ ({\bf top}) for R22\_0.2.2 (solid,
  light blue) and R22\_0.2.6 (dashed, dark red) and the ratio $
  M_{\mathrm{rms,3Y}} / M_{\mathrm{rms,Y}}$ ({\bf bottom}).}
\label{fig:m_tis_y_3y}
\end{figure}
\begin{figure}[tp]
\centerline{\includegraphics[width=9.0cm]{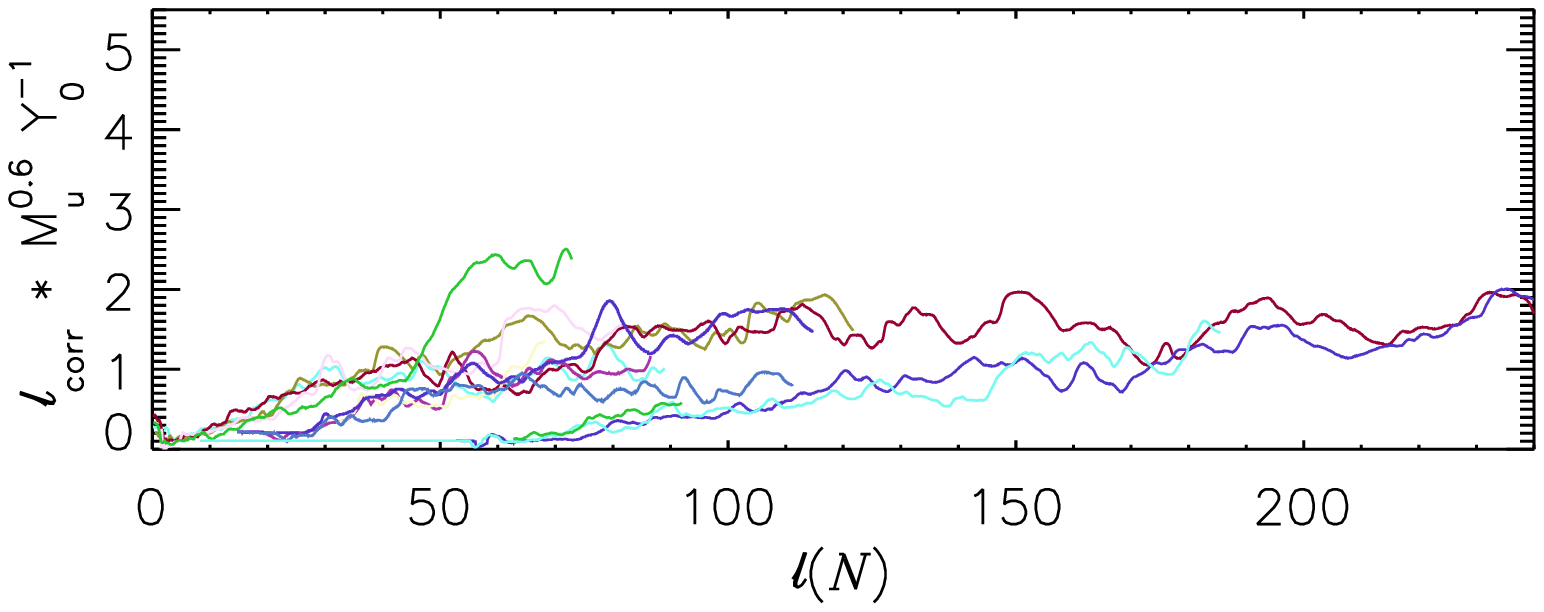}}
\centerline{\includegraphics[width=9.0cm]{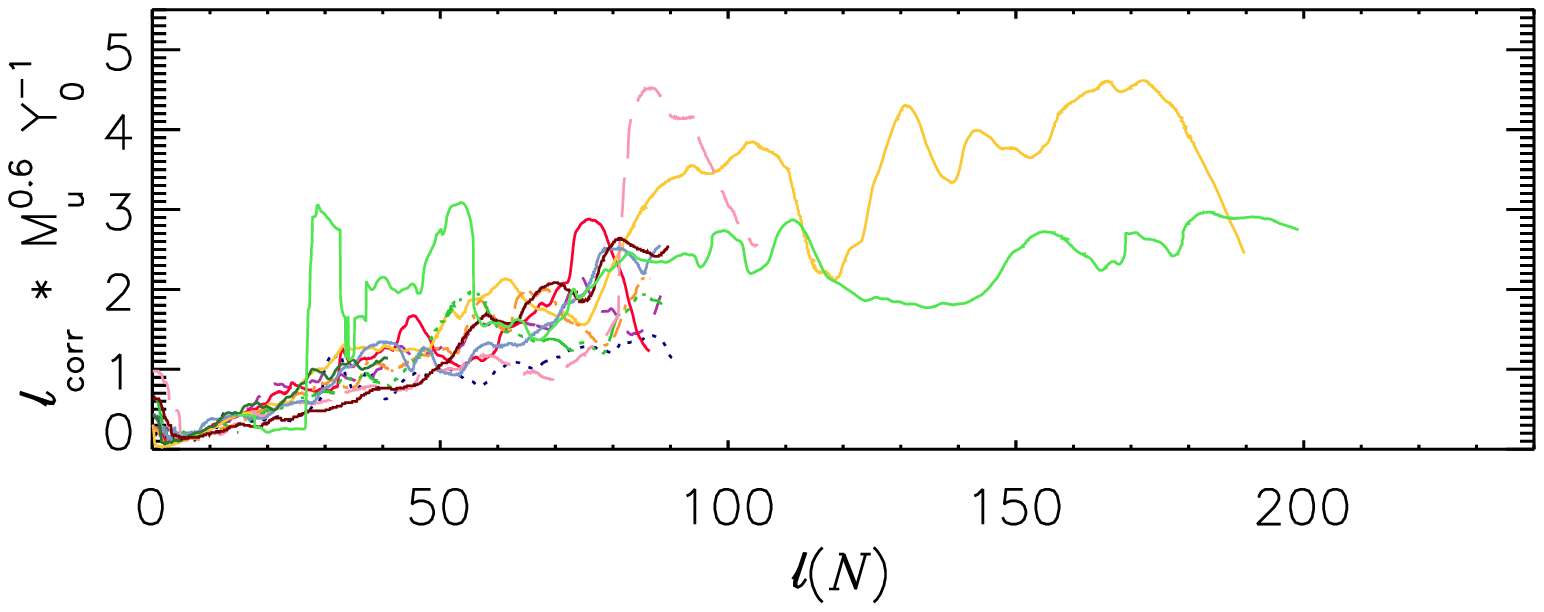}}
\caption{Scaled auto-correlation lengths
  of all symmetric simulations on domains with a y-extent less or equal
  to $2 \mathrm{Y}_{\mathrm{0}}$ ({\bf top}) and a y-extent greater or
  equal to $4 \mathrm{Y}_{\mathrm{0}}$ ({\bf bottom}).}
\label{fig:lcorr_all_y_2y}
\end{figure}
\subsection{Grid and domain studies}
\label{sec:griddomain}
The numerical results presented in Sect.~\ref{sec:symmetric_nocdl}
were all based on simulations with a domain $\mathrm{Y}=4
\mathrm{Y}_{\mathrm{0}}$ and a discretization of $1.5 \cdot 10^{-3}
\mathrm{Y}_{\mathrm{0}} $ (R=2) or 2560 cells in the y-direction. Here we
want to check whether these choices have any systematical effect on
the numerical results of Sect.~\ref{sec:symmetric_nocdl}.
\subsubsection{Different y-extent}
\label{sec:diff_y_ext}
To check whether the size of the computational domain has any
systematic effect on the results of Sect.~\ref{sec:symmetric_nocdl},
we performed some of the simulations again, but this time on smaller
domains of $\mathrm{Y}=2 \mathrm{Y}_{\mathrm{0}}$ and
$\mathrm{Y}=\mathrm{Y}_{\mathrm{0}}$. We also performed one simulation 
on a larger domain $\mathrm{Y}=6 \mathrm{Y}_{\mathrm{0}}$.

Figure~\ref{fig:m_feff_lcorr_tis_y_2y} illustrates our findings for
simulations on domains $\mathrm{Y}=2 \mathrm{Y}_{\mathrm{0}}$ and
$\mathrm{Y}=4 \mathrm{Y}_{\mathrm{0}}$.  $M_{\mathrm{rms}}$ shows no
systematic effect and is, as such, representative of other volume-related
quantities (Fig.~\ref{fig:m_feff_lcorr_tis_y_2y}, top panel).  As a typical
representative for interface-related quantities, $f_{\mathrm{eff}}$ also shows
no clear overall effect of the domain size
(Fig.~\ref{fig:m_feff_lcorr_tis_y_2y}, middle panel).  The quantity for which
we find the most clear effect is the auto-correlation length
$\ell_{\mathrm{corr}}$ (Fig.~\ref{fig:m_feff_lcorr_tis_y_2y}, bottom panel).
However, even for $\ell_{\mathrm{corr}}$ the effect sets in only for two of
the four runs and only for $\ell( N ) \gapprox 30$, i.e. once the CDL
extension reaches about half the size of the smaller domain.  For the
numerical results in Sect.~\ref{sec:symmetric_nocdl}, $\ell_{\mathrm{cdl}}
\approx \mathrm{Y}/2$ corresponds to $\ell( N )= 60$. We conclude that the
y-extent of the computational domain has no apparent systematic effect on
these results up to $\ell( N ) \lapprox 30$ and probably even up to $\ell( N )
\lapprox 60$.

A systematic effect of the computational domain on the numerical solution does
become apparent if the simulations are carried on much longer. One pair of
runs, R22\_0.2.2 and R22\_0.2.6, were carried on much longer, till $\ell ( N )
\approx 200$. For this pair of runs, Fig.~\ref{fig:m_tis_y_3y} shows the
evolution of $M_{\mathrm{rms}}$ for each run, as well as their ratio,
$M_{\mathrm{rms,3y}}/M_{\mathrm{rms,y}}$. The run on the smaller domain
apparently shows a faster decay in $M_{\mathrm{rms}}$ after $\ell ( N )
\approx 100 $.  From Fig.~\ref{fig:lcorr_all_y_2y} we take that the behavior
of this one pair of runs is most likely the rule, and not the exception.  The
top panel of Fig.~\ref{fig:lcorr_all_y_2y} shows $\ell_{\mathrm{corr}}$,
scaled, for all the symmetric runs we have performed and whose domain has a
y-extent $\le 2 \mathrm{Y}_{\mathrm{0}}$. The bottom panel of
Fig.~\ref{fig:lcorr_all_y_2y} gives the same quantity for all the runs with a
domain extention $\ge 4\mathrm{Y}_{\mathrm{0}}$.  Comparison of the two
figures shows that runs on a domain $\le 2 \mathrm{Y}_{\mathrm{0}}$ saturate
around $\ell_{\mathrm{corr}} M_{\mathrm{u}}^{0.6} \approx 1.6
\mathrm{Y}_{\mathrm{0}}$.  For runs on a domain $\ge
4\mathrm{Y}_{\mathrm{0}}$, $\ell_{\mathrm{corr}}$ reaches much higher values.
\begin{figure}[tp]
  \centerline{\includegraphics[width=9.0cm]{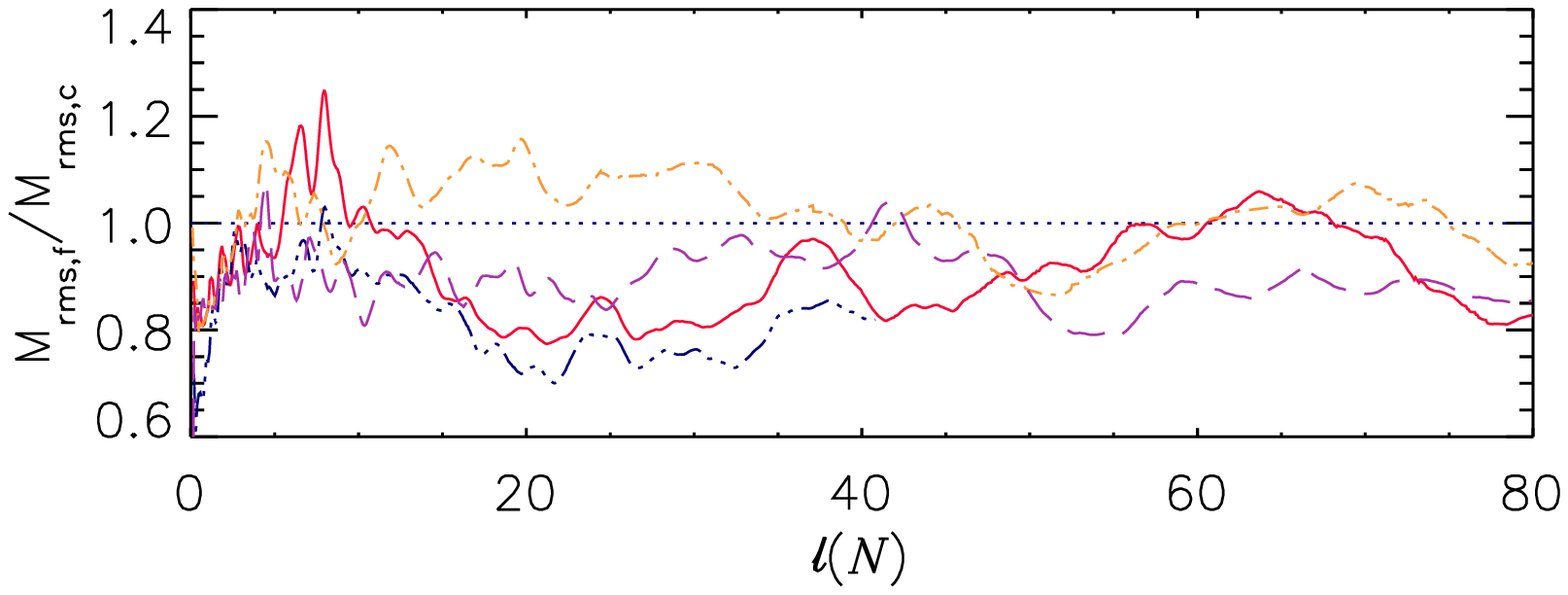}}
\caption{Comparison of $M_{\mathrm{rms}}$ for runs whose spatial resolution
  differs by a factor of 2 (subscript c = coarse, f = fine).  Shown
  are (giving only the name of the finer run) runs R22\_0.2.4 (solid,
  red), R22\_0.4.4 (dash-three-dots, blue), R43\_0.2.4 (long dashes,
  purple), and R11\_0.2.4 (dash-dotted, orange).}
\label{fig:diff_disc}
\end{figure}
\begin{figure}[tp]
  \centerline{\includegraphics[width=8.0cm]{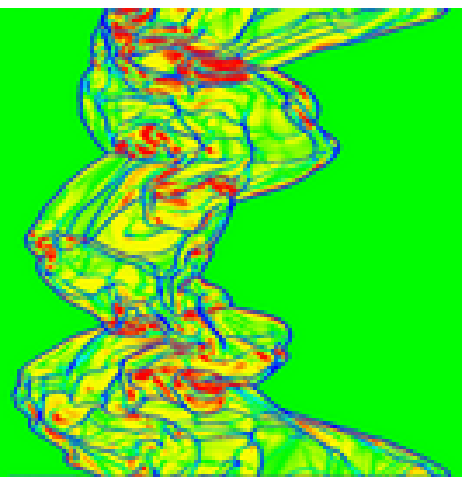}}
\vspace{.2cm}
  \centerline{\includegraphics[width=8.0cm]{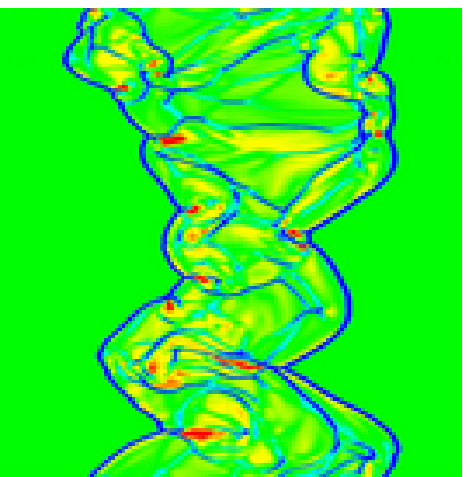}}
\caption{Plots of $\mathrm{div}(\vec{v})$ for two runs that are identical 
  to run R11\_0.2.4, shown in Fig.~\ref{fig:div-ti1.5-tihf}, except for
  their discretization.  The runs shown here were computed with two times lower
  (top) and at four times lower (bottom) resolution.  Blue (dark lines)
  indicates convergence, red (dark patches) divergence. As can be seen, the
  number of convergent regions within an average CDL column decreases with
  decreasing resolution.}
\label{fig:div-tihf-c-tihf-cc}
\end{figure}
\subsubsection{Different discretization}
\label{sec:diff_resol}
The results presented in Sect.~\ref{sec:symmetric_nocdl} were all based on
simulations with a discretization of $1.5 \cdot 10^{-3}
\mathrm{Y}_{\mathrm{0}} $ (R=2) or 2560 cells in the y-direction. To check the
effect of the discretization on our results, we repeated several
simulations with coarser and/or finer discretization.  These simulations
indeed reveal a systematic effect of the discretization on the values of
average quantities.  Nevertheless, the general properties of the solution, its
approximate self-similarity and Mach-number dependences, remain unaltered.
Only the numerical constants $\eta_{\mathrm{i}}$ are affected. The changes
are, however, small when compared to the differences between the 1D and 2D
solution (for example, $\rho_{\mathrm{m}} = \eta_{\mathrm{1}}
\rho_{\mathrm{u}}$ in 2D, while $\rho_{\mathrm{m}} = M_{\mathrm{u}}^{2}
\rho_{\mathrm{u}}$ in 1D).

We find that finer discretization generally leads to reduced turbulence.
Using finer meshes we obtained larger mean densities and lower values of
$M_{\mathrm{rms}}$, as shown in Fig.~\ref{fig:diff_disc}. The driving
efficiency gets lower and the shocks become more inclined with respect to the
direction of the upstream flows, and the angle distribution is shifted to
lower values.  The characteristic length scale
$\ell_{\mathrm{e_{\mathrm{kin}}}}$ remains about constant if taken in units of
$\ell_{\mathrm{cdl}}$.

A possible explanation for the reduction of turbulence (smaller
$M_{\mathrm{rms}}$) on finer grids could be the dominance of shocks
for the energy dissipation in the CDL. On a coarser grid, the network
of shocks within the CDL is less dense. The divergence plots shown in
Fig.~\ref{fig:div-tihf-c-tihf-cc} illustrate this effect.  A closer
analysis of this idea is, however, beyond the scope of the present
paper.

We stress that, so far, no convergence has been reached in our discretization
studies. Looking at the comparison of the three runs R22\_0.1.4,
R22\_0.2.4, and R22\_0.4.4 in Fig.\ref{fig:diff_disc} shows that each
reduction of the cell size by a factor of two leads to a reduction of
about 20\% in $M_{\mathrm{rms}}$. This indicates that the resolution
of 2560 cells in y-direction in our standard runs (R*\_0.2.4) and of
5120 cells in the y-direction in the refined runs is still not
sufficient. This should be kept in mind when interpreting these
results or any results on shock bound turbulent structures
in 2D, let alone 3D.

Also, no clear picture emerges regarding the deviation of $M_{\mathrm{rms}}$
from the constant value predicted by Eq.~\ref{eq:ansatz_v}. A linear fit to
$M_{\mathrm{rms}}$ for $10 \le \ell ( N ) \le 70$ yields -12\% for run
R22\_0.2.4 and -23\% for the two times coarser run R22\_0.1.4. For runs
R43\_0.*.4, the grid dependence is the other way round: R43\_0.2.4 shows a
decrease of -25\%, the twice coarser run R43\_0.1.4 decreases by only -15\%.
\section{Discussion}
\label{sec:discussion}
We want to address four points in this section. First, we sketch
possible reasons for the slight difference between the numerical
solution and the relations we derived in Sect.~\ref{sec:anal-scaling}.
Second, we look once more at the driving of the turbulence and, in
particular, the back-coupling between interface and volume properties.
Third, we briefly consider our results in an astrophysical context, in
particular with regard to molecular clouds. Finally, based on
preliminary numerical results, we sketch the effect of some additional
physics.
\subsection{The numerical solution versus the analytical solution}
In Sect.~\ref{sec:anal-scaling_2d} we suggested that a self-similar
solution to our 2D model problem may still exist for the limiting case
where the system approaches infinity.  The relations derived in that
section give a reasonable estimate for the numerical results of
Sect.~\ref{sec:num_results}.  However, while $M_{\mathrm{rms}}$ is
constant in Sect.~\ref{sec:anal-scaling_2d}, the numerical simulations
show a gradual decrease in $M_{\mathrm{rms}}$ already for small CDLs,
$\ell_{\mathrm{cdl}} \lapprox \mathrm{Y}/2$ (15\% decrease of
$M_{\mathrm{rms}}$ as $\ell (N)$ increases from 10 to 70,
Sect.~\ref{sec:symmetric_nocdl}). We have no firm explanation for this
difference. We sketch three possible effects in the following, but
stress that the available data do not allow us to clearly distinguish
between them.

A first, obvious reason could be the finite y-extent of the computational
domain, $\mathrm{Y}$.  It sets an upper limit on the total energy input into
the CDL, thus on the amount of mass within the CDL that can be driven. Once
the CDL has accumulated too much mass, the driving per unit mass weakens and
the turbulence starts to weaken.  The spatial growth of the CDL slows down
while the average density increases. The following considerations on time
scales may illustrate this point further.

An upper limit to the time at which $\mathrm{Y}$ starts to affect the solution
is given by the time $t_{\mathrm{y}}$ at which $\ell_{\mathrm{cdl}} =
\mathrm{Y}$.  At later times structures may still grow in the x-direction (up
to $\ell_{\mathrm{cdl}}$ at most) but cannot grow any more in the y-direction
(where $\mathrm{Y}$ sets an upper limit). For the runs in
Sect.~\ref{sec:symmetric_nocdl}, $\ell_{\mathrm{cdl}} = \mathrm{Y}$
corresponds to $\ell ( N ) \approx 120$ or $t_{\mathrm{y}} = 12
\mathrm{Y}_{\mathrm{0}}/v_{\mathrm{rms}}$.  A lower limit for the decay time
scale of the turbulence may be obtained as follows.  For the case of uniformly
driven isothermal hydrodynamic turbulence in a 3D periodic box,
\citet{maclow:99} has shown that the typical decay time once the driving is
turned off, $t_{\mathrm{0}}$, and the initial driving wave length,
$\lambda_{\mathrm{drv}}$, are related by $t_{\mathrm{0}} \approx
\lambda_{\mathrm{drv}} / v_{\mathrm{rms}}$.  Assuming that this result also
holds for our slab, that $\lambda_{\mathrm{drv}} = \mathrm{Y}$, and that
driving is turned off completely, it follows that $t_{\mathrm{0}} \approx
\mathrm{Y} / v_{\mathrm{rms}}$, or $t_{\mathrm{0}} \approx 4
\mathrm{Y_{\mathrm{0}}} / v_{\mathrm{rms}}$ for the runs in
Sect.~\ref{sec:symmetric_nocdl}. However, driving continues in our simulations
and so the effective decay time scale of the turbulence is likely to be much
longer than $t_{\mathrm{0}}$. Finally, for the runs in
Sect.~\ref{sec:symmetric_nocdl}, and a typical integration time of $\ell ( N )
= 60$ corresponds to about $\tau = 6 \mathrm{Y}_{\mathrm{0}} /
v_{\mathrm{rms}}$, a typical turbulent crossing time at $\ell ( N ) = 60$) is
$\tau_{\mathrm{cross}} = \ell_{\mathrm{cdl}} / v_{\mathrm{rms}} \approx 2
\mathrm{Y}_{\mathrm{0}}/ v_{\mathrm{rms}}$.  Comparing these different time
scales makes it seem likely that at $\ell ( N ) =60$, turbulence in the center
of the CDL is still essentially driven, not essentially decaying.

Our simulation data do not allow us to either clearly confirm or reject the
hypothesis that the finite y-extent of the computational domain is responsible
for the slight decrease in $M_{\mathrm{rms}}$ that we observe at early times,
$\ell ( N ) \lapprox 70$.  If the finite domain size were responsible,
$M_{\mathrm{rms}}$ should decay differently on different domains.  Comparison
of simulations on different domains up to $\ell ( N ) \approx 70$
(Sect.~\ref{sec:diff_y_ext}), however, gives no clear picture. The data are
rather noisy, and simulations on domains $2 \mathrm{Y}_{\mathrm{0}}$ and $4
\mathrm{Y}_{\mathrm{0}}$ show no systematic differences as long as $\ell ( N )
\lapprox 30$ ($\ell_{\mathrm{cdl}} < \mathrm{Y}/2$ on the smaller domain).
Only for much later times, $\ell ( N ) >> 70$, well beyond the range for the
results in Sect.~\ref{sec:symmetric_nocdl}, does $\mathrm{Y}$ have a clear
effect and $M_{\mathrm{rms}}$ decreases faster on smaller domains
(Fig.~\ref{fig:m_tis_y_3y}).

A second, more speculative, reason might be numerical dissipation,
provided that its effect were to increase with $\ell_{\mathrm{cdl}}$.
While we have no evidence that the latter is really the case, it may
also be hasty to discard this possibility right away.
\citet{porter-woodward:94} found, by observing how simple 2D
hydrodynamical flows (shear flows and sound waves of definite wave
number, their section 3.3) damp with time, that the decay rate due to
numerical dissipation alone is a non-linear function of the wave
number. Their results are certainly not directly applicable to the
present case. But in view of these results, and given the change in
structure size with $\ell_{\mathrm{cdl}}$ as suggested by
Fig.~\ref{fig:pheno_dens}, it might be possible that the effect of
numerical dissipation indeed changes with $\ell_{\mathrm{cdl}}$.  Note
that this would also imply that the MILES approach, outlined in
Sect.~\ref{sec:simulating}, were not strictly valid for the problem we
consider.  The currently available data do not allow us to clearly
reject or confirm the effect.
 
A third reason, or rather an amplifying mechanism, could be back-coupling
between $M_{\mathrm{rms}}$ and the driving efficiency.  Once the turbulence
within the CDL is slightly reduced (for whatever reason), the reduction is
further amplified by the back-coupling between turbulence and driving,
$f_{\mathrm{eff}} = (1 - M_{\mathrm{rms}}^{-0.6})$.  The decrease in
$M_{\mathrm{rms}}$ results in larger inclination of the shocks with respect to
the direction of the upstream flows, more energy is dissipated at the
confining shocks of the CDL, and less driving energy enters the CDL.  For the
observed 15\% reduction of $M_{\mathrm{rms}}$, the reduced driving may, in
fact, play a dominant role: as $\ell ( N )$ increases from 10 to 70,
$\dot{\cal E}_{\mathrm{drv}} / \dot{\cal E}_{\mathrm{drv}}^{\mathrm{th}}$
decreases by 13\% (Sect.~\ref{sec:driving_efficiency}). But to really estimate
the relative importance of the three effects just sketched, further studies
are certainly necessary.

Two more points seem noteworthy to us in this section. One concerns the near
independence of $\dot{\cal E}_{\mathrm{diss}}$ on $\ell_{\mathrm{cdl}}$.  From
Fig.~\ref{fig:pheno_dens} (increase in structure size with increasing
$\ell_{\mathrm{cdl}}$), we take that it is rather the increasing average
distance between shocks that allows $\dot{\cal E}_{\mathrm{diss}}$ to be
essentially independent of $\ell_{\mathrm{cdl}}$ and not so much the, on
average, decreasing strength of shocks (Sect.~\ref{sec:energy_dissipation}).
Whether this is indeed true, only a closer analysis of the structure within
the CDL along the lines of~\citet{maclow-ossenkopf:00} can tell, which is,
however, beyond the scope of the present paper. Such an analysis could also
shed light on whether (or in which sense) $\ell_{\mathrm{e_{\mathrm{kin}}}}$
(see Sect.~\ref{sec:driving_efficiency}) is indeed a measure of the average
distance between shocks. It would also allow us to quantify our impression that
small scale structures are preferably located close to the confining
interfaces. If true, this would fit with the result by~\citet{smith-et-al:00}
that the high-frequency part of the shock spectrum is lost most efficiently.

The other point concerns run R5\_0.2.4. With corr$(\rho,v) \approx
-0.4$ $M_{\mathrm{rms}} \approx 0.9$, it violates two of the basic
assumptions we made in Sect.~\ref{sec:anal-scaling_2d}. Its mean
density is close to the isothermal value for strong shocks,
$\rho_{\mathrm{m}} \approx 22 \rho_{\mathrm{u}} \approx 0.9
\rho_{\mathrm{u}} M_{\mathrm{u}}^{2}$.  Both $\dot{\cal
  E}_{\mathrm{diss}}$ and $\dot{\cal E}_{\mathrm{drv}}$ increase with
$\ell_{\mathrm{cdl}}$.  With these characteristics, R5\_0.2.4 may mark
the transition from compressible supersonic turbulence, the topic of
this paper, to compressible subsonic turbulence.
\subsection{CDL and confining shocks: a coupled system}
The turbulence within the CDL is `naturally driven' in the sense that
we control neither what fraction of the total upstream kinetic energy,
$\rho_{\mathrm{u}} M_{\mathrm{u}}^{2}$, really enters the CDL nor the
spatial scale on which this energy input varies. Both are directly
determined by the confining shocks instead and indirectly depend on
the system as a whole. The driving efficiency at each confining shock
scales with $M_{\mathrm{rms}}$, even for situations where
$M_{\mathrm{l}} \ne M_{\mathrm{r}}$ (see
Sect.~\ref{sec:results_asym}). The auto-correlation length of the
confining shocks and the characteristic length scale of the turbulence
within the CDL are proportional to each other, both scaling as
$\ell_{\mathrm{cdl}} M_{\mathrm{u}}^{-0.6}$. We take these facts as
evidence that the CDL as a whole, its interface and volume properties,
forms a tightly coupled, quasi-stationary, and self-regulating system.
Back coupling between post shock flow and shock is also described in
other contexts, for example by~\citet{foglizzo:02} for the case of
Bondi-Hoyle accretion.

An aspect that remained elusive in Sect.~\ref{sec:num_results} is the
spatial scale on which the energy input varies, the energy injection
scale. To really tackle this issue, it would be necessary to analyze
the energy spectrum of the CDL. This task requires, however, some
caution because of the highly irregular boundary of the CDL, and we
postpone it for the moment. Nevertheless, we would like to present a
few thoughts on the subject.

A first question is whether it is justified to speak at all of only
one injection scale, of monochromatic driving. The homogeneous
upstream flow is modulated by the confining shocks. These are wiggled
on a variety of spatial scales at any given moment. This strongly
suggests that the kinetic energy input into the CDL is most likely not
monochromatic but occurs at a whole spectral range instead.
Consequences of such non-monochromatic driving have been studied, for
example, by~\citet{norman-ferrara:96}.

It also seems worthwhile to briefly look at monochromatically-driven
turbulence, in particular at the numerical simulations by~\citet{maclow:99}.
For the case of artificially, monochromatically driven hydrodynamic turbulence
in a 3D box with periodic boundaries, he found that the characteristic length
of the turbulence is proportional to the driving wave length, independent of
the Mach-number: $\lambda / \ell^{\mathrm{3d}}_{\mathrm{e_{\mathrm{kin}}}} =
1.42$, where $\lambda$ is the (known) driving wave length and
$\ell^{\mathrm{3d}}_{\mathrm{e_{\mathrm{kin}}}}$ is the 3D analogon of
$\ell_{\mathrm{e_{\mathrm{kin}}}}$ in Eq.~\ref{eq:lambda_ekin}.  In addition,
\citet{maclow:99} observed that
$\ell^{\mathrm{3d}}_{\mathrm{e_{\mathrm{kin}}}}$ increases with $\lambda$,
which is mirrored in the apparent increase in the structure size (patches,
filaments).

Although our setting clearly differs from that of~\citet{maclow:99},
two thoughts come to mind. The first is an actual observation,
namely that we also observe an increase in structure size with
$\ell_{\mathrm{e_{\mathrm{kin}}}}$. The second thought is more of a
question or speculation. \citet{maclow:99} determines the
proportionality constant between the characteristic scale of the
turbulence and the monochromatic driving wave length.  One may wonder
about the implications of this finding if not one driving wave length
is present but a whole spectrum.  How will the characteristic length
scale of the turbulence, which can still be determined following
Eq.~\ref{eq:lambda_ekin}, depend on this spectrum? And, given our
finding that $\ell_{\mathrm{e_{\mathrm{kin}}}} \propto
\ell_{\mathrm{corr}}$, what does $\ell_{\mathrm{corr}}$ tell us about
this spectrum? Both questions should become tractable once the energy
spectrum of the CDL is determined.
\subsection{A glimpse at astrophysics}
With regard to astrophysics, the presented work basically suggests
that, within the frame of isothermal hydrodynamics and a roughly plane
parallel setting, larger Mach-numbers of the colliding flows results
in a finer and finer network of higher and higher density contrast
within the interaction zone. In different types of wind-driven
structures, this connection between Mach-number and structure may be
directly observable.

For the clumping of line-driven hot-star winds, our results suggest
that the sheets or clumps formed by the instability of the
line-driving are not homogeneous but possess fine-scale substructure
with high density contrast.

Concerning molecular clouds, we first mention that recent arguments
support the idea, originally brought forward by~\citet{hunter:79} and
~\citet{larson:81}, that molecular clouds result from the collision of
large-scale flows in the ISM. \citet{basu-murali:01} make the point
that small-scale driving ($\approx$ 0.1 - 1 pc) of molecular clouds is
incompatible with observed total luminosities, unless the energy
dissipation rates derived from MHD simulations are seriously
overestimated. Using a principal component analysis of
$^{\mathrm{12}}$CO (J=1-0) emission, \citet{brunt:03} identifies
large-scale flows of atomic material in which the globally turbulent
molecular clouds are embedded. Similar observational results were
reported by~\citet{ballesteros-hartmann-vazquez:99}.

Driven supersonic turbulence as a structuring agent for the interior
of molecular clouds was examined by many authors
\citep{hunter-et-al:86, elmegreen:93, vazquez-passot-pouquet:95,
  maclow-klessen-burkert:98, ballesteros-hartmann-vazquez:99,
  ballesteros-et-al:99, maclow:99, hartmann-et-al:01, joung-maclow:04,
  burkert-hartmann:04, maclow-klessen:04, audit-hennebelle:05,
  heitsch-et-al:05, kim-ryu:05, vazquez-semadeni-et-al:06,
  ballesteros-paredes-et-al:06}.  The driving wave length of the
turbulence, and thus the largest structure size~\citep{maclow:99,
  ballesteros-maclow:02}, is usually a free parameter.  Our results
show instead that, at least for the case of an isothermal, shock
compressed, supersonically turbulent 2D slab, the structure size
rather depends on the size of the slab or cloud.
\subsection{Additional physics: an outlook}
The model presented in this paper covers only some very basic physics. To
  obtain results with a more direct relation to reality, additional physics
  must be included in the future, among these the following.  Strongly
asymmetric flows, where $M_{\mathrm{l}} \ne M_{\mathrm{r}}$, lead to more
complicated dependences, as we will demonstrate in a forthcoming paper. 
  Inclusion of radiative cooling, instead of assuming isothermal conditions,
  can affect the problem in different ways. Thermal instability can lead to
  additional dynamical effects~\citep{chevalier-imamura:82, gaetz-et-al:88,
    strickland-blondin:95, walder-folini:96, hennebelle-perault:99,
    hennebelle-perault:00, vazquez-semadeni-et-al:00, koyama-inutsuka:02,
    audit-hennebelle:05, heitsch-et-al:05, pittard-et-al:05, mignone:05}.  Extended
  cooling layers, on the other hand, tend to act as a cushion.  Simulations
  by~\citet{hyp:98} and~\citet{walder-folini:00}, which include radiative
  cooling but have otherwise similar parameters as some of the simulations
  presented here, show comparatively more small scale structure and even
  roll-ups at the interfaces confining the CDL. The CDL as a whole evolves
  less violently, and mean densities are about a factor of four to eight higher
  that what we found here for the isothermal case. Strongly asymmetric flows,
  where $M_{\mathrm{l}} \ne M_{\mathrm{r}}$, lead to a qualitatively different
  solution if radiative cooling is included~\citep{walder-folini:98} and to
  more complicated dependences on the upwind Mach-numbers in the isothermal
  case, as we will demonstrate in a forthcoming paper. The role of thermal
  conduction has only been considered by relatively few publications so
  far~\citep{begelman-mckee:90, myasnikov-zhekov:98, koyama-inutsuka:04}.
Global bending of the interaction zone affects the stability properties of the
interaction zone as a whole and thus probably also its interior properties.
In colliding wind binaries, for example, matter is transported out of the
central part of the system and diluted in the outer part.  Simulations of bow
shocks and colliding winds in binaries show strong traveling waves, together
with a systematic change of the mean properties in the flow off from the
stagnation point~\citep{stevens-et-al:92,rolf-doris:95,blondin-koerwer:98}.
\section{Summary and conclusions}
\label{sec:conc}
We looked at symmetric, supersonic ($5 \lapprox M_{\mathrm{u}}
\lapprox 90$), isothermal, plane-parallel, colliding flows in 2D.  The
resulting shock-confined interaction zone (CDL) is supersonically
turbulent ($1 \lapprox M_{\mathrm{rms}} \lapprox 10$). We investigated
the CDL and its interplay with the upstream flows by dimensional
analysis and numerical simulations.  The latter we generally stopped
when $\ell_{\mathrm{cdl}} \approx \mathrm{Y} / 2$. The results are
interesting not only with regard to flow collisions, but also shed new
light on the properties of supersonic turbulence in general.

The numerical simulations show that the CDL has an irregular shape and a
patchy, supersonically turbulent interior.  The driving of the turbulence is
natural in that it depends on the shape of the confining shocks.  The
dimensional analysis is based on isothermal Euler equations in infinite space.
Within this frame, a self-similar solution may exist that would depend on
$M_{\mathrm{u}}$ but must not depend on $\ell_{\mathrm{cdl}}$. Relations for
average quantities are obtained under some further simplifying assumptions
(Sect.\ref{sec:expectedrelations}).

Based on both the analytical and numerical results, we arrive at the following
conclusions.

1) Comparison of the numerical and the self-similar solution shows
generally good agreement if $M_{\mathrm{rms}} \gapprox 1$. The modest
deviation between the numerical and the self-similar solutions
increases with $\ell_{\mathrm{cdl}}$. We suggest some explanations for
the deviation, but our data do not allow any clear conclusions on the
issue. For $M_{\mathrm{rms}} \lapprox 1$, we have but one simulation.
It shows clear differences to the other runs and may be more
characteristic of compressible subsonic turbulence than of
supersonic turbulence.

2) The CDL is characterized by $M_{\mathrm{rms}} \approx
\eta_{\mathrm{1}}^{-1/2} M_{\mathrm{u}}$ and $\rho_{\mathrm{m}}
\approx \eta_{\mathrm{1}} \rho_{\mathrm{u}}$.  The average compression
ratio of the CDL is thus independent of $M_{\mathrm{u}}$. This is in
sharp contrast to the 1D case, where $\rho_{\mathrm{m,1d}} =
M_{\mathrm{u}}^{2}\rho_{\mathrm{u}}$. From the numerical simulations,
we find $\eta_{\mathrm{1}} \approx 30$.

3) The turbulence within the CDL and the driving efficiency are related by
$f_{\mathrm{eff}} = 1 - M_{\mathrm{rms}}^{-0.6}$. The relation also holds for
asymmetric settings, where $M_{\mathrm{l}} \ne M_{\mathrm{r}}$, emphasizing
the mutual coupling between volume and interface properties. For larger
upstream Mach-numbers, the shocks confining the interaction zone are less
strongly inclined with respect to the direction of the upstream flows. The
driving is more efficient, a larger fraction of the upstream kinetic energy is
dissipated only within the CDL and not already at the confining shocks.

4) The characteristic length scale of the turbulence,
$\ell_{\mathrm{e_{\mathrm{kin}}}}$, and the auto-correlation length of
the confining shocks, $\ell_{\mathrm{corr}}$, are proportional to each
other. Both scale as $\ell_{\mathrm{cdl}} M_{\mathrm{u}}^{-0.6}$, this
although the former is based on volume quantities while the latter is
derived from interface properties.

5) The separation of filaments and the size of patches within the CDL
both get larger as $\ell_{\mathrm{cdl}}$ increases and/or
$M_{\mathrm{u}}$ decreases.

For increasing upstream Mach-numbers in summary we thus expect a faster
expanding CDL with less strongly inclined confining interfaces with respect to
the direction of the upstream flows, similar mean density, finer interior
structure relative to the CDL size, and a gradual shift of the energy
dissipation from the confining shocks to internal shocks within the CDL.  We
expect to observe these general dependencies in real objects where
shock-confined slabs play a role, like molecular clouds, wind driven
structures, supernova remnants, or $\gamma$-ray bursts.
\begin{acknowledgements}
  The authors wish to thank the crew running the Cray SV1 at ETH
  Z\"{u}rich, where the simulations were performed, the system
  administrator of the institute for astronomy, ETH Z\"{u}rich, P.
  Steiner, for steady support, and J. Favre from the Swiss Center of
  Scientific Computing CSCS/SCSC, Manno, for graphics support. The
  authors also would like to thank the referee, E. Vazquez-Semadeni,
  for the detailed and engaged report.
\end{acknowledgements}
\newpage
%
%
%
%
%%%%%%%%%%%%%%%%%%%%%%%%%%%%%  Bibliography  %%%%%%%%%%%%%%%%%%%%%%%%%%%
%
% 
%
\bibliographystyle{apj} 
\bibliography{3898} 
\appendix
\section{Numerical computation of obliqueness angle}
\label{app:alpha}
While shocks are smeared over approximately 3 grid cells in our
simulations, the confining shocks in our analysis are specified as a
series of discrete x,y-coordinate pairs only (see
Sect.~\ref{sec:confshocks}).  This information is sufficient to
compute most shock-related quantities to good accuracy, for example
the shock length $\ell_{\mathrm{sh}}$. The only quantity that requires
a more careful proceeding is the obliqueness angle $\alpha$.  If it
were computed directly from the discrete shock positions, only
discrete values would be obtained, for example 0$^{\circ}$,
45$^{\circ}$, 63.4$^{\circ}$ etc. for one-sided differences.
  
To compute the obliqueness angle $\alpha_{\mathrm{i}}(y_{\mathrm{j}})$ (see
Fig.~\ref{fig:sketch2d} and Sect.~\ref{sec:confshocks}) at each position
$y_{\mathrm{j}}$, $1\le j \le J$, of the left and right shock
($s_{\mathrm{l}}$ and $s_{\mathrm{r}}$), we proceed as follows. In a first
step, we use spline interpolation to double the number of points in the
y-direction along the shock front.  Next, we smooth the shock front slightly,
using a running mean with an averaging window of $\pm 5$ points (this
corresponds to an averaging window of $\pm 2.5$ points in the original data.
Then we compute the derivative at each point of this smoothed shock front,
using a 3-point Lagrangian interpolation.  To avoid abrupt changes in the
derivative from one point to the next, we smooth it again by a running mean with
averaging window $\pm 5$ points. We finally obtain the obliqueness angle
$\alpha_{\mathrm{i}}(y_{\mathrm{j}})$, $1 \le j \le 2J$, as the arctan of the
derivative.

We checked that the size of the averaging window ($\pm 3$ points or $\pm 7$
points) has only a marginal effect on the angle distribution and the driving
efficiency. For the latter, which is an integral over both shocks, tests show
that $\alpha$ can even be computed directly from the discrete positions.
\section{List of runs, their parameters, and naming schemes}
\label{app:list_of_runs}
\begin{table*}[h]
\caption{List of performed simulations.}
\begin{center}
\begin{tabular}{lcccccccc} 
\hline
\hline 
 label                            & $M_{\mathrm{u}}$                          & $\ell( N )$                                   & 
 $\ell_{\mathrm{cdl}}/\mathrm{Y}$ & $M_{\mathrm{rms}}$ & $\frac{\rho_{\mathrm{m}}}{\rho_{\mathrm{u}}}$      & 
 $\frac{\ell_{\mathrm{sh}}}{Y}$   & $f_{\mathrm{eff}}$                        \\
\hline 
\multicolumn{8}{c}{Symmetric runs, no CDL at t=0} \\
\hline
R5\_0.2.4     & 5.42  &  91 & 1.07 &  0.90 &    24 &   1.1 &   0.16  \\ % tihhf\_2y       \\
R11\_0.2.4    & 10.85 &  88 & 0.59 &   2.2 &    33 &   1.5 &   0.35  \\ % tihf\_2y        \\
R22\_0.2.4    & 21.7  &  86 & 0.30 &   4.6 &    30 &   2.6 &   0.59  \\ % ti\_2y          \\
R33\_0.2.4    & 32.4  &  86 & 0.50 &   6.9 &    26 &   3.6 &   0.70  \\ % ti1.5f\_2y      \\
R43\_0.2.4    & 43.4  &  88 & 0.55 &   9.1 &    29 &   4.6 &   0.76  \\ % ti2f\_2y        \\
R87\_0.2.4    & 86.8  & 105 & 0.82 &   15. &    23 &  12.1 &   0.89  \\ % ti4f\_2y        \\
\hline
R22\_0.4.4    & 21.7  &  41 & 0.25 &   4.3 &    35 &   2.3 &   0.55  \\ % ti\_2y\_fine    \\
R22\_0.1.4    & 21.7  &  88 & 0.74 &   5.0 &    26 &   2.7 &   0.62  \\ % ti\_2y\_coarse  \\    
R43\_0.1.4    & 43.4  &  90 & 0.59 &   8.9 &    32 &   4.1 &   0.76  \\ % ti2f\_2y\_coarse\\    
\hline
R11\_0.2.2    & 21.7  &  89 & 1.10 &   2.2 &    33 &   1.4 &   0.33  \\ % tihf            \\
R22\_0.2.2    & 21.7  & 307 & 0.79 &   4.7 &    28 &   2.6 &   0.59  \\ % ti\_new         \\
R33\_0.2.2    & 32.4  &  82 & 1.45 &   6.7 &    30 &   3.6 &   0.70  \\ % ti\_1.5fast     \\
R43\_0.2.2    & 43.4  &  73 & 1.09 &   9.4 &    27 &   4.7 &   0.76  \\ % ti\_2fast       \\
\hline
R22\_0.2.6    & 21.7  & 190 & 0.84 &   4.7 &    29 &   2.6 &   0.60  \\ % ti\_3y          \\
\hline 
\multicolumn{8}{c}{Symmetric runs, with CDL at t=0} \\
\hline
R22\_1.2.2    & 21.7  &  87 & 0.83 &   3.3 &    61 &   1.9 &   0.50  \\ % cci              \\
R22\_1.2.1    & 21.7  & 111 & 1.33 &   3.2 &    68 &   1.8 &   0.46  \\ % cci\_half        \\
R22\_1.4.4    & 21.7  & 199 & 0.72 &   3.4 &    59 &   1.9 &   0.49  \\ % cci\_2y          \\
R22\_1.4.2    & 21.7  &  68 & 0.40 &   2.8 &    91 &   1.6 &   0.39  \\ % cci\_fine        \\
R22\_1.1.2    & 21.7  & 115 & 1.21 &   3.9 &    42 &   2.4 &   0.59  \\ % cci\_coarse      \\
\hline  
R22\_2.2.2    & 21.7  & 313 & 1.44 & (2.4) & (109) & (1.5) & (0.34)  \\ % dddoubl\_WI      \\
R22\_2.4.2    & 21.7  & 186 & 0.37 & (1.8) & (253) & (1.2) & (0.24)  \\ % dddoubl\_wi\_.hr2\\
R22\_2.8.2    & 21.7  &  92 & 0.14 & (1.4) & (281) & (1.2) & (0.21)  \\ % dddoubl\_wi\_.hr4\\
\hline 
%R22\_0.2.2    & 21.7  & 122 & 1.65 &   4.7 &    28 &   2.5 &   0.58 &  0.58  \\ % ti              \\
%
%
\end{tabular}
\end{center}
\label{tab:list_of_runs}
\end{table*}

\end{document}